\documentclass[a4paper,11pt,openany]{book}

\usepackage{glo}
\usepackage[dvips]{graphicx}
\usepackage{makeidx}
\usepackage{rotating}
\usepackage{subfigure}    
\usepackage{latexsym}
\usepackage{array} 
\usepackage{cite} 
\usepackage{index}
\usepackage{times}
\usepackage{pifont}
\usepackage{color}
\usepackage{citesort}
\usepackage{comment}
\usepackage{glossaries}
\usepackage{tikz}
\usepackage{calc}
\usetikzlibrary{calc}
\usepackage{setspace}
\usepackage{bm}
\usepackage{mathtools}

\usepackage[hang,small,bf]{caption}
\usepackage[Lenny]{fncychap}
\usepackage{fancyhdr} 
\usepackage{pstricks}
\usepackage{psfrag}
\usepackage{epic}
\usepackage{eepic}
\usepackage{eepicemu}
\usepackage{chgbar}
\usepackage{times}

\usepackage{float}
\usepackage{amssymb}
\usepackage{amsmath}
\usepackage[english]{babel}
\usepackage{ifthen}
\usepackage{graphics}
\usepackage{afterpage}

\usepackage{graphicx}		
\usepackage{multirow}
\usepackage{array}
\usepackage{amssymb, amsmath}
\usepackage{cite}	
\usepackage{arydshln}
\usepackage{comment}
\usepackage{enumerate}
\usepackage{amsmath}

\usepackage{color}
\usepackage{tabularx}
\usepackage{tabulary}

\usepackage{epsf}
\usepackage{subfigure}
\usepackage{array}
\usepackage[amsmath,thmmarks]{ntheorem}
\usepackage{amsfonts}
\usepackage{dsfont}
\usepackage{xfrac}
\usepackage{bbm}
\usepackage{rotating}
\usepackage{lipsum}

\newcommand{\be}{\begin{equation}}
\newcommand{\ee}{\end{equation}}
\newcommand{\bea}{\begin{eqnarray}}
\newcommand{\eea}{\end{eqnarray}}

\renewcommand{\baselinestretch}{1.0}

\makeatletter
\newcommand{\chapterauthor}[1]{%
  {\parindent0pt\vspace*{-25pt}%
  \linespread{1.1}\large\scshape#1%
  \par\nobreak\vspace*{35pt}}
  \@afterheading%
}
\makeatother

\theoremseparator{.}
\usepackage[ruled, lined, linesnumbered, commentsnumbered, longend]{algorithm2e}
\usepackage{algorithmic}

\newtheorem{example}{Example}
\newtheorem*{solution}{Solution}

\newtheorem{remark}{Remark}

\DeclarePairedDelimiterX\braket[2]{\langle}{\rangle}{#1\,\delimsize\vert\,\mathopen{}#2}

\makeindex

\makeglossary

\newindex{aut}{adx}{and}{Author Index}

\newindex{cite}{cdx}{cnd}{Citation Index}            

\newcommand{\ncite}[1]{\cite{#1}\index[cite]{#1}}

\title{\bf Doppler Effect: Analyses and Applications in Wireless Sensing and Communications}

\author{\\~~\\{Lie-Liang Yang}\footnote{Lie-Liang Yang is the professor of wireless communications, with the School of Electronics and Computer Science (ECS) in the University of Southampton, UK. This document is a chapter of my next book to be published. If you have any comments, please email: lly@ecs.soton.ac.uk, which is highly appreciated.}}
\date{~}

\begin{document}
\pagenumbering{roman}

\maketitle

\renewcommand{\baselinestretch}{1.}


\newpage

{\bf Abstract}: This chapter is motivated by the need for a rigorous and comprehensive
analysis of the Doppler effects encountered by electromagnetic and
acoustic signals across a diverse spectrum of modern
applications. These include land mobile communications, various
Internet of Things (IoT) networks, machine-type communications (MTC),
and various radar and satellite-based systems for navigation and
sensing, as well as the emerging regime of integrated sensing and
communications (ISAC). A wide array of kinematic profiles is
investigated, ranging from uniform motion and constant acceleration to
more complex general motion. Consequently, the multi-faceted factors
influencing the Doppler shift are addressed in detail, encompassing
classical kinematics, special and general relativity, atmospheric
dynamics, and the properties of the propagation medium. This work is
intended to establish a definitive theoretical foundation for both the
general enthusiast and the specialized researcher seeking to master
the complexities of signal frequency shifts in modern wireless sensing
and communications systems.\\


\newpage

\tableofcontents
\renewcommand{\baselinestretch}{1.}

\clearpage


\pagenumbering{arabic}


\section{Introduction}\label{section-Doppler-FS-Introduction}

Traditionally, radar and wireless communications are operated in
different frequency bands and implemented with different hardware
platforms, resulting in the separation of their research and
development.  Correspondingly, in the majority of textbooks and
research-oriented books in wireless communications, such as
~\cite{book:Wireless-Rappaport-2ed,book:Wireless-Parsons,book:Mobile-Hanzo,book:Gordon-Stuber,Andrea-Goldsmith-A,Proakis-5th,Henry-Propagation-book,5635453,book-W-C-Y-Lee},
to list a few, the principles of the Doppler effect are mainly
presented in association with one to three scenarios. The first one is
the non-relativistic uniform motion of signal source or/and receiver,
which is the most common scenario considered. The second scenario is
the relativistic uniform motion between source and receiver, which is
often learned from physics classes. The third scenario is the
non-relativistic uniform motion in medium like water. The fact is
that, the practical scenarios generating Doppler effect are much
richer and their analyses are also more challenging.

This was not an issue in the era when wireless communications was operated in relatively low frequency bands, such as sub-6GHz, and separated from radar, which may more generally be referred to as
wireless sensing. This is because the design and operation of
traditional wireless communication systems concern mostly about the
maximum Doppler shift and Doppler spectrum experienced in a
system. However, as wireless communications enter the era of
information transmission in millimeter wave (mmWave), terahertz, and
even optical bands, especially when communication and sensing
functions are jointly implemented in the same system, the knowledge
about Doppler effect at the ray level becomes increasingly
critical. The reason behind is that in mmWave, terahertz and optical
bands, the number of propagation paths from a transmitter to its aimed
receiver is very limited, resulting in a sparse channel where signal
rays conveyed over different paths are well separated in space and
time. In this kind of channels, it is desirable that the Doppler
effect experienced by individual rays can be estimated. Moreover, when
sensing, such as velocity estimation of environmental moving objects,
is implemented with communications, the Doppler shifts induced by the
individual environmental objects need to be distinguished and
estimated.

For example, in satellite navigation, there are various factors
resulting in Doppler effect. The kinematics of satellite and ground
device result in not only the classic Doppler shift but also the
time-dilation induced Doppler shift. The difference of gravity
potentials between satellite and ground device also induces
time-dilation, adding another Doppler shift. Furthermore, a satellite
navigation signal passing through the atmosphere experiences the path
bending resulted from the density increasing air, the ion's dynamics
resulted by solar flares, geomagnetic storms, etc., and many other
random events, including wind, cloud, rain and varying-pressure. To
achieve the most accurate sensing for the velocity of ground device,
errors introduced by all the above-mentioned factors, except that by
the device itself, must be corrected to a best level, which demands
knowledge of the Doppler effects by these different factors.

To bridge this knowledge gap within the wireless communications field,
this chapter provides a comprehensive analysis of the Doppler effect
across a broad spectrum of scenarios. We transition from conventional
models to modern wireless contexts, examining motions ranging from
uniform to general, and cases from non-relativistic to relativistic
regimes. The analysis further scales from individual Doppler events to
the complex interplay of simultaneous multi-event phenomena. Given the
movement towards integrated space-air-ground networks
(ISAGN)~\ncite{10745905,8368236,8961915}, we also characterize the
specific Doppler effects induced by the Earth’s
atmosphere. Furthermore, we explore two practical applications to
demonstrate the utility of Doppler analysis in wireless
sensing. Finally, a comparative study of the acoustic Doppler effect
is presented, emphasizing its fundamental physical distinctions from
the effects experienced by radio signals.

\section{Theoretical Basis of the Doppler Effect}\label{section-Doppler-FS.01a}

In physics, frequency is defined as the number of occurrences of a
periodic, such as a harmonic, wave within a unit of time, typically in
one second. In electromagnetic (EM) wave based communications,
frequency can be measured in different ways, which can be explained
from the basic EM wave functions. Specifically, a transverse electric wave
propagating in $z$-direction while polarized in $x$-direction can be
written as
\begin{align}
\bm{E}_x(z,t)=&A\cos\left(\omega t-\kappa z \right)\bm{i}\nonumber\\
\label{eq:Doppler-FS-15x7}
=&A\cos\left(2\pi ft-\frac{2\pi }{\lambda}z \right)\bm{i}\\
=&A\cos\left(2\pi ft-\frac{2\pi }{cT}z \right)\bm{i}\nonumber
\end{align}
which is a function of both time $t$ and position $z$. In
\eqref{eq:Doppler-FS-15x7}, $A$ is magnitude, $\bm{i}$ is a unit
vector pointing in $x$-direction, $\omega=2\pi f$ is angular frequency
representing the rate of phase change with time $t$,
$\kappa=2\pi/\lambda$ is wave number representing the rate of phase
change with distance $z$, $f$, $\lambda$ and $T$ are frequency,
wavelength and period, respectively. The wave propagation speed in
$z$-direction is
\begin{align}\label{eq:Doppler-FS-15x8}
c=\lambda f
\end{align}
Furthermore, $T$ and $\lambda$ are connected via the
formula $\lambda=cT$. Substituting it into \eqref{eq:Doppler-FS-15x8}
gives $f=1/T$. Hence, \eqref{eq:Doppler-FS-15x7} also has the form of
\begin{align}\label{eq:Doppler-FS-15x9}
\bm{E}_x(z,t)=A\cos\left(2\pi ft-\frac{2\pi f z}{c} \right)\bm{i}
\end{align}

Note that in the above formulas, the constant $c$ usually representing
by fault the speed of light in vacuum free space is directly used. It
will be continuously used in this way in this chapter, provided that
it does not generate confusion. However, we should emphasize that most
EM-wave based signals propagate in media, such as air, water, where
the speed of wave propagation is smaller than the speed of light in
vacuum. For example, the speed of light in vacuum is $c=299,792,458$
meters per second (m/s), while the speed of EM waves, including light,
in air is about $c'=299,702,547$~m/s, giving a difference of about
$89,911$~m/s. Hence, in the following contents, when distinction
between the speed of wave in media and that in vacuum is needed, $c'$
is used to indicate the speed of wave in media, while $c$ is kept as
the constant of speed of light in vacuum.

According to \eqref{eq:Doppler-FS-15x7} and \eqref{eq:Doppler-FS-15x9},
to estimate frequency $f$, in addition to directly estimating $f$,
we can instead estimate $T$. Furthermore, when $c$ is given, we can
estimate frequency $f$ via estimating the wavelength $\lambda$.

Write the phase in \eqref{eq:Doppler-FS-15x9} as
\begin{align}\label{eq:Doppler-FS-16x0}
\phi(z,t)=2\pi ft-\frac{2\pi f z}{c}
\end{align}
Then, $f$ is also given by the formulas
\begin{align}\label{eq:Doppler-FS-16x1}
f=\frac{1}{2\pi}\frac{\partial\phi(z,t)}{\partial t}=-\frac{c}{2\pi}\frac{\partial\phi(z,t)}{\partial z}
\end{align}
Hence, information about frequency $f$ can be obtained via observing the phase change with respect to time  or space. 

Doppler effect accounts for the phenomenon - first described by an
Austrian scientist Christian Doppler in the 19th century - that the
wave frequency or period perceived by a receiver or an
observer\footnote{Receiver and observer may be alternatively used in
the forthcoming analysis of Doppler effect without distinction. Also,
transmitter and source may alternatively be used.} is different
from the frequency or period of source wave. Doppler effect occurs
when wave source, where wave emits, and wave receiver (observer),
where wave is measured, are in relative motion, or/and when the
propagation environment between wave source and receiver, such as due
to reflectors, is time-varying. For example, if a stationary observer
stays at $z=r_0$, the phase of \eqref{eq:Doppler-FS-16x0} is
$\phi(r_0,t)=2\pi ft-\frac{2\pi f r_0}{c}$ and the frequency $f$ can
be ideally recovered by the first equation in
\eqref{eq:Doppler-FS-16x1}. If observer moves by following the equation
$z=r_0+vt$, where $v$ is velocity, the phase varies according to
\begin{align}\label{eq:Doppler-FS-3.6}
\phi(z,t)=2\pi ft-\frac{2\pi f (r_0+vt)}{c}
\end{align}
Then, the measured frequency by observer is
\begin{align}\label{eq:Doppler-FS-3.7}
f'=\frac{1}{2\pi}\frac{\partial\phi(z,t)}{\partial t}=f-\frac{fv}{c}
\end{align}

Furthermore, if the motion of observer follows the equation $z=r_0+v_0t+\frac{1}{2}at^2$, where $v_0$ is initial velocity and constant $a$ is acceleration, by following the steps of \eqref{eq:Doppler-FS-3.6} and \eqref{eq:Doppler-FS-3.7}, we obtain
\begin{align}\label{eq:Doppler-FS-3.8}
f(t)=f-\frac{f(v_0+at)}{c}
\end{align}
which is time dependent. In both cases of \eqref{eq:Doppler-FS-3.7} and \eqref{eq:Doppler-FS-3.8}, the frequency measured by observer is different from that emitted by source.

In general, let the geometric propagation path from transmitter to receiver be expressed as $P(t)$. Then, when without considering relativity, and when the propagation path length $r(t)=|P(t)|$ from transmitter via propagation environment to receiver is time-variant, then the frequency measured at the receiver follows the formulas
\begin{subequations}\label{eq:Doppler-FS-3.9t}
\begin{align}
\label{eq:Doppler-FS-3.9u}
f'=&f-\frac{f}{c}\times\frac{dr(t)}{dt}\\
\label{eq:Doppler-FS-3.9v}
=&f-\frac{1}{\lambda}\times\frac{dr(t)}{dt}\\
\label{eq:Doppler-FS-3.9q}
=&f-\frac{fv(t)}{c}
\end{align}
\end{subequations}
where $v(t)=dr(t)/dt$ is the relative moving speed between transmitter and receiver.

Assume that source emits a harmonic wave within one uniform medium or
in free space. Then, the level of Doppler effect can be reflected by
either of the ratios expressed as
\begin{align}\label{eq:Doppler-FS-16x2}
\frac{T}{T'},~~\frac{f'}{f}
\end{align}
where $T$ and $f$ are respectively the period and frequency of source emitted wave, and correspondingly, $T'$ and $f'$ are the perceived period and frequency by
receiver. Hence, for convenience, ${T}/{T'}$ and ${f'}/{f}$ can either be referred to as the Doppler effect
factor (DEF) or simply Doppler effect. Furthermore,
\begin{align}\label{eq:Doppler-FS-16x3}
f_D=f'-f
\end{align}
represents the direct measurement of the Doppler frequency shift,
which is often referred to as Doppler frequency or Doppler shift for
short. Following \eqref{eq:Doppler-FS-3.9t}, we have the corresponding
formulas for Doppler frequency represented as
\begin{subequations}\label{eq:Doppler-FS-3.12t}
\begin{align}
\label{eq:Doppler-FS-3.12u}
f_D=&-\frac{f}{c}\times\frac{dr(t)}{dt}\\
\label{eq:Doppler-FS-3.12v}
=&-\frac{1}{\lambda}\times\frac{dr(t)}{dt}\\
\label{eq:Doppler-FS-3.12q}
=&-\frac{fv(t)}{c}
\end{align}
\end{subequations}
where positive $v(t)=dr(t)/dt$ explains that the length of propagation path from transmitter to receiver increases with time, hence resulting in negative Doppler shift.

From the above analysis we should emphasize that it is the time-varying distance, i.e., $r(t)$, of the propagation path, i.e., $P(t)$, from transmitter to receiver that generates the Doppler effect. If $r(t)=r_0$ is fixed, there is no Doppler effect. Otherwise, any phenomenas making $P(t)$, and hence $r(t)$, time-dependent generate Doppler effect. To reflect this, $P(t)$ is terminologically referred to as the {\em phase path} from transmitter to receiver. In further detail, a phase path from transmitter to receiver can be represented as ${P}(n,t)$ to account for both the physical signal propagation path, expressed as $\mathcal{P}(t)$, and the optical properties of propagation media on the path. Here, $n(p,t)$ is medium's refractive index, which may be a function of $p$ of the points on $\mathcal{P}(t)$, and may also be time-variant. The reason for having such an expression of ${P}(n,t)$ can be explained with the aid of the $z$-relied component in \eqref{eq:Doppler-FS-16x0}. Specifically, when transmitting in the media with refractive index $n(t)$ that is independent of $\mathcal{P}(t)$, the phase contributed by this component is
\begin{align}\label{eq:Doppler-FS-3.13u}
-\frac{2\pi fr(t)}{c/n(t)}=-\frac{2\pi fn(t)r(t)}{c}=-\frac{2\pi fP(n,t)}{c}
\end{align}
where $c/n(t)$ is the phase velocity of wave in medium and $P(n,t)=n(t)r(t)$. Correspondingly, following \eqref{eq:Doppler-FS-3.12t}, the Doppler frequency is
\begin{align}\label{eq:Doppler-FS-3.14u}
f_D=&-\frac{f}{c}\frac{dP(n,t)}{dt}=-\frac{f}{c}\left[n(t)\frac{dr(t)}{dt}+r(t)\frac{dn(t)}{dt}\right]
\end{align}
demonstrating that Doppler effect may be generated by the dynamics of physical propagation path, propagation medium, or both.   

In practice, the refractive index $n$ is often correlated with the physical position in medium. For example, when a satellites sends a radio signal to a ground station, the refractive index $n$ experienced by a ray of signal increases in the propagation direction, as the result that the density of air increases. In this case, $P(n,t)$ should be represented as\footnote{In \eqref{eq:Doppler-FS-3.15u}, Tx and Rx are for transmitter (Tx) and receiver (Rx), respectively.}
\begin{align}\label{eq:Doppler-FS-3.15u}
P(n,t)=&\lim_{\Delta p\rightarrow 0}\sum_{m=\textrm{Tx}}^{\textrm{Rx}}n(m\Delta p,t)\Delta p\nonumber\\
=&\int_{\mathcal{P}(t)}n(p,t)dp
\end{align}
where the integration is along the signal's propagation path $\mathcal{P}(t)$ from transmitter to receiver. Here, we should emphasize that the path $\mathcal{P}(t)$ is not necessary a straight line, but may bend due to refraction. Then, corresponding to \eqref{eq:Doppler-FS-3.14u}, 
\begin{align}\label{eq:Doppler-FS-3.16u}
f_D=&-\frac{f}{c}\frac{dP(n,t)}{dt}=-\frac{f}{c}\frac{d}{dt}\int_{\mathcal{P}(t)}n(p,t)dp
\end{align}
We will return this issue in Section~\ref{section-Doppler-FS.5} when the Doppler effect by Earth's atmosphere is analyzed.

It is worth noting that although we have the formulas for deriving the
Doppler effect from $dr/dt$, as seen in \eqref{eq:Doppler-FS-3.9t}
and \eqref{eq:Doppler-FS-3.12t}, or more generally, from $dP/dt$, as shown in  \eqref{eq:Doppler-FS-3.14u} and \eqref{eq:Doppler-FS-3.16u}, the results obtained in such ways for some cases may not be accurate. This is because $dr/dt$ (or $dP/dt$) implies $dt\rightarrow
0$. However, Doppler effect is explained by the change of wave's
period, which has the minimum time interval of $T$. Hence, unless $T$
is indeed close to zero, the actual Doppler effect would appear
difference from that predicted by applying $dr/dt$ (or $dP/dt$). Generally, in principle, if $T\times dr/dt$ (or/and $T\times dP/dt$) is close to zero, the Doppler effect predicted
from the $dr/dt$  (or $dP/dt$) related formulas should provide sufficiently close approximation for the true Doppler effect.

Below we analyze the Doppler effects in some scenarios, when
classic and relativistic kinematics, gravity, atmosphere, etc., are considered. The analysis
mainly assumes EM waves, while the Doppler effect of acoustic waves is
finally treated in Section~\ref{section-Doppler-FS.5}.

Note that, in some motion scenarios, close zone and remote zone
are distinguished. When receiver (or observer) falls in the remote zone of
source, wave propagation is approximated to be in parallel, resulting
in plane wave, and wave analysis does not involve explicitly the
distance between source and receiver.  By contrast, when receiver
falls in the close zone of source, wave cannot be treated as plane
wave but is spherical. In this case, the distance between source and
receiver must be taken into account in wave analysis. Note furthermore
that, close zone and remote zone can be divided according to the
distance $r$ between source and receiver relative to wavelength
$\lambda$. For example, $3\%$-zoning means that in the close zone,
$\lambda/r\geq 3\%$, and in the remote zone, $\lambda/r< 3\%$. As the
results, when source and receiver are in each other's remote zones,
their relative motion of one wavelength does not generate noticeable
effect on the involved vectors, including moving velocity and the
position vector of receiver referenced to source, or vice versa.

\section{Doppler Effect in Uniform Motion}\label{section-Doppler-FS.1}

In this section, several methods for analyzing the Doppler effect are
introduced by considering the simplest uniform motion between wave
source and its receiver (observer). For clarity of explanation, a unit vector,
$\hat{\bm{r}}$, is defined to point in the direction from source to
receiver, the velocity of source or receiver is
$\bm{v}=v\hat{\bm{v}}$, where $\hat{\bm{v}}$ is a unit vector in
the direction of $\bm{v}$ and $v$ is the magnitude, i.e.,
speed. Furthermore, the angle between $\hat{\bm{r}}$ and
$\hat{\bm{v}}$ is expressed as $\theta$, governed by the relation of
\begin{align}\label{eq:Doppler-FS-16x4}
\cos\theta=\hat{\bm{r}}\cdot \hat{\bm{v}},~0\leq \theta <\pi
\end{align}
where $\cdot$ is inner product operation.  

Note that, when considering Doppler effect, a wave source may be the
communication transmitter or a scatter (reflector) in communication
channel. Similarly, a wave receiver may be the communication receiver
or a scatter (reflector) in communication channel.

Also note that in the analysis of classic Doppler effect, a rest
reference frame, in addition to transmitter and receiver, is implied,
which can usually be the medium, such as air, water, in which wave
propagates. The motion states of transmitter, receiver or objects in
environments are defined relative to this rest frame.

\subsection{Non-Relativistic Doppler Effect in Remote Zones}\label{subsubsection-Doppler-FS.1.1}

\begin{figure}[tb]
  \begin{center}
 \includegraphics[angle=0,width=.99\linewidth]{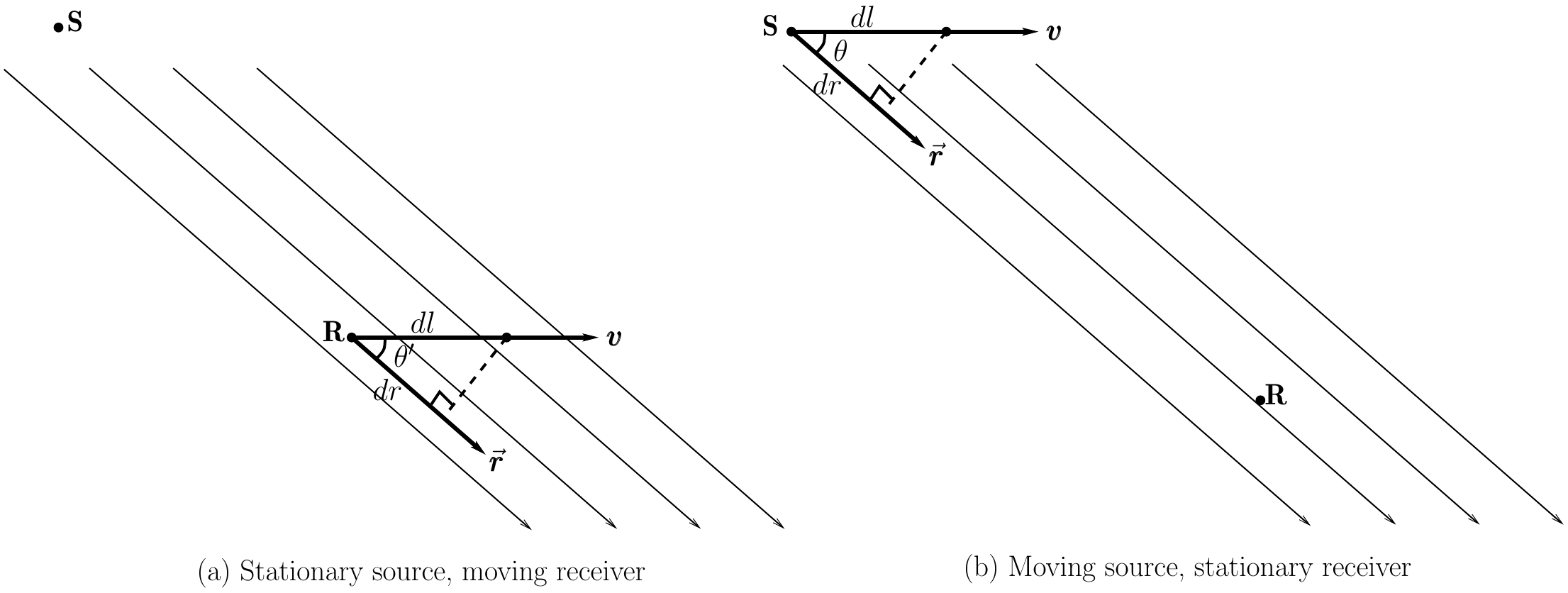}
  \end{center}
  \caption{A plane wave generated by a remote-zone static (or moving)
    source is received by a moving (or static) receiver. }
  \label{figure-Doppler-effect-UV-far-field}
\end{figure}

\subsubsection{Stationary Source and Moving Receiver}\label{subsubsection-Doppler-FS.1.1.1}

First, assume that wave source is stationary and receiver is moving
with velocity $\bm{v}$, as shown in
Fig.~\ref{figure-Doppler-effect-UV-far-field}(a). Consider that source
emits two adjacent maxima (wave crests) at $t_1=t_0$ and
$t_2=t_0+T$. Correspondingly, the receiver receives two maxima at
$t'_1=t_0'$ and $t'_2=t_0'+T'$. During $T'$ of a period of received
wave, the distance travelled by receiver in $\hat{\bm{r}}$ direction,
which is wave's propagation direction, is given by $dr=(\bm{v}\cdot
\hat{\bm{r}})T'=v\cos\theta' T'$. Hence, the period of received wave
satisfies
\begin{align}\label{eq:Doppler-FS-165}
T'=t'_2-t'_1=T+\frac{dr}{c}=T+\frac{v\cos\theta' T'}{c}
\end{align}
making
\begin{align}\label{eq:Doppler-FS-16x6}
\frac{T}{T'}=1-\frac{v\cos\theta'}{c}
\end{align}
and the received wave frequency and wavelength be
\begin{align}\label{eq:Doppler-FS-16x7}
  f'=&f\left(1-\frac{v\cos\theta'}{c}\right)\\
  \label{eq:Doppler-FS-16x8}
  \lambda'=&{\lambda}\left({1-\frac{v\cos\theta'}{c}}\right)^{-1}
\end{align}
respectively. \eqref{eq:Doppler-FS-16x8} is obtained under the constraint of $c=f'\lambda'=f\lambda$.  Accordingly, the Doppler shift is
\begin{align}\label{eq:Doppler-FS-16x9}
f_D(v,\theta')=& -\frac{fv\cos\theta'}{c}=-\frac{v\cos\theta'}{\lambda}
\end{align}

As shown in Fig.~\ref{figure-Doppler-effect-UV-far-field}(a), when
receiver moves away from source, $0\leq \theta<\pi/2$, the received
frequency is lower than the source emitted frequency. Otherwise, if
receiver moves towards source, $\pi/2\leq \theta<\pi$, the received
frequency is increased by a Doppler shift.

\subsubsection{Moving Source and Stationary Receiver}\label{subsubsection-Doppler-FS.1.1.2}

In the case that the wave source is moving and receiver is stationary,
as shown in Fig.~\ref{figure-Doppler-effect-UV-far-field}(b), let us
again assume that source emits two adjacent maxima at $t_1=t_0$ and
$t_2=t_0+T$. Correspondingly, the receiver observes two maxima at
$t'_1=t_0'$ and $t'_2=t_0'+T'$.  As the source moves towards the receiver
by a distance of $dr=v\cos\theta T$ during the period of $T$, $T'$ satisfies
\begin{align}\label{eq:Doppler-FS-17x0}
T'=T-\frac{dr}{c}=T\left(1-\frac{v\cos\theta }{c}\right)
\end{align}
Accordingly, we have the formulas
\begin{align}\label{eq:Doppler-FS-17x1}
  \frac{T}{T'}=&\left(1-\frac{v\cos\theta }{c}\right)^{-1}\\
  \label{eq:Doppler-FS-17x2}
  f'=&f\left(1-\frac{v\cos\theta }{c}\right)^{-1}\\
 \label{eq:Doppler-FS-17x3}
  \lambda'=&\lambda\left(1-\frac{v\cos\theta }{c}\right)
\end{align}
The Doppler shift is
\begin{subequations}\label{eq:Doppler-FS-17x4}
\begin{align}\label{eq:Doppler-FS-17x4a}
  f_D(v,\theta)=& f'-f=f\left(\frac{\frac{v\cos\theta}{c}}{1-\frac{v\cos\theta }{c}}\right)\\
  \label{eq:Doppler-FS-17x4b}
  \approx& \frac{fv\cos\theta}{c}=\frac{v\cos\theta}{\lambda}
\end{align}
\end{subequations}
where the approximation is hold, when $v<<c$. Notice from
Fig.~\ref{figure-Doppler-effect-UV-far-field}(b) that $0\leq \theta
<\pi/2$ idicates source moving towards receiver, while $\pi/2\leq
\theta <\pi$ is the situation of source moving away from receiver.

When comparing \eqref{eq:Doppler-FS-16x6} - \eqref{eq:Doppler-FS-16x9}
with \eqref{eq:Doppler-FS-17x1} - \eqref{eq:Doppler-FS-17x4}, we can see
that the Doppler effect in the case of moving source stationary
receiver and that of stationary source moving receiver are not exactly
the same on the received wave. In other words, the Doppler effects are
asymmetric.  While the Doppler shift is linearly related to the moving
speed of receiver, it is non-linearly dependent on the moving speed of
source. The Doppler effects due to moving source and receiver become
the same, only when $v<<c$. Otherwise, source moving towards receiver
strengthens the Dopper effect due to $1-\frac{v\cos\theta }{c}<1$ in
\eqref{eq:Doppler-FS-17x4}, and source moving away from receiver
weakens the Doppler effect, as the result of $1-\frac{v\cos\theta
}{c}>1$ in \eqref{eq:Doppler-FS-17x4}.

\begin{example}\label{Example-DE-1} 
Assume that two imaginary children, S and R, are playing ball
catching. Relative to ground, assume that S is stationary and stands
at the origin, while R is running away from S at a speed of
$v=5$~m/s. The horizontal flying speed of ball is $c=10$~m/s. Two
balls are throwing with an interval of $T=1$ second. At the
time when the first ball is thrown at $t=0$, R is $d=30$ meters away
from S. Find the interval $T'$ between the two catches in the
cases:
\begin{enumerate}

\item[A)] S is throwing the ball to R.
  
\item[B)] R is trowing the ball to S.
\end{enumerate}

\begin{solution}
  
  \begin{enumerate}
    
\item[A)] In this case, the arrival time of the first ball follows
      $ct_1'=d+vt_1'$, from which we obtain $t_1'=6$ second. The
      arrival time of the second ball follows $c(t_2'-1)=d+vt_2'$,
      from which we obtain $t_2'=8$ second. Hence, the interval
      between the two catches of R is $T'=t_2'-t_1'=2$ seconds.

\item[B)] In this case, the first ball arrives at $t_1'=d/c=3$
  second. After $T=1$ second, R moves away from S by $1v=5$ meters and
  throws the second ball. Hence, the second ball needs to fly $d+5=35$
  meters, and arrives at S at $t_2'=35/10+1=4.5$ second. Therefore,
  the interval between the two catches of S is $T'=t_2'-t_1'=1.5$
  seconds.
  \end{enumerate}
\end{solution}
\end{example}

Similarly, let us consider an example where balls are replaced by EM
waves. The results are obtained directly from the formulas derived so
far. Furthermore, instead of considering time intervals - which can be
explained as periods - the resulted frequencies are aimed.

\begin{example}\label{Example-DE-2}
Consider a rocket escaping or approaching the Earth at a velocity of
about 12~km/s communicates with a ground station on the 10~GHz
x-band. Assume that the radio wave propagation is in line with the
line connecting the rocket and ground station. Find the Doppler
frequencies in the following cases.
\begin{enumerate}

  \item[A)] When rocket is escaping the Earth, the Doppler shift on
    the signal sent by rocket and received by ground station, and the
    Doppler shift on the signal sent by ground station and received by
    rocket.
    
  \item[B)] When rocket is approaching the Earth, the Doppler shift on
    the signal sent by rocket and received by ground station, and the
    Doppler shift on the signal sent by ground station and received by
    rocket.
  
  \end{enumerate}

\begin{solution}

  \begin{enumerate}

    \item[A)] When rocker is escaping the Earth, the Doppler shift on
      the signal sent by rocket and received by ground station is
      calculated from \eqref{eq:Doppler-FS-17x4a}, giving
      $f_{D,RG}=-399984$~KHz. The Doppler shift on the signal sent by
      ground station and received by rocket is calculated by
      \eqref{eq:Doppler-FS-16x9}, giving $f_{D,GR}=-400000$~Hz.

\item[B)] When a rocket approaching the Earth, the Doppler shift on
  the signal sent by rocket and received by ground station is
  calculated by \eqref{eq:Doppler-FS-17x4a}, which is
  $f_{D,RG}=400016$~Hz, while the Doppler shift on the signal sent by
  ground station and received by rocket is obtained from
  \eqref{eq:Doppler-FS-16x9}, yielding $f_{D,GR}=400000$~Hz.

\end{enumerate}
\end{solution}
\end{example}

As shown in Example~\ref{Example-DE-2}, in both cases, the Doppler
shift is dependent on whether source (receiver) are moving
(stationary) or stationary (moving).  They also show that the
differences are small, even when the moving speed is as high as
12~km/s at the Earth escaping speed. The reason is that relative to $c$,
the speed of light, this speed is still very small, the ratio is only
about $v/c=1.112\times 10^{-8}$. Hence, in most practical cases, the
formula $f_D={fv\cos\theta}/{c}$ can be applied, regardless of moving
source stationary receiver, or stationary source moving receiver, or
other relative motion.

\subsubsection{Geometric Analysis of Doppler Effect}\label{subsubsection-Doppler-FS.1.1.3}

\begin{figure}[tb]
  \begin{center}
 \includegraphics[angle=0,width=.99\linewidth]{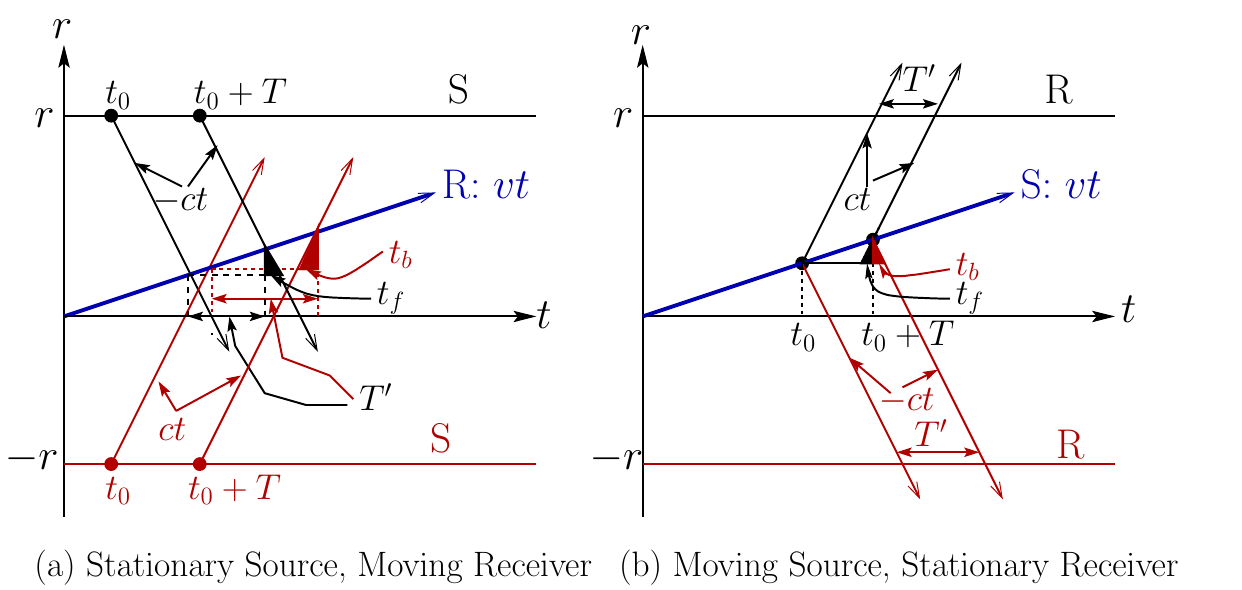}
  \end{center}
  \caption{Graphic explanation of the Doppler effect occurred in remote zones. S=Source, R=Receiver. }
  \label{figure-Doppler-effect-UV-far-field-Graphic-Explain}
\end{figure}

Now an alternative geometric
method~\ncite{10.1119/1.1990758,10.1119/1.2342706,10.1119/1.2731281,10.1119/10.0004145,Worner_2017}
is introduced to derive the formulas obtained in the two cases as
above-considered, i.e., the case of stationary source moving receiver
and the case of moving source stationary receiver.  For simplicity, we
analyze only the longitudinal Doppler effect. This involves assuming
that the source and receiver move purely along the line connecting
them. Mathematically, this means that the angle between the velocity vector
and the propagation vector is $0^{\circ }$ or $180^{\circ }$. For the
general non-longitudinal case, one must use the velocity component
projected onto the line of sight, i.e., replacing the scalar speed $v$
with $\bm{v}\cdot\hat{\bm{r}}$.

The Doppler effect in the above-analyzed two cases can be analyzed
with the aid of graphic drawings, as shown in
Fig.~\ref{figure-Doppler-effect-UV-far-field-Graphic-Explain}, where
Fig.~\ref{figure-Doppler-effect-UV-far-field-Graphic-Explain}(a) is
for the case that source is stationary and receiver is moving, while
Fig.~\ref{figure-Doppler-effect-UV-far-field-Graphic-Explain}(b)
corresponds to the case that receiver is stationary and source is
moving. In
Fig.~\ref{figure-Doppler-effect-UV-far-field-Graphic-Explain}, $r$ is
distance: positive $r$ indicates that source (or receiver) moves
towards receiver (or source), while negative $r$ means that source (or
receiver) moves away from receiver (or source).  Then, from
Fig.~\ref{figure-Doppler-effect-UV-far-field-Graphic-Explain}(a) - upper part - and
considering that receiver moves towards source, from the geometry, we
have
\begin{align}\label{eq:Doppler-FS-3.19}
T'=T-t_f=T-\frac{v(T'+t_0')-vt_0'}{c}=T-\frac{vT'}{c}
\end{align}
yielding
\begin{align}\label{eq:Doppler-FS-3.20}
\frac{T}{T'}=1+\frac{v}{c}~~~~\textrm{or}~~~~ f'=f\left(1+\frac{v}{c}\right)
\end{align}
which is \eqref{eq:Doppler-FS-16x6} or \eqref{eq:Doppler-FS-16x7} with $\cos\theta=-1$. When receiver moves away from source, from Fig.~\ref{figure-Doppler-effect-UV-far-field-Graphic-Explain}(a) - lower part,  we have the relationship of
\begin{align}\label{eq:Doppler-FS-3.21}
T'=T+t_b=T+\frac{v(t_0'+T')-vt_0'}{c}=T+\frac{vT'}{c}
\end{align}
Therefore, we obtain
\begin{align}\label{eq:Doppler-FS-3.22}
\frac{T}{T'}=1-\frac{v}{c}~~~~\textrm{or}~~~~ f'=f\left(1-\frac{v}{c}\right)
\end{align}
which is \eqref{eq:Doppler-FS-16x6} or \eqref{eq:Doppler-FS-16x7} with $\cos\theta=1$.

When considering Fig.~\ref{figure-Doppler-effect-UV-far-field-Graphic-Explain}(b) - upper part, if source moves towards receiver, as shown in the drawing, $T'$ and $T$ satisfy the relation of
\begin{align}\label{eq:Doppler-FS-3.23}
T'=T-t_f=T-\frac{v(t_0+T)-vt_0}{c}=T-\frac{vT}{c}
\end{align}
Hence, 
\begin{align}\label{eq:Doppler-FS-3.24}
\frac{T}{T'}=\left(1-\frac{v}{c}\right)^{-1}~~~~\textrm{or}~~~~ f'=f\left(1-\frac{v}{c}\right)^{-1}
\end{align}
Otherwise, when source moves away from receiver, as seen in Fig.~\ref{figure-Doppler-effect-UV-far-field-Graphic-Explain}(b) - lower part, $T'$ and $T$ are related by
\begin{align}\label{eq:Doppler-FS-3.25}
T'=T+t_b=T+\frac{v(t_0+T)-vt_0}{c}=T+\frac{vT}{c}
\end{align}
Accordingly, 
\begin{align}\label{eq:Doppler-FS-3.26}
\frac{T}{T'}=\left(1+\frac{v}{c}\right)^{-1}~~~~\textrm{or}~~~~ f'=f\left(1+\frac{v}{c}\right)^{-1}
\end{align}
Explicitly, \eqref{eq:Doppler-FS-3.24} and \eqref{eq:Doppler-FS-3.26} are the same as \eqref{eq:Doppler-FS-17x1} with $\cos\theta=1$ and $-1$, respectively.

\subsubsection{Moving Source and Moving Receiver}\label{subsubsection-Doppler-FS.1.1.4}

A more general case where both source and receiver are moving
(relative to a fixed reference frame) can be analyzed as
follows. Assume that source's moving velocity is $v$, the angle
between which and $\hat{\bm{r}}$ is $\theta$. Assume that receiver's
moving velocity is $v'$, the angle between which and $\hat{\bm{r}}$
is $\theta'$.  Assume that the wave transmitted by source has two
adjacent crests occurring at $t_1=t_0$ and
$t_2=t_0+T$. Correspondingly, the crests in the received wave occur
at $t_1'=t_0'$ and $t_2'=t_0'+T'$.  During $T=t_2-t_1$, source moves
towards receiver by a distance $dr=v\cos\theta T$. During
$T'=t_2'-t_1'$, receiver moves away from source by a distance
$dr'=v'\cos\theta' T'$. Hence, $T'$ and $T$ have the relationship of
\begin{align}\label{eq:Doppler-FS-3.27}
T'=T-\frac{v\cos\theta T}{c}+\frac{v'\cos\theta' T'}{c}
\end{align}
Re-arranging them gives
\begin{align}\label{eq:Doppler-FS-3.28}
\frac{T}{T'}=\frac{1-\displaystyle\frac{v'\cos\theta'}{c}}{1-\displaystyle\frac{v\cos\theta}{c}}
\end{align}
Explicitly, when $v=0$, \eqref{eq:Doppler-FS-3.28} is reduced to
\eqref{eq:Doppler-FS-16x6} of the case of source stationary and reciver
moving. When $v'=0$, \eqref{eq:Doppler-FS-3.28} is reduced to
\eqref{eq:Doppler-FS-17x1} of the case of source moving while reciver
stationary. From \eqref{eq:Doppler-FS-3.28}, the other related
formulas are
\begin{align}\label{eq:Doppler-FS-3.29}
  f'=&f\left(\frac{1-\displaystyle\frac{v'\cos\theta'}{c}}{1-\displaystyle\frac{v\cos\theta}{c}}\right)\\
 \label{eq:Doppler-FS-3.31}
   f_D(v,v',\theta,\theta')=&f'-f\nonumber\\
  =&f\left(\frac{\displaystyle\frac{v\cos\theta}{c}-\displaystyle\frac{v'\cos\theta'}{c}}{1-\displaystyle\frac{v\cos\theta}{c}}\right)\nonumber\\
  \approx & f\left({\frac{v\cos\theta}{c}-\frac{v'\cos\theta'}{c}}\right)
\end{align}

\subsection{Non-Relativistic Doppler Effect in Close Zones}\label{subsubsection-Doppler-FS.1.2}

\begin{figure}[tb]
  \begin{center}
 \includegraphics[angle=0,width=.7\linewidth]{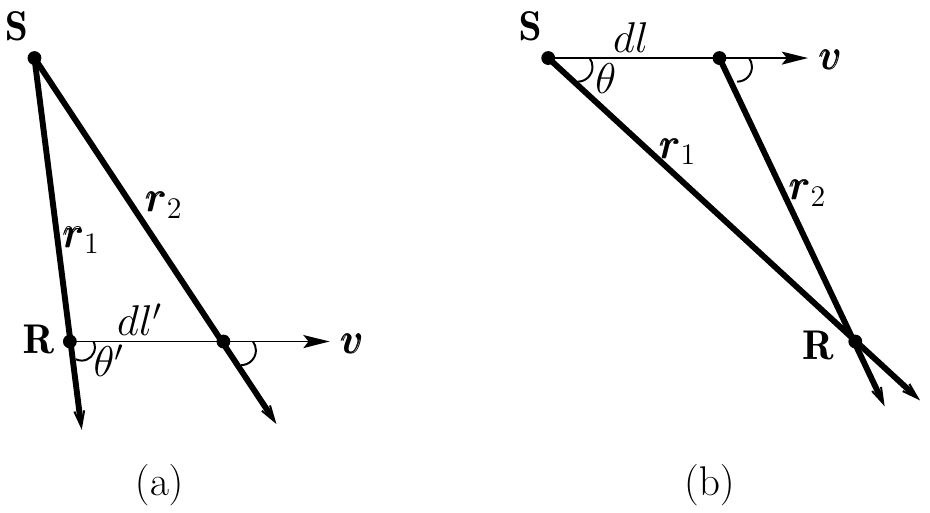}
  \end{center}
  \caption{A source sends radio signals to a receiver in its
    close-zone: (a) stationary source and moving receiver, (b) moving
    source and stationary receiver. }
  \label{figure-Doppler-effect-UV-close-field}
\end{figure}

When source and receiver fall in the close zone, spherical wave has to
be considered. Fig.~\ref{figure-Doppler-effect-UV-close-field}
illustrates the two cases of stationary souce and moving receiver
(Fig.~\ref{figure-Doppler-effect-UV-close-field}(a)), and moving source
and stationary receiver
(Fig.~\ref{figure-Doppler-effect-UV-close-field}(b)). Let us first
analyze the case of Fig.~\ref{figure-Doppler-effect-UV-close-field}(a).

Assume that source emits two adjacent wave crests at $t_1=t_0$ and
$t_2=t_0+T$. Correspondingly, receiver measures the first wave crest
at $t_1'=t_0'$, and the second wave crest at $t_2'=t_0'+T'$. As shown
in Fig.~\ref{figure-Doppler-effect-UV-close-field}(a), during
$T'=t_2'-t_1'$, receiver moves in $\hat{\bm{v}}$-direction a distance
of $dl'=vT'$. The position vectors of receiver with respect to source
at $t_1'$ and $t_2'$ are denoted as $\bm{r}_1$ and $\bm{r}_2$,
respectively.  The angle between $\bm{v}$ and the position vector
$\bm{r}_1$ is represented as $\theta'$. Express
$\bm{r}_i=r_i\hat{\bm{r}}_i$, $i=1,2$, where $r_i$ is the distance
between source and receiver. Then, with the aid of the cosine rule,
\begin{align}\label{eq:Doppler-FS-3.32}
  dr=r_2-r_1=&\sqrt{r_1^2+(dl')^2-2r_1dl'\cos(\pi-\theta')}-r_1\nonumber\\
  =&\sqrt{r_1^2+(vT')^2+2r_1vT'\cos(\theta')}-r_1
\end{align}
Therefore, $T'$ and $T$ have the relation of
\begin{align}\label{eq:Doppler-FS-3.33}
  T'=&T+\frac{dr}{c}\nonumber\\
  =&T+\frac{\sqrt{r_1^2+(vT')^2+2r_1vT'\cos(\theta')}-r_1}{c}
\end{align}
which can be re-arranged to obtain an quadratic equation
\begin{align}\label{eq:Doppler-FS-3.34}
  a_0T'^2+a_1T'+a_2=0
\end{align}
with
\begin{align}
  a_0=&c^2-v^2\nonumber\\
  a_1=&-2(c\lambda-cr_1+vr_1\cos\theta')\nonumber\\
  a_2=&\lambda^2-2\lambda r_1\nonumber 
\end{align}
where $\lambda=cT$. The closed-form solution to \eqref{eq:Doppler-FS-3.34} is\footnote{Note that the second solution to \eqref{eq:Doppler-FS-3.34} has no practical meaning.}
\begin{align}\label{eq:Doppler-FS-3.35}
  T'=&\frac{-a_1+\sqrt{a_1^2-4a_0a_2}}{2a_0}\nonumber\\
  =&\frac{c\lambda-cr_1+vr_1\cos\theta'+\sqrt{(c\lambda-cr_1+vr_1\cos\theta')^2-(c^2-v^2)(\lambda^2-2\lambda r_1)}}{c^2-v^2}
\end{align}
which is a function of $r_1$ - the distance between source and receiver at $t_0'$, in addition to its dependance on $v\cos\theta'$.

Here are three special cases corresponding to $\theta'=0^o$, $\theta'=90^o$ and $\theta'=180^o$: 
\begin{subequations}
\begin{align}\label{eq:Doppler-FS-3.36}
T'(0^o)=&T\left(1-\frac{v}{c}\right)^{-1}\\
\label{eq:Doppler-FS-3.37}
T'(90^o)=&\frac{c\lambda-cr_1+\sqrt{(c\lambda-cr_1)^2-(c^2-v^2)(\lambda^2-2\lambda r_1)}}{c^2-v^2}\nonumber\\
=&\frac{c^2T-cr_1+\sqrt{(c^2T-cr_1)^2-(c^2-v^2)(c^2T^2-2cT r_1)}}{c^2-v^2}
\\
\label{eq:Doppler-FS-3.38}
T'(180^o)=&T\left(1+\frac{v}{c}\right)^{-1}
\end{align}
\end{subequations}
Equation \eqref{eq:Doppler-FS-3.36} is the same as
\eqref{eq:Doppler-FS-16x6} with $\theta'=0^o$,
\eqref{eq:Doppler-FS-3.38} is the same as \eqref{eq:Doppler-FS-16x6}
with $\theta'=180^o$. Hence, when receiver moves in the direction or
opposite direction of the position vector from source to receiver,
there is no distinction between close- and remote zones. In contrast,
when $\theta'=90^o$, the Doppler effect in remote-zone is zero, while
that in close-zone is non-zero, as shown in
\eqref{eq:Doppler-FS-3.37}.

Notice that \eqref{eq:Doppler-FS-3.33} can be written as
\begin{align}\label{eq:Doppler-FS-3.39}
  T' =&T+\frac{(r_1+vT')\sqrt{1-\displaystyle\frac{2r_1vT'(1-\cos\theta')}{(r_1+vT')^2}}-r_1}{c}
\end{align}
If the first-order Taylor expansion is used to approximate the square-root term, an approximation to \eqref{eq:Doppler-FS-3.39} is
\begin{align}\label{eq:Doppler-FS-3.40}
  T' \approx&T+\frac{v^2T'^2+r_1v\cos\theta'T'}{c(r_1+vT')}
\end{align}
From \eqref{eq:Doppler-FS-3.40}, we obtain
\begin{align}\label{eq:Doppler-FS-3.x40}
  \frac{f'}{f}=\frac{T}{T'}\approx&1-\frac{v^2T'+r_1v\cos\theta'}{c(r_1+vT')}\nonumber\\
  =&1-\frac{vdl'/r_1+v\cos\theta'}{c(1+dl'/r_1)}
\end{align}
Explicitly, if $r_1>>vT'=dl'$, implying that source and receiver are in remote zone,
\eqref{eq:Doppler-FS-3.x40} becomes \eqref{eq:Doppler-FS-16x6}. 

Note that \eqref{eq:Doppler-FS-3.x40} can be computed by receiver, as it knows its velocity $v$ and the distance $dl'$ it travels in a duration of $T'$, in addition to $r_1$ and $\theta'$.

In the context of the scenario where source is moving and receiver is
stationary, and they are in the close zone of each other, based on
Fig.~\ref{figure-Doppler-effect-UV-close-field}(b), applying the cosine
rule gives
\begin{align}\label{eq:Doppler-FS-3.41}
r_2=&\sqrt{r_1^2+dl^2-2r_1dl\cos\theta}\nonumber\\
=&\sqrt{r_1^2+v^2T^2-2r_1vT\cos\theta}
\end{align}
Hence,
\begin{align}\label{eq:Doppler-FS-3.42}
T'=T+\frac{\sqrt{r_1^2+v^2T^2-2r_1vT\cos\theta}-r_1}{c}
\end{align}
which is dependent on $r_1$ of the distance between source and receiver at $t_0$. 

For $\theta=0^o,~90^o$ and $180^o$,
\begin{subequations}
\begin{align}\label{eq:Doppler-FS-3.43}
T'(0^o)=&T\left(1-\frac{v}{c}\right)\\
\label{eq:Doppler-FS-3.44}
T'(90^o)=&T+\frac{\sqrt{r_1^2+v^2T^2}-r_1}{c}\\
\label{eq:Doppler-FS-3.45}
T'(180^o)=&T\left(1+\frac{v}{c}\right)
\end{align}
\end{subequations}
Again, \eqref{eq:Doppler-FS-3.43} is the same as
\eqref{eq:Doppler-FS-17x0} with $\theta'=0^{\circ}$,
\eqref{eq:Doppler-FS-3.45} is the same as \eqref{eq:Doppler-FS-17x0}
with $\theta'=180^\circ$. In these two cases, there is no distinction
between close and remote zone. However, when $\theta=90^\circ$, the
Doppler effect in remote zone is zero, while that in close zone is
non-zero, as shown in \eqref{eq:Doppler-FS-3.37}.

Notice from \eqref{eq:Doppler-FS-3.37} and \eqref{eq:Doppler-FS-3.44} that $T'$ converges to $T$ when $r_1$ increases, meaning that, when the moving direction is perpendicular to the line connecting source and receiver, the Doppler effect becomes weaker, as the distance between source and receiver increases. 

From \eqref{eq:Doppler-FS-3.42}, after neglecting $-r_1$, an approximation of
\begin{align}\label{eq:Doppler-FS-3.x42}
\frac{f'}{f}=\frac{T}{T'}\approx \left(1+\frac{\sqrt{v^2+(fr_1)^2-2vfr_1\cos\theta}}{c}\right)^{-1}
\end{align}
can be obtained. Reader can show that when assuming $c>>fr_1>>v$, \eqref{eq:Doppler-FS-3.x42} reduces to \eqref{eq:Doppler-FS-17x0}.

Note that, to use \eqref{eq:Doppler-FS-3.x42} to sense the velocity of source, receiver requires to first know the moving direction of source relative to $\bm{r}_1$.   

Let us now extend the above two special cases to a more general case
where source moves at velocity $\bm{v}$ and receiver moves at
velocity $\bm{v}'$. Assume at reference time $t_0=0$ that the
position vectors of source and receiver are $\bm{r}_0$ and
$\bm{r}_0'$, respectively. All position vectors are relative to the
origin of the considered coordinate system.  Then, at time $t$,
\begin{align}\label{eq:Doppler-FS-X3.49}
\bm{r}(t)=&\bm{r}_0+\bm{v}t\nonumber\\
\bm{r}'(t)=&\bm{r}'_0+\bm{v}'t
\end{align}
Again, assume that source emits two adjacent wave crests at $t_1=t_0$ and
$t_2=t_0+T$. Accordingly, the position vectors are registered as $\bm{r}(t_0)$ and $\bm{r}(t_0+T)$, respectively. Corresponding to the two crests emitted, assume that receiver observes the first wave crest at $t_1'=t_0'$ and the second wave crest at $t_2'=t_0'+T'$, and the position vectors are $\bm{r}'(t'_0)$ and $\bm{r}'(t'_0+T')$. Then, the Doppler effect can be retrieved from the relationship of
\begin{align}\label{eq:Doppler-FS-X3.50}
T'=T+\frac{\|\bm{r}'(t'_0+T')-\bm{r}(t_0+T)\|-\|\bm{r}'(t'_0)-\bm{r}(t_0)\|}{c}
\end{align}
where $\|\bm{r}'(t'_0)-\bm{r}(t_0)\|$ is the distance between source at $t_0$ and receiver at $t'_0$, while $\|\bm{r}'(t'_0+T')-\bm{r}(t_0+T)\|$ is the distance between source at $t_0+T$ and receiver at $t'_0+T'$, both of which can be derived once the velocities and the initial position vectors are known. 

\begin{remark}
From previous analysis and, in particular \eqref{eq:Doppler-FS-X3.50}, we can conceive that for communications, the Doppler related issues are relatively easy to deal with. Typically, in communications, the signal processing at transmitter and/or at receiver may need to address the challenges generated by the Doppler effect. However, in both cases, only the Doppler shift itself or its statistics is needed, regardless of the dynamics of transmitter, receiver and communication channels, although the Doppler effect is actually generated by these dynamics. 

By contrast, the Doppler effect in sensing applications is highly
challenging. The objective of sensing is to derive the dynamics of
transmitter, receiver and/or environments via the measurements of
Doppler shifts at observers. While it is not difficult to find the
Doppler effect based on \eqref{eq:Doppler-FS-X3.50} by imposing
appropriate assumptions, it may be extremely involved to find the
mechanical quantities based on the measured Doppler effects at
observers. For example, as seen in
Fig.~\ref{figure-Doppler-effect-UV-close-field}(b), while receiver may
easily obtain the Doppler shift via measuring the change of its
received frequency, it is very challenging to derive the velocity of
source, as to achieve this, it also needs the knowledge of $r_1$, $r_2$,
$dl$, as well as $\theta$.
  
\end{remark}

\subsection{Relativistic Doppler Effect}\label{subsubsection-Doppler-FS.1.3}

The {\em special theory of relativity} is based on two postulates proposed by Albert Einstein in 1950, which are stated as~\ncite{Book-Albert-Einstein-Relativity}:
\begin{enumerate} 

\item The laws of physics are the same in all inertial reference frames;

\item The speed of light in free space has the same value $c$ in all inertial reference frames. 

\end{enumerate}
Related to Doppler effect, the consequences of the special theory of
relativity are time dilation and the effect on the analysis of Doppler
shift. In this section, the Doppler effect is analyzed on the
constraint of the above two postulates by Einstein.  For further
explanation of the special theory of relativity, the interested reader
is referred to references, including
\ncite{Book-Albert-Einstein-Relativity,Book-University-Physics-Young,Book-Kenneth-Modern-Physics,Zanchini_2024}.

Note that, starting from now on and when EM wave is considered, we
will mainly state a receiver as an observer to follow the term widely
used in description of the special/general theory of
relativity. Furthermore, for simplicity of presentation, the following
constant and variables are defined:
\begin{itemize}

\item $c$: speed of light in free space.

\item $c'$: speed of light in medium.

\item $\bm{v}$ and $\bm{v}'$: velocity vectors of source and
  observer, $|\bm{v}|$ and $|\bm{v}'|$ represent speed of source or
  observer moving. Note that, $\bm{v}$ is always used if there is
  only one velocity involved. 

  \item $\hat{\bm{r}}$ and $\hat{\bm{r}}'$: unit position vector
    pointing in the direction from observer to source, and unit
    position vector pointing in the direction from source to observer.

    \item $v=\bm{v} \cdot \hat{\bm{r}}$ and $v'=\bm{v}' \cdot
      \hat{\bm{r}}'$: velocity components on the line connecting
      source and observer. Note that, $v$ is always used if there is
      only one velocity involved. Positive $v$ is for moving away from
      each other and negative $v$ is for moving towards each other.
  
\end{itemize}

\begin{remark}
 In the following formulas, only $v$ or $v'$ appears for accounting of the
 classic Doppler effect generated by relative motion.  More
 general-form formulas can be obtained by replacing $v$ with
 $\bm{v}\cdot\hat{\bm{r}}$ or $v'=\bm{v}' \cdot \hat{\bm{r}}'$.
 Note also that in the analysis of relativistic Doppler effect, it is
 the magnitude of $\bm{v}$, i.e., $|\bm{v}|$, that generates time
 dilation, not $v$, the speed along the line connecting source with
 observer.  Additionally, unless specified, source and observer are
 assumed in each other's remote zone.
\end{remark}

The analysis of relativistic Doppler effect may be found in any
textbook on modern physics, including in
\ncite{Book-University-Physics-Young,Book-Kenneth-Modern-Physics},
where the special theory of relativity is addressed. However, in most
textbooks, only the longitudinal Doppler effect is analyzed, which
assumes that the moving direction aligns with the line connecting source
and observer, resulting in that the speed $|\bm{v}|$ generating time
dilation and the speed $|v=\bm{v}\cdot\hat{\bm{r}}|$, which
generates the (classic) Doppler effect\footnote{To distinguish the
Doppler effect generated in classic way and that by other effect due
to special and general relativities, in following description, the
phrase of `(classic) Doppler effect' is used to indicate the Doppler
effect generated in classic way to avoid confusion. }, are equal, i.e,
$|\bm{v}|=|\bm{v}\cdot\hat{\bm{r}}|$. Explicitly, this is not
necessary true and, in fact, they are not the same in many practical
scenarios. Furthermore, in most textbooks, the relativistic Doppler
effect is analyzed in free space, without including propagation
medium. Therefore, in this section, the longitudinal Doppler effect is
first analyzed by following textbooks. Then, the analysis is
generalized to consider: a) $|\bm{v}\cdot\hat{\bm{r}}|\neq
|\bm{v}|$, and b) EM wave propagates in medium.

\subsubsection{Longitudinal Relativistic Doppler Effect}\label{subsubsection-Doppler-FS.1.3.1}

Consider that in a free space, a system includes a source emitting
EM-wave signals to an observer.  The relative velocity between source
and observer is $v$, the direction of which aligns with the line
connecting them. This system involves two reference frames, the
reference frame of source and the reference frame of observer. To
analyze the Doppler effect, three sets of times, periods and
frequencies are defined. The first set includes the proper time $t$,
proper period $T$ and proper frequency $f$ at the source in the
reference frame of source. The second set includes the proper time
$t'$, proper period $T'$ and proper frequency $f'$ at the observer in
the reference frame of observer. The third set includes the time
$t''$, period $T''$ and the frequency $f''$ either in the reference
frame of source or in the reference frame of observer, depending on
which of them is considered. This should become clear in the forthcoming
analysis.

First, let us analyze the Doppler effect in the observer's reference
frame. In this case, time $t''$, period $T''$ and frequency $f''$ are
associated with source but in the reference frame of observer.  Also,
source's moving velocity is $v$ relative to observer. Assume that in
the reference frame of observer, source emits two adjacent wave crests
at $t_1''=t_0''$ and $t_2''=t_0''+T''$. Then, following the analysis
in Section~\ref{subsubsection-Doppler-FS.1.1}, first, $T'$ and $T''$
follow the relation of
\begin{align}\label{eq:Doppler-FS-XW-3.49}
T'=T''+\frac{vT''}{c}=T''\left(1+\frac{v}{c}\right)
\end{align}
which tributes to the (classic) Doppler effect. Second, since in the
reference frame of observer, source is moving at a velocity $v$, its
clock ticks slower than the clock in the reference frame of observer,
yielding
\begin{align}\label{eq:Doppler-FS-XW-3.50}
T''=\frac{1}{\sqrt{1-v^2/c^2}}T
\end{align}
where ${1}/{\sqrt{1-v^2/c^2}}$ is the time dilation factor or Lorentz
factor~\ncite{Book-Kenneth-Modern-Physics}. Substituting this into
\eqref{eq:Doppler-FS-XW-3.49} gives
\begin{align}\label{eq:Doppler-FS-XW-3.51}
  \frac{f'}{f}=\frac{T}{T'}=\frac{\sqrt{1-v^2/c^2}}{1+v/c}\nonumber\\
  =\sqrt{\frac{1-v/c}{1+v/c}}
\end{align}
or
\begin{subequations}\label{eq:Doppler-FS-XW-3.52}
\begin{align}\label{eq:Doppler-FS-XW-3.52-a}
  f'=&f\times\frac{\sqrt{1-v^2/c^2}}{1+v/c}\\
\label{eq:Doppler-FS-XW-3.52-b}
  =&f\times\sqrt{\frac{1-v/c}{1+v/c}}
\end{align}
\end{subequations}

Now let us turn to derive the Doppler effect in the reference frame of
source. Under this assumption, time $t''$, period $T''$ and frequency
$f''$ are associated with observer in the reference frame of source.
Relative to source, observer's moving velocity is $v$. Accordingly,
$T$ and $T''$ are related by the formula
\begin{align}\label{eq:Doppler-FS-XW-3.53}
T''=T+\frac{vT''}{c}
\end{align}
giving
\begin{align}\label{eq:Doppler-FS-XW-3.54}
T=T''\left(1-\frac{v}{c}\right)
\end{align}
Since in the source's reference frame, observer is moving at a
velocity $v$. According the Lorentz
transformation~\ncite{Book-Kenneth-Modern-Physics},
\begin{align}\label{eq:Doppler-FS-XW-3.55}
T''=\gamma T'=\frac{1}{\sqrt{1-v^2/c^2}}T'
\end{align}
Substituting it into \eqref{eq:Doppler-FS-XW-3.54} and after some
arrangement, the Doppler effect can be obtained as
\begin{subequations}\label{eq:Doppler-FS-XW-3.56}
\begin{align}\label{eq:Doppler-FS-XW-3.56-a}
  f'=&f\times\frac{1-v/c}{\sqrt{1-v^2/c^2}}\\
\label{eq:Doppler-FS-XW-3.56-b}
  =&f\times\sqrt{\frac{1-v/c}{1+v/c}}
\end{align}
\end{subequations}

Equations \eqref{eq:Doppler-FS-XW-3.52-b} and
\eqref{eq:Doppler-FS-XW-3.56-b} show that the Doppler effect formula
derived in view of source and that in view of observer are the same,
fully agreeing with the postulates by Einstein. In fact, in the free space
where two objects experience relative motion, any object can only
observe that the other one is moving. Hence, to observer, it is always
the case that source is moving, and the signals received from source
experience both time dilation and (classic) Doppler effect. In other words, to analyze the relativistic Doppler effect in free space, it is sufficient to consider the scenario of source moving observer stationary.

\begin{remark}\label{Remark-DEAM-3}
In the analysis of the classic Doppler effect in
Section~\ref{subsubsection-Doppler-FS.1.1}, two different formulas are
obtained respectively for the cases of source rest observer moving and
source moving observer rest, which are
\eqref{eq:Doppler-FS-3.20}/\eqref{eq:Doppler-FS-3.22}, and
\eqref{eq:Doppler-FS-3.24}/\eqref{eq:Doppler-FS-3.26} for moving
towards or away from each other. Since $c>>v$ in most applications,
one may attempt to get an approximate formula for application in both
cases. One way to reach this objective is first multiplying
\eqref{eq:Doppler-FS-3.20} with \eqref{eq:Doppler-FS-3.24} in the case
of moving towards each other, or multiplying
\eqref{eq:Doppler-FS-3.22} with \eqref{eq:Doppler-FS-3.26} in the case
of moving away from each other, and then taking the square-root on
both sides of the resulted equations. After integrating the $+$ and
$-$ signs explaining moving states into $v$, these operations result
in the formula
\begin{align}\label{eq:Doppler-FS-XW-3.57}
f'=f\times \sqrt{\frac{1- {v}/{c}}{1+ {v}/{c}}}
\end{align}
However, it is surprise to see that this formula is exactly the one in
\eqref{eq:Doppler-FS-XW-3.52-b}, or \eqref{eq:Doppler-FS-XW-3.56-b},
derived under the constraints of the special theory of
relativity~\ncite{Book-University-Physics-Young,Book-Kenneth-Modern-Physics},
with both time dilation and classic Doppler effect included. This is in fact
not a coincidence, but the consequence of the special theory of
relativity.  The time dilation effect inherent in special relativity
precisely bridges the gap between the two asymmetric classical
scenarios, forcing a symmetric outcome that is accurately captured by
their geometric mean.  This can also be illustrated by multiplying
\eqref{eq:Doppler-FS-XW-3.52-a} with \eqref{eq:Doppler-FS-XW-3.56-a}
and taking the square-root of the product, giving
\begin{subequations}\label{eq:Doppler-FS-XW-3.58}
\begin{align}\label{eq:Doppler-FS-XW-3.58-a}
  f'=&f\times\sqrt{\frac{\sqrt{1-v^2/c^2}}{1+v/c}\times \frac{1-v/c}{\sqrt{1-v^2/c^2}}}\\
\label{eq:Doppler-FS-XW-3.58-b}
  =&f\times\sqrt{\frac{1-v/c}{1+v/c}}
\end{align}
\end{subequations}
It is shown in \eqref{eq:Doppler-FS-XW-3.58-a} that the time dilation
factors yielded by the two reference frames are ideally cancelled,
leaving only the asymmetric (classic) Doppler effects viewing
respectively from the two reference frames.
\end{remark}

\begin{example}\label{Example-DE-3}
Considering the situations and assumptions in
Example~\ref{Example-DE-2}, find the relativistic Doppler shifts,
respectively, when rocket is moving towards or moving away from the
Earth.
\begin{solution}
  When the rocket is moving towards the Earth, the Doppler shift
  $f_D=f'-f$ can be found from \eqref{eq:Doppler-FS-XW-3.52-b} with
  $v=-12$~km/s, giving $f_D=400008$~Hz.

  When the rocket is moving away from the Earth, the Doppler shift can
  be found from \eqref{eq:Doppler-FS-XW-3.52-b} with $v=12$~km/s,
  which is $f_D=-399992$~Hz.
\end{solution}
\end{example}

Comparing the results in Example~\ref{Example-DE-3} with the
corresponding results in Example~\ref{Example-DE-2} tells that the
time dilation in Example~\ref{Example-DE-3} has only very light effect
on the Doppler shift, even when the rocket flies at the Earth escaping
speed.

\subsubsection{Relativistic Doppler Effect in Uniform Medium}\label{subsubsection-Doppler-FS.1.3.2}

We now generalize the previous analysis, expanding the context from
the longitudinal free space scenario to more general cases. This
generalization involves two aspects.

First, considering that most EM waves propagate in media in practical
applications - such as wireless communications, optical
communications, and radar sensing, to name a few - the analysis is
extended to include wave propagation media. Accordingly, the wave
propagation speed is expressed as $c'$ to distinguish it from $c$, the
speed of light in a vacuum.

Second, in the longitudinal scenario, the velocity that generates time
dilation is the same as the velocity that generates the (classic)
Doppler effect. However, in general cases, these two velocities can be
different. Hence, the previous analysis is also generalized to
encompass this situation. The speed responsible for time dilation is
denoted as $|\bm{v}|$, the magnitude of the velocity vector
$\bm{v}$, while the speed responsible for the (classic) Doppler
effect is expressed as $v$ (or $v'$), which is positive when source
and observer move away from each other and negative when they move
towards each other.\\

\noindent{\bf Moving Source and Rest Observer:}\\

Consider a system that involves three reference frames, the reference
frame of source, the reference frame of observer and the reference
frame of medium. The motion states of source and observer are relative
to the medium.  Assume that a source moving at a velocity $\bm{v}$
sends EM signals via a propagation medium to a rest observer.  The
wave propagation speed in medium is denoted as $c'$. The velocity
along the line connecting source with observer is expressed as $v$.
Again, to analyze the Doppler effect, three sets of times, periods and
frequencies are defined. The first set is associated with source,
expressed as $t$, $T$ and $f$, the second set is associated with
observer, denoted as $t'$, $T'$ and $f'$, and finally, the third set is
associated with the medium, expressed as $t''$, $T''$ and $f''$.

Assume that source in its reference frame emits two adjacent wave
crests at time $t_1=t_0$ and $t_2=t_0+T$. Due to the time dilation,
the clock resting in the frame of medium ticks faster than the clock
resting in the reference frame of source. According to
\ncite{Book-Albert-Einstein-Relativity,Book-University-Physics-Young},
in the reference frame of media, source emits two adjacent wave crests
at time $t_1''=t_0''$ and $t_2''=t_0''+T''$, where $T''$ is related to
$T$ by the Lorentz transformation~\ncite{Book-Kenneth-Modern-Physics}
\begin{align}\label{eq:Doppler-FS-3.46}
T''=\frac{1}{\sqrt{1-|\bm{v}|^2/c^2}}T
\end{align}
The distance source travelled during $T''$ is $dr=vT''$.  Accordingly,
the observer resting in the frame of medium measures two adjacent
crests at time $t_1'=t'_0$ and $t_2'=t'_0+T'$ in its reference frame,
which is the same as the medium's frame.  Hence, $T'$ and $T''$ have
the relation of
\begin{align}\label{eq:Doppler-FS-3.47}
T'=T''+\frac{vT''}{c'}=T''\left(1+\frac{v}{c'}\right)
\end{align}
Substituting \eqref{eq:Doppler-FS-3.46} into
\eqref{eq:Doppler-FS-3.47} gives
\begin{align}\label{eq:Doppler-FS-3.48}
T'=&T\frac{1+{v}/{c'}}{\sqrt{1-|\bm{v}|^2/c^2}}
\end{align}
from which 
\begin{align}\label{eq:Doppler-FS-3.48af}
f'=&f\times\left(\frac{\sqrt{1-|\bm{v}|^2/c^2}}{1+{v}/{c'}}\right)
\end{align}

Eq.\eqref{eq:Doppler-FS-3.48af} is the generalized formula for the
relativistic Doppler effect, which has rarely seen in textbooks, such
as,
\ncite{Book-Albert-Einstein-Relativity,Book-University-Physics-Young,Book-Kenneth-Modern-Physics}. In
addition to including propagation media, this generalized formula
allows the velocity component generating (classic) Doppler effect and
the velocity accounting for time dilation to be different.
\begin{example}\label{Example-DE-4}
Consider that a station fixed on the Earth is receiving EM signals
from a spaceship directly above the station and flying over with a
velocity $\bm{v}$. Explain its resulted Doppler effect.
\begin{solution}
As in the considered case $\bm{v}$ is perpendicular to the line
connecting spaceship and station, $v=0$, there is no (classic) Doppler
effect. However, the time dilation coefficient
$\gamma=1/\sqrt{1-|\bm{v}|^2/c^2}$ in \eqref{eq:Doppler-FS-3.48af}
results in $f'<f$, making the received signals by the station
experience red-shift.
\end{solution}
\end{example}

In \eqref{eq:Doppler-FS-3.48af}, if $c'=c$ and $v=\pm |\bm{v}|$, it
reduced to \eqref{eq:Doppler-FS-XW-3.52-b}, the relativistic Doppler
effect in the longitudinal scenario.\\

\noindent{\bf Rest Source and Moving Observer:}\\

When source rests and observer moves in the reference frame of medium,
time dilation occurs from observer's frame to medium's frame. Hence,
we let $t''$, $T''$ and $f''$ represent the observed time, period and
frequency of the received signals by observer in the reference frame
of medium. Assume that source emits two adjacent wave crests at time
$t_1=t_0$ and $t_2=t_0+T$. Correspondingly, in the reference frame of
medium, observer receives these two adjacent wave crests at time
$t_1''=t_0''$ and $t_2''=t_0''+T''$. Considering that observer moves a
distance of $dr=vT''$ during $T''$ in the reference frame of medium,
the relationship of
\begin{align}\label{eq:Doppler-FS-XW-3.63}
T''=T+\frac{vT''}{c'}
\end{align}
holds. Arranging it gives
\begin{align}\label{eq:Doppler-FS-XW-3.64}
T=T''\left(1-\frac{v}{c'}\right)
\end{align}
Since observer is moving relative to the frame of medium, the clock in
medium's frame ticks faster than the clock in observer's frame. Hence, 
the Lorentz transformation gives
\begin{align}\label{eq:Doppler-FS-XW-3.65}
T''=\frac{T'}{\sqrt{1-|\bm{v}|^2/c^2}}
\end{align}
Substituting it into \eqref{eq:Doppler-FS-XW-3.63} and applying
$T'=1/f'$ and $T=1/f$, we obtain
\begin{align}\label{eq:Doppler-FS-XW-3.66}
f'=f\times\left(\frac{1-v/c'}{\sqrt{1-|\bm{v}|^2/c^2}}\right)
\end{align}

Comparing \eqref{eq:Doppler-FS-3.48af} with
\eqref{eq:Doppler-FS-XW-3.66} explains that, when EM waves propagate
in media, the Doppler effect resulted from moving source stationary
observer and that resulted from stationary source moving observer are
different and, hence, asymmetric. Furthermore, even in free space, if
$v\neq |\bm{v}|$, the above two Doppler effects are asymmetric. As
demonstrated by the results shown in Example~\ref{Example-DE-4} and the following
Example~\ref{Example-DE-5}.
\begin{example}\label{Example-DE-5}
Following Example~\ref{Example-DE-4}, consider that a spaceship
directly above a station fixed on the Earth flies over the station
with a velocity $\bm{v}$, and is receiving signals from the
station. Explain its resulted Doppler effect.
\begin{solution}
Again, $v=0$ and hence there is no (classic) Doppler effect. However,
from \eqref{eq:Doppler-FS-XW-3.66}, the time dilation
makes $f'>f$, making the received signals by spaceship experience
blue-shift.
\end{solution}
\end{example}

Again, if $c'=c$ and $v=|\bm{v}|$, \eqref{eq:Doppler-FS-XW-3.66} and
\eqref{eq:Doppler-FS-3.48af} become the same formula of
\eqref{eq:Doppler-FS-XW-3.52-b}, and hence, the relativistic Doppler effects are
symmetric.\\

\noindent{\bf Moving Source and Moving Observer:}\\

Assume that, along the line connecting source and observer, source moves at velocity $v$ and observer moves at
velocity $v'$. Accordingly, their velocities are $\bm{v}$ and $\bm{v}'$. All velocities are relative to the reference frame of medium. Assume that
in source's reference frame, source emitted two adjacent wave crests
at $t_1=t_0$ and $t_2=t_0+T$. These time instants become $t_1''=t_0''$ and
$t_2''=t_0''+T''$ in medium's reference frame. Hence, we have 
\begin{align}\label{eq:Doppler-FS-XW-3.67}
T''=\frac{T}{\sqrt{1-|\bm{v}|^2/c^2}}
\end{align}
In medium's reference frame, source moves a distance $dr=vT''$ during a period
$T''$.

In medium's reference frame, observer receives the two crests at
$t_1^*=t_0^*$ and $t_2^*=t_0^*+T^*$, and during $T^*$, observer moves
a distance $dr'=v'T^*$. Hence, $T^*$ is related to $T''$ by
\begin{align}\label{eq:Doppler-FS-XW-3.68}
  T^*=T''+\frac{dr+dr'}{c'}=T''+\frac{vT''+v'T^*}{c'}
\end{align}
Rearranging it gives
\begin{align}\label{eq:Doppler-FS-XW-3.69}
  T^*\left(1-\frac{v'}{c'}\right)=T''\left(1+\frac{v}{c'}\right)
\end{align}

As observer moves at $v'$ ($\bm{v}'$) relative to medium, $T^*$ is related to $T'$
by
\begin{align}\label{eq:Doppler-FS-XW-3.70}
T^*=\frac{T'}{\sqrt{1-|\bm{v'}|^2/c^2}}
\end{align}

Finally, substituting \eqref{eq:Doppler-FS-XW-3.67} and
\eqref{eq:Doppler-FS-XW-3.70} into \eqref{eq:Doppler-FS-XW-3.69}, the
Doppler effect follows an expression of
\begin{align}\label{eq:Doppler-FS-XW-3.71}
f'=f\times\left(\frac{1-v'/c'}{1+v/c'}\right)\times \left(\frac{\sqrt{1-|\bm{v}|^2/c^2}}{\sqrt{1-|\bm{v'}|^2/c^2}}\right)
\end{align}

Notice that \eqref{eq:Doppler-FS-XW-3.71} is a general formula of
Doppler effect in uniform motion, which includes all the formulas
derived so far for the Doppler effect in this section, except only those in close zone. First, it is straightforward to reduce \eqref{eq:Doppler-FS-XW-3.71} to the formulas for the classic Doppler effect considered in Section~\ref{subsubsection-Doppler-FS.1.1}. Second, by setting $v'=0$ or $v=0$, the formula is reduced to that obtained above in this section for the case of source moving observer rest or the case of source rest observer moving. Third, if we set $v=|\bm{v}|$, $v'=|\bm{v}'|$, $c=c'$ and assume free space, whether is the formula returned to the form of \eqref{eq:Doppler-FS-XW-3.56-b} - the longitudinal relativistic Doppler effect that is only relied on the relative speed between source and observer. The answer is confirmative, as detailed now.

When $v=|\bm{v}|$, $v'=|\bm{v}'|$ and $c=c'$, it can be readily analyzed to obtain  
\begin{align}\label{eq:Doppler-FS-XW-3.71a}
f'=f\times\sqrt{\frac{1-v/c-v'/c+vv'/c^2}{1+v/c+v'/c+vv'/c^2}}
\end{align}
In the square-root, dividing both the numerator and denominator by $1+vv'/c^2$ yields
\begin{align}\label{eq:Doppler-FS-XW-3.71b}
f'=f\times\sqrt{\frac{1-\displaystyle\frac{1}{c}\cdot\frac{v+v'}{1+vv'/c^2}}{1+\displaystyle\frac{1}{c}\cdot\frac{v+v'}{1+vv'/c^2}}}
\end{align}
Define
\begin{align}\label{eq:Doppler-FS-XW-3.71c}
u=\frac{v+v'}{1+vv'/c^2}
\end{align}
which is the relative speed between source and observer, obtained by the addition of $v$ and $v'$ under the special relativity. Then, \eqref{eq:Doppler-FS-XW-3.71b} becomes  
\begin{align}\label{eq:Doppler-FS-XW-3.71d}
f'=f\times\sqrt{\frac{1-u/c}{1+u/c}}
\end{align}
which is \eqref{eq:Doppler-FS-XW-3.56-b} for the longitudinal relativistic Doppler effect, when the relative speed between source and observer is $u$.

\section{Doppler Effect in General Motion}\label{section-Doppler-FS.2}

In this section, non-relativistic Doppler-effect in general motion is
analyzed in Section~\ref{subsection-Doppler-FS.2.1} and relativistic
Doppler-effect in general motion is addressed in
Section~\ref{subsection-Doppler-FS.2.2}. 


\subsection{Non-relativistic Doppler Effect in General Motion}\label{subsection-Doppler-FS.2.1}

In this section, the non-relativistic Doppler effect, or (classic)
Doppler effect, in general motion is analyzed. First, the Doppler
effect is analyzed when source or observer, or both are in general
motion. Then, the Doppler effects in several special scenarios are
analyzed.

\subsubsection{General Motiving Source and Stationary Observer}\label{subsection-Doppler-FS.2.1.1}

%
\begin{figure}[tb]
  \begin{center}
 \includegraphics[angle=0,width=.5\linewidth]{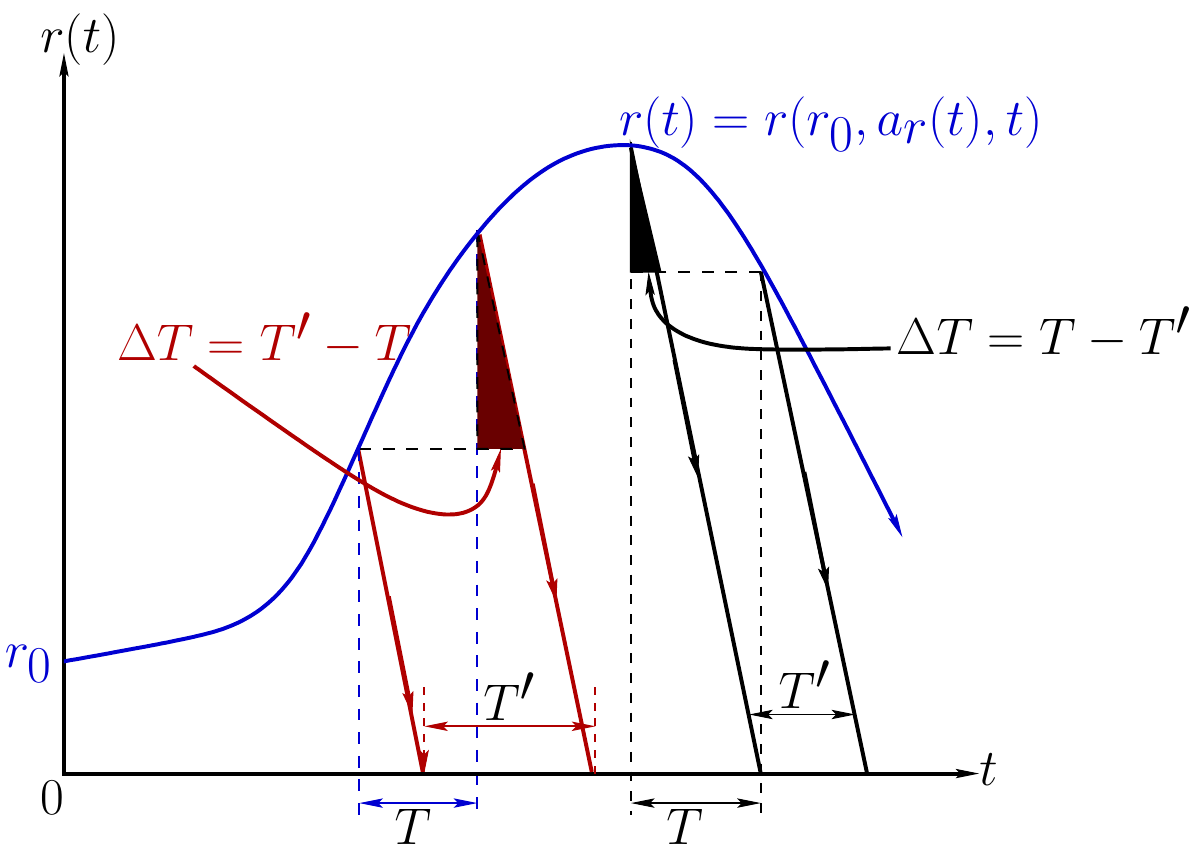}
  \end{center}
  \caption{Graphic representation of the Doppler effect in
    general motion. }
  \label{figure-Doppler-effect-Acceleration}
\end{figure}

Assume that source has the distance equation expressed as $r(t)$ with respect to an observer. For example, it may be represented as
\begin{align}\label{eq:DEAM-61}
r(t)=r(r_0,a_r(t),t)
\end{align}
which, in addition to time, is a function of the initial distance
$r_0$ and the acceleration $a_r(t)$, which may also be a
time-dependent function.  For example, $r(t)$ may have a function as
shown in Fig.~\ref{figure-Doppler-effect-Acceleration}, which assumes
that an observer seats at the origin.

Based on \eqref{eq:Doppler-FS-3.12v}, we can immediately have a
formula for the Doppler shift as
\begin{align}\label{eq:DEAM-62}
f_D(t)\approx-\frac{1}{\lambda}\frac{dr(t)}{dt}=-\frac{1}{\lambda}\left[\frac{\partial r(t)}{\partial t}+\frac{\partial r(t)}{\partial a_r(t)}\times \frac{d(a(t))}{dt}\right]
 \end{align}
However, as mentioned with \eqref{eq:Doppler-FS-3.12v}, the formula
obtained from \eqref{eq:DEAM-62} is only an approximation. Below we
derive the accurate formula for the Doppler effect based on the
graphic representation~\ncite{10.1119/1.2731281,10.1119/10.0004145},
as shown, for example, in
Fig.~\ref{figure-Doppler-effect-Acceleration}.

As shown in Fig.~\ref{figure-Doppler-effect-Acceleration}, it is
assumed that source has a distance equation with respect to observer
expressed as $r(t)$. First, consider the case when source moves away
from observer, as shown by the left-side scenario in
Fig.~\ref{figure-Doppler-effect-Acceleration}. When assuming that
source sends two adjacent maxima (crests) at time instants of $t_1=t$
and $t_2=t+T$, from the graph, we can readily know that the period of
received signal by observer is~\ncite{10.1119/10.0004145}
\begin{align}\label{eq:DEAM-63}
T'(t)=T+\frac{1}{c}\left[r(t+T)-r(t)\right]
\end{align}
where $T'(t)$ is surely time-dependent if there is acceleration.

In the case that source moves towards observer, as shown by the
right-side scenario in Fig.~\ref{figure-Doppler-effect-Acceleration},
it can be shown that the period of received signal by observer has the
same expression of \eqref{eq:DEAM-63}.

From \eqref{eq:DEAM-63}, the Doppler frequency-shift can be derived to
be
\begin{subequations}\label{eq:DEAM-64}
\begin{align} \label{eq:DEAM-64-a}
  f'(t)=&\frac{f}{1+\frac{1}{cT}\left[r(t+T)-r(t)\right]}\\
  \approx & f\left(1-\frac{1}{cT}\left[r(t+T)-r(t)\right]\right)\nonumber\\
  \label{eq:DEAM-64-b}
  f_D(t)=&-\frac{f'(t)}{c}\times\frac{1}{T}\left[r(t+T)-r(t)\right]\\
  \approx& -\frac{f}{c}\times\frac{1}{T}\left[r(t+T)-r(t)\right]
\end{align}
\end{subequations}
where $f'(t)=1/T'(t)$, $f=1/T$ and $f_D(t)=f'(t)-f$. The
approximations require that
$\left(\left[r(t+T)-r(t)\right]/cT\right)^2$ is ignorable.
Explicitly, $\bar{v}(t)=\left[r(t+T)-r(t)\right]/T$ is the average
speed over one period $T$ of source relative to observer. Accordingly,
\eqref{eq:DEAM-64-a}  and \eqref{eq:DEAM-64-b} can be represented as
\begin{subequations}
\begin{align}\label{eq:DEAM-64p}
  f'(t)=&\frac{f}{1+\displaystyle\frac{\bar{v}(t)}{c}}\\
  \label{eq:DEAM-64p-b}
  f_D(t)=& -\displaystyle\frac{f\bar{v}(t)}{c+\bar{v}(t)}
\end{align}
\end{subequations}

Comparing \eqref{eq:DEAM-64} with \eqref{eq:DEAM-62} shows that, to
use \eqref{eq:DEAM-62}, $f'(t)\approx f$ and $T\rightarrow 0$ are
expected. Alternatively, using $f'(t)=1/T'(t)$ in \eqref{eq:DEAM-64-b}
and then substituting $T'(t)$ by that from \eqref{eq:DEAM-63}, the Doppler
frequency can be expressed as
\begin{align}\label{eq:DEAM-65}
f_D(t)=-\frac{1}{\lambda}\times\frac{\displaystyle\frac{1}{T}\left[r(t+T)-r(t)\right]}{1+\displaystyle\frac{1}{\lambda}\left[r(t+T)-r(t)\right]}
\end{align}
which shows that, to use \eqref{eq:DEAM-62}, $T\rightarrow 0$, and
$\frac{1}{\lambda}\left[r(t+T)-r(t)\right]\approx 0$, meaning that the
distance of source moving within one period should be much smaller
than wavelength. This in turn means that source's moving velocity
should be small, yielding $f'(t)\approx f$.

\subsubsection{ Stationary Source and General Motiving Observer}\label{subsection-Doppler-FS.2.1.2}

%
\begin{figure}[tb]
  \begin{center}
 \includegraphics[angle=0,width=.5\linewidth]{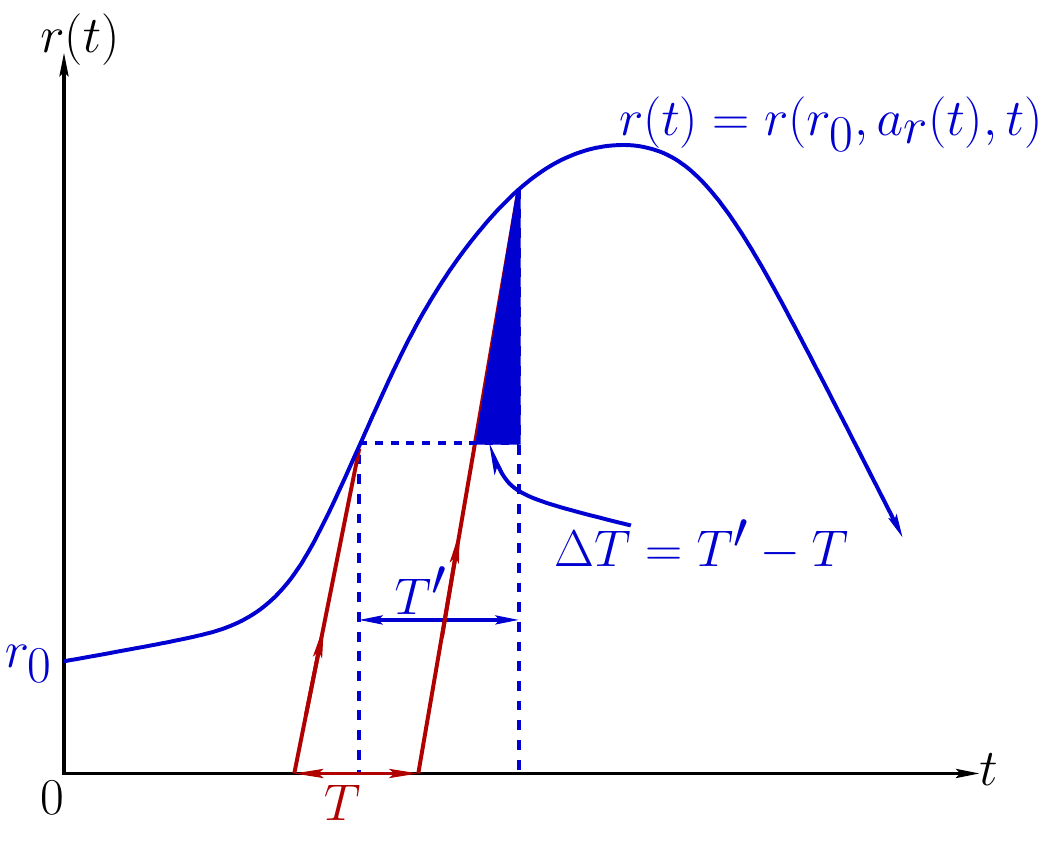}
  \end{center}
  \caption{Graphic representation of the Doppler effect in
    general motion. }
  \label{figure-Doppler-effect-Acceleration-2}
\end{figure}

As an example, Fig.~\ref{figure-Doppler-effect-Acceleration-2} shows
the graphic representation of the scenario where a stationary source
is located at origin, while an observer moves following a motion
equation $r(t)$ with respect to source. Following the previous
description for \eqref{eq:DEAM-63}, there is a relationship of
\begin{align}\label{eq:DEAMad-67}
T'(t')=T+\frac{1}{c}\left[r(t'+T'(t'))-r(t')\right]
\end{align}
where $T'(t')$ is a function of $t'$, which is the time instant when
observer receives the first maxima. From \eqref{eq:DEAMad-67}, we can
obtain
\begin{subequations}
\begin{align}\label{eq:DEAMad-68a}
  f'(t')=&\frac{f}{1+\displaystyle\frac{1}{cT}\left[r(t'+T'(t))-r(t')\right]}\\
  \label{eq:DEAMad-68b}
  f_D(t')=&-\frac{f'(t')}{c}\times\frac{1}{T}\left[r(t'+T'(t))-r(t')\right]
\end{align}
\end{subequations}

Comparing \eqref{eq:DEAMad-68a} and \eqref{eq:DEAMad-68b} respectively
with \eqref{eq:DEAM-64-a} and \eqref{eq:DEAM-64-b} reveals that, when
source is moving and observer is rest, the average speed
$\bar{v}(t)=\left[r(t+T)-r(t)\right]/T$ is source's average speed in
$[t,t+T]$. In contrast, in the scenario where source is stationary
while observer is moving, in \eqref{eq:DEAMad-68a} and
\eqref{eq:DEAMad-68b}, the involved average speed
$\bar{v}(t')=\left[r(t'+T'(t))-r(t')\right]/T$ is the distance variant
over a period of observed signal but averaged by the period of
source's emitted signal. If we make
$\bar{v}'(t')=\left[r(t'+T'(t))-r(t')\right]/T'(t)$ the average speed
of observer in $[t',t'+T'(t)]$, \eqref{eq:DEAMad-68a} and
\eqref{eq:DEAMad-68b} are represented (easier to derive from
\eqref{eq:DEAMad-67}) as
\begin{subequations}
\begin{align}\label{eq:DEAMad-69a}
  f'(t')=&f\left(1-\frac{\bar{v}'(t')}{c}\right) \\
  \label{eq:DEAMad-69b}
  f_D(t')=&-\frac{f\bar{v}'(t')}{c}
\end{align}
\end{subequations}

It is worth noting that when there is acceleration, $\bar{v}(t)$ and
$\bar{v}'(t')$ are different, although both are calculated from the
change of the distance between source and observer made over one
period.

\subsubsection{General Moving Source and Observer}\label{subsection-Doppler-FS.2.1.3}

Following the two scenarios considered above, the Doppler effect in
the scenario where both source and observer are in general motion can
be straightforwardly derived. The analysis is under the assumption
that there is a third reference frame, which seats at a point on the
line connecting source and observer.  With this in mind, assume source
moves according to $r(t)$ and observer moves according to $r'(t')$,
which represent the distances of, respectively, source and observer
relative to this reference frame. Then, following the previous
definitions, we have
\begin{align}\label{eq:DEAMad-71}
T'(t')=T+\frac{1}{c}\left(\left[r(t+T)-r(t)\right]+\left[r'(t'+T'(t'))-r(t')\right]\right)
\end{align}
where, in the second term at the right-hand side, the first bracket is
the distance added by the moving source in $T$, and the second bracket
is the distance added by the moving observer in $T'(t)$. Using
$f'(t')=1/T'(t')$ and $f=1/T$ in \eqref{eq:DEAMad-71} obtains
\begin{align}\label{eq:DEAMad-72}
  f'(t')=f\left[1+\frac{1}{c}\frac{r(t+T)-r(t)}{T}+\frac{1}{c}\frac{f}{f'(t')}\frac{r'(t'+T'(t'))-r(t')}{T'(t')}\right]^{-1}
\end{align}
Let $\bar{v}(t)={r(t+T)-r(t)}/{T}$ and
$\bar{v}'(t')={r'(t'+T'(t'))-r(t')}/{T'(t')}$ represent the average
speed of source in $[t,t+T]$ and the average speed of observer in
$[t',t'+T'(t')]$, respectively. Then, \eqref{eq:DEAMad-72} becomes
\begin{align}\label{eq:DEAMad-73}
  f'(t')=f\left[1+\frac{\bar{v}(t)}{c}+\frac{\bar{v}'(t')}{c}\frac{f}{f'(t')}\right]^{-1}
\end{align}
Simplifying it yields\footnote{Eq.~\eqref{eq:DEAMad-74} can also be
directly derived from \eqref{eq:DEAMad-71} after applying the average
speeds as defined. }
\begin{align}\label{eq:DEAMad-74}
  f'(t')=f\times\displaystyle\frac{1-\displaystyle\frac{\bar{v}'(t')}{c}}{1+\displaystyle\frac{\bar{v}(t)}{c}}
\end{align}
showing that, without considering $f$, it is the product of
\eqref{eq:DEAM-64p} and \eqref{eq:DEAMad-69a}. 

Furthermore, if $({\bar{v}'(t')}/{c})({\bar{v}(t)}/{c})$ and
$({\bar{v}(t)}/{c})^2$ are negligible, an approximation equation is
\begin{align}\label{eq:DEAMad-75}
  f'(t')=f\left(1-\displaystyle\frac{\bar{v}'(t')}{c}-\frac{\bar{v}(t)}{c}\right)
\end{align}
giving the Doppler frequency of
\begin{align}\label{eq:DEAMad-76}
  f_D(t')=\displaystyle -\frac{f\bar{v}'(t')}{c}-\frac{f\bar{v}(t)}{c}
\end{align}

\begin{remark}\label{Remark-DEAM-5}
As considered in Remark~\ref{Remark-DEAM-3}, we may consider to use
one formula of the Doppler effect for all the scenarios. For this
sake, we can let $v(t)=v'(t')$, which represents the relative speed
between source and observer, and then, take the square-root of the
product of \eqref{eq:DEAM-64p} and \eqref{eq:DEAMad-69a}, giving
\begin{align}\label{eq:DEAMad-77}
  f'(t)=&f\sqrt{\displaystyle\frac{1-\frac{\bar{v}(t)}{c}}{1+\frac{\bar{v}(t)}{c}}}
\end{align}

Note again that, if in free space, this formula incorporates the time
dilation resulted from the relative motion between source and
observer, but only in terms of their average speed. Further details on
the relativistic Doppler effect in general motion will be provided in
Section~\ref{subsection-Doppler-FS.2.2}.
\end{remark}

Below the Doppler effect in two application scenarios is considered, where EM waves are emitted from acceleration objects.

\subsubsection{Linearly Uniform Acceleration}\label{subsubsection-Doppler-FS.2.1.1}

%
\begin{figure}[tb]
  \begin{center}
 \includegraphics[angle=0,width=.5\linewidth]{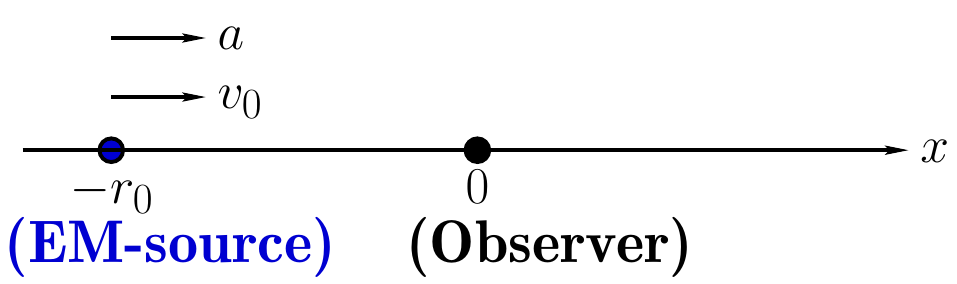}
  \end{center}
  \caption{Uniformly accelerating EM-source moves in $x$-axis direction.}
  \label{figure-Classic-Uniform-Acceler}
\end{figure}

Assume on a straight line represented by $x$-axis, as shown in
Fig.~\ref{figure-Classic-Uniform-Acceler}, an observe seats at the
origin, while an EM-wave source moves in the $x$-direction. Assume that,
at $t=0$, source is at $-r_0$ and moves at the speed of $v_0$. The
acceleration is a constant $a$. Then, the distance from source to
observer is give by the equation
\begin{align}\label{eq:DEAM-66}
r(t)=\left|x(t)=-r_0+v_0t+\frac{at^2}{2}\right|,~t\geq 0
\end{align}
Accordingly, following \eqref{eq:DEAM-63}, when source is
approaching observer, meaning that both $x(t)<0$ and $x(t+T)<0$, the
Doppler effect is
\begin{align}\label{eq:DEAM-67}
  T'(t)=&T+\frac{1}{c}\left[r(t+T)-r(t)\right] \nonumber\\
  =&T+\frac{1}{c}\left[-x(t+T)+x(t)\right]\nonumber\\
  =&\frac{T}{c}\left[c-v_0-at-aT/2\right]
\end{align}
When source is leaving observer, meaning that both $x(t)>0$ and
$x(t+T)>0$, the Doppler effect is then
\begin{align}\label{eq:DEAM-68}
  T'(t) =&\frac{T}{c}\left[c+v_0+at+aT/2\right]
\end{align}

From \eqref{eq:DEAM-67} and \eqref{eq:DEAM-68},
the Doppler frequency can be found to be
\begin{align}\label{eq:DEAM-68k}
  f'(t) =&f\left(1 \pm \frac{\bar{v}(t)}{c}\right)^{-1}
\end{align}
where $\bar{v}(t)=v_0+at+aT/2$ is the average speed in the time
$[t,t+T]$, while `$+$' and `$-$' are for leaving and approaching observer,
respectively.

In addition to the above cases, there is a situation that source
passes observer between the transmissions of two adjacent crests,
which correspond to $x(t)<0$ but $x(t+T)>0$. In this situation, the
period of received signal by observer can still be found using
\eqref{eq:DEAM-63}, giving
\begin{align}\label{eq:DEAM-69}
  T'(t)=&T+\frac{1}{c}\left[r(t+T)-r(t)\right] \nonumber\\
  =&T+\frac{1}{c}\left[x(t+T)+x(t)\right]\nonumber\\
  =&\frac{1}{c}\left[cT-2r_0+2v_0t+v_0T+at^2+aTt+aT^2/2\right]
\end{align}
Due to the involvement of $-2r_0$, the Doppler effect can be positive
or negative, depending on $t$ when the first crest is emitted.

\begin{figure}[tb]
  \begin{center}
 \includegraphics[angle=0,width=.65\linewidth]{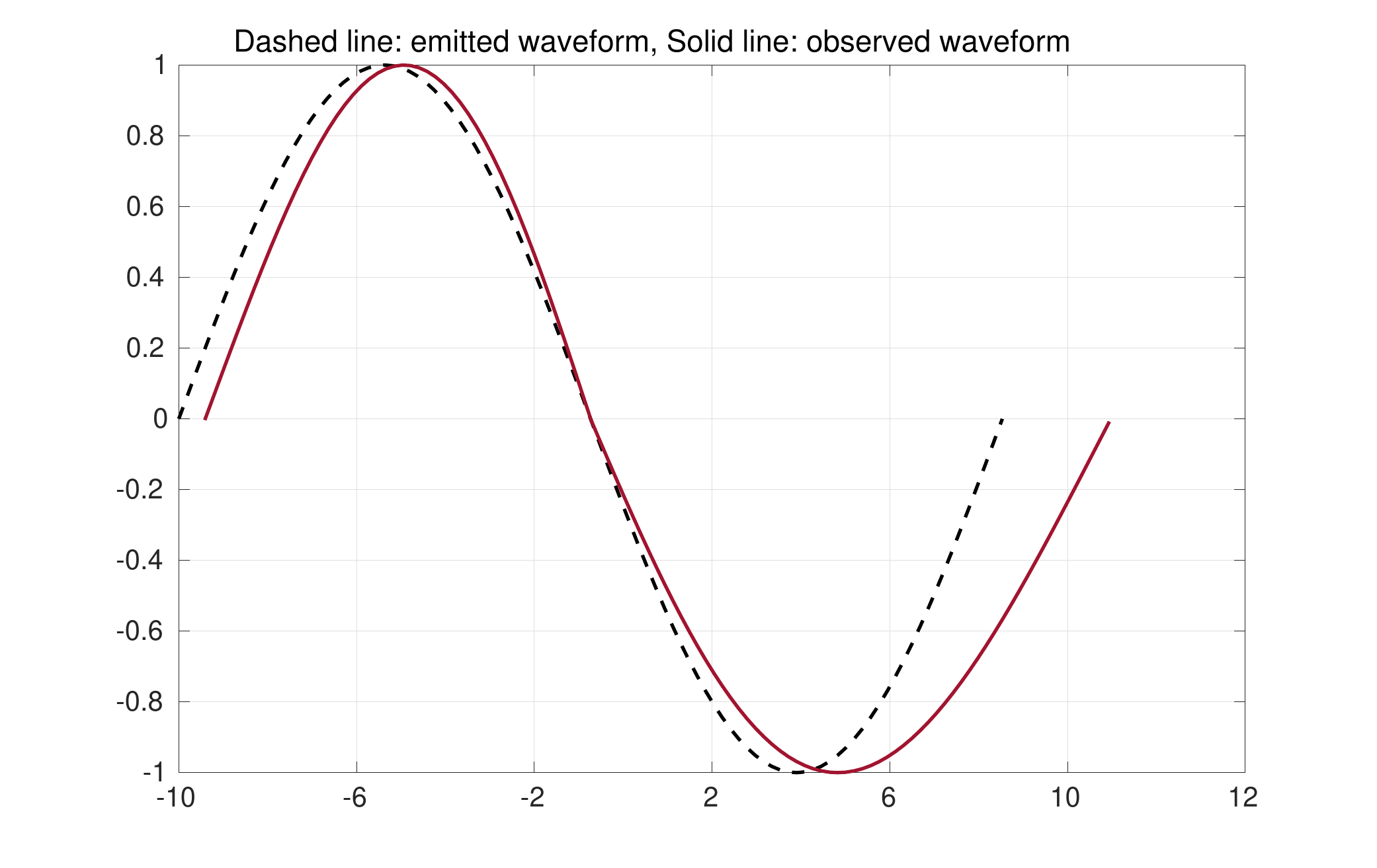}
  \end{center}
  \caption{Ilustration of Doppler distortion, where the frond part is
    contracted and the rear part is dilated.}
  \label{figure-Doppler_distortion}
\end{figure}

While the period in this situation can be found by \eqref{eq:DEAM-69},
we should note that the waveform of a period at the instant of source
passing observer is different from that in the first two cases
considered above.  In the first and second cases, the Doppler effect
is either positive or negative over a whole period, predicated
by \eqref{eq:DEAM-67} or \eqref{eq:DEAM-68}.  By contrast, in the
third case, over one period, the front part waveform experiences
positive Doppler shift, while the rear part receives negative Doppler
shift.

For example, assume that a source emitting single-tone sine-waveform
passes the observer exactly in the middle of a waveform, i.e., at the
phase of $(2n+1)\pi$. Then, the observed waveform containing the
passing instant may look like the one shown in
Fig.~\ref{figure-Doppler_distortion}. The front half of
original waveform is squeezed, while the rear half is stretched. If
acceleration is positive, the amount of stretching is more than that
of squeezing.

Due to the involvement of $T^2$, it is cumbersome to directly obtain
the closed-from equations for the Doppler frequency from
\eqref{eq:DEAM-67} and \eqref{eq:DEAM-68}. For that, let us express
the distance travelled with an initial speed $0$ in a period $T$ be
expressed as $H=aT^2/2$, which is a constant. From it we obtain
$T=\sqrt{2H/a}$. Replacing $T$ by this in the brackets of \eqref{eq:DEAM-67}
and \eqref{eq:DEAM-68}, replacing $T'(t)=1/f'(t)$ and the other $T$ by
$T=1/f$, and combining \eqref{eq:DEAM-67} and \eqref{eq:DEAM-68} into
one give
\begin{align}\label{eq:DEAM-E70}
  f'(t) =&\frac{f}{\left(1\pm\frac{v_0+at+\sqrt{\frac{aH}{2}}}{c}\right)}
  =\frac{f\left(1\mp\frac{v_0+at+\sqrt{\frac{aH}{2}}}{c}\right)}{\left(1\pm\frac{v_0+at+\sqrt{\frac{aH}{2}}}{c}\right)^2}\nonumber\\
  \approx & f\left(1\mp\frac{v_0+at+\sqrt{\frac{aH}{2}}}{c}\right)
\end{align}
where the last approximation is due to the assumption of
$c>>\left|v_0+at+\sqrt{\frac{aH}{2}}\right|$. From this last approximation, the
Doppler frequency is
\begin{align}\label{eq:DEAM-E71}
  f_D(t) 
  \approx & \pm \frac{f\left(v_0+at+\sqrt{\frac{aH}{2}}\right)}{c}
\end{align}
Specifically, when source starts moving at $t=0$ with an initial speed
$v_0=0$, the Doppler frequency at $t=0$ is
\begin{align}\label{eq:DEAM-E72}
  f_D \approx &  \pm \frac{f\sqrt{\frac{aH}{2}}}{c}
\end{align}
which shows explicitly the contribution of acceleration to Doppler
shift.

Note that an enquiry of \eqref{eq:DEAM-67}, \eqref{eq:DEAM-68} and
\eqref{eq:DEAM-E71} discovers that there is a problem to apply these
formulas, as for a constant acceleration $a$, speed $at$ - and hence the Doppler frequency - goes towards infinity with the increase of $t$. Therefore, these formulas can only be used for a resulted speed that is small enough to be considered ignorable when relative to $c$ of the speed of light. Otherwise, the relativistic
Doppler effect discussed in Section~\ref{subsection-Doppler-FS.2.2}
should be considered.

\subsubsection{Circularly Uniform Motion}\label{subsubsection-Doppler-FS.2.1.2}

%
\begin{figure}[tb]
  \begin{center}
 \includegraphics[angle=0,width=.65\linewidth]{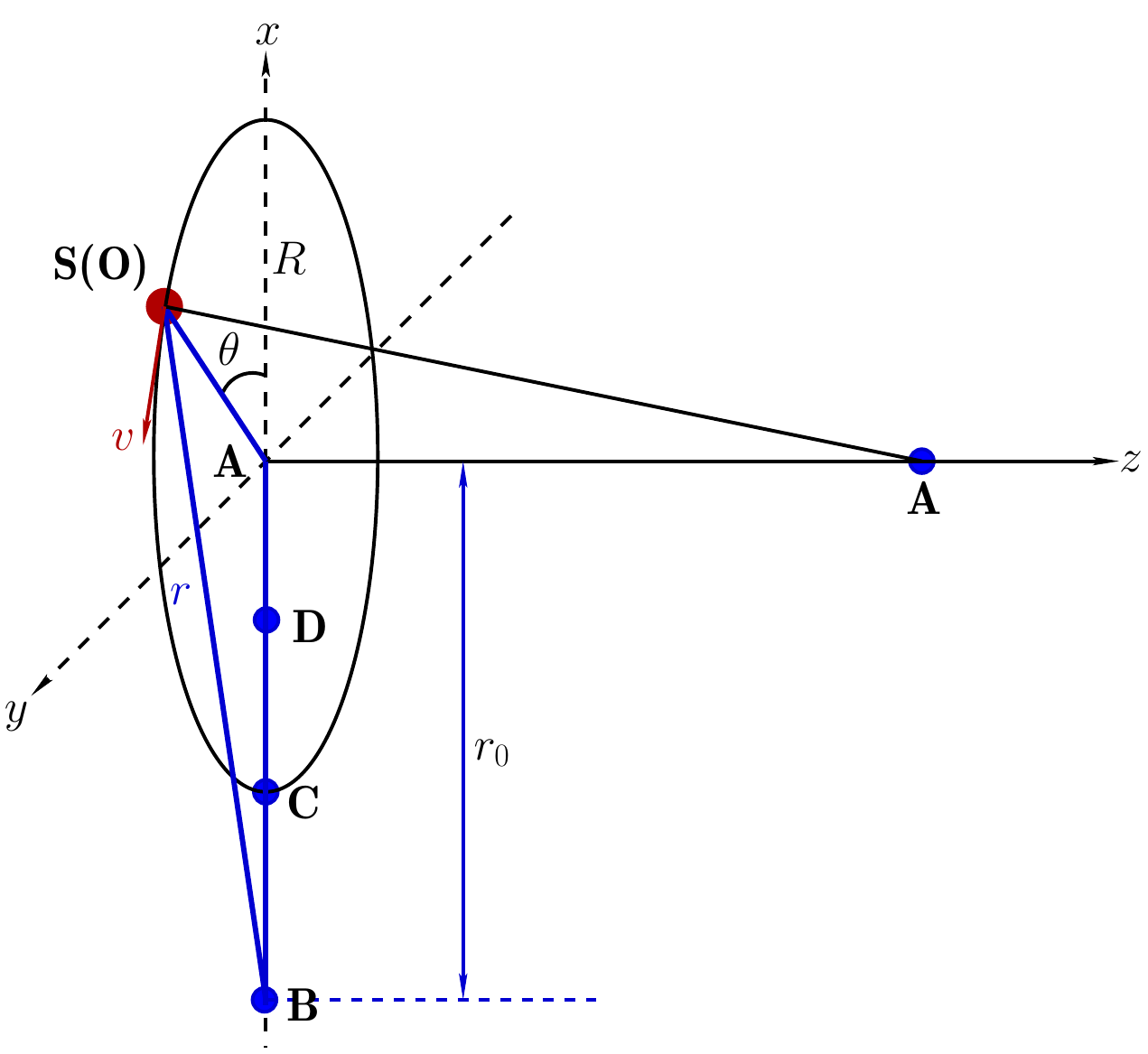}
  \end{center}
  \caption{Illustration of circularly uniform acceleration, where {\bf
      A}, {\bf B}, {\bf C} and {\bf D} are source/observer locations,
    and {\bf S} ({\bf O}) is for moving source (observer).}
  \label{figure-Circular-Acceleration}
\end{figure}

It is well known that the circularly uniform motion exists an
centripetal acceleration, whose magnitude is constant while direction
is always toward the center of the circle. To analyze the Doppler
effect of this system, as shown in
Fig.~\ref{figure-Circular-Acceleration}, we assume that node {\bf
  S}({\bf O}) - which may be a satellite, spaceship, a sensor on a
flying wheel, etc., and may be source or observer - moves circularly
on the $xy$-plane with a constant speed $v$. The radius of the circle
is represented by $R$. Then, assuming that {\bf S}({\bf O}) is
positioned at the positive direction of $x$-axis at $t=0$, the phase
$\theta$ is
\begin{align}\label{eq:DEAM-70}
\theta(t)=\frac{vt}{R}=\omega t,~t\geq 0
\end{align}
where $\omega=v/R$ is the angular speed. The period and frequency of
{\bf S}({\bf O}) moving are $T_0=2\pi/\omega$ and $f_0=\omega/2\pi$,
respectively.

Now let us analyze the Doppler effect, when another node seats on the
$z$-axis or $x$-axis, such as at location {\bf A}, {\bf B}, {\bf C} or
{\bf D}. This node may be a ground station in satellite systems, a
sensor in industrial applications, etc.

First, consider the situation that {\bf S}({\bf O}) is the signal
source and, correspondingly, the nodes at {\bf A}, {\bf B}, {\bf C} or
{\bf D} are observers. Hence, {\bf S}({\bf O}) is reduced to {\bf S}
in description. The analysis of Doppler effect is as follows.

When an observer is on the $z$-axis, such as, at the location of {\bf
  A}, the distance from any point on the circle to {\bf A} is
equal. Hence, without considering the effect from acceleration itself
in terms of general relativity, the Doppler effect is zero. This can
also be verified by the fact that the velocity
$\bm{v}=v\hat{\bm{v}}$ is always perpendicular with the line
connecting source and an observer, provided that it is on the $z$-axis,
including the circle's center.

When an observer is on the $x$-axis, we only need to consider the case
that observer is on the negative side of $x$-axis, such as at location
{\bf B}, {\bf C} or {\bf D}. Furthermore, owing to symmetry, this
scenario is general, provided that observer is in the
$xy$-plane\footnote{This model can mirror many application scenarios,
for example, a sensor on a satellite (ground station) observes the
signals sent from a ground station (satellite), or a stationary
observer receives the signals sent by a sensor installed on a wheel in
industrial applications.}. Assume that the distance between an
observer and the origin of circle is $r_0$, as shown in
Fig.~\ref{figure-Circular-Acceleration}. Accordingly, the distance
between source and observer is
\begin{align}\label{eq:DEAM-71}
  r(t)=&\sqrt{R^2\sin^2 \omega t+\left(R\cos \omega t+r_0\right)^2}\nonumber\\
  =&R\sqrt{\left(1-\frac{r_0}{R}\right)^2+\frac{4r_0}{R}\cos^2\left(\frac{\omega t}{2}\right)},~t\geq 0\\
  \label{eq:DEAM-72}
  =&r_0\sqrt{\left(1-\frac{R}{r_0}\right)^2+\frac{4R}{r_0}\cos^2\left(\frac{\omega t}{2}\right)},~t\geq 0
\end{align}
Taking the derivative of, say \eqref{eq:DEAM-71}, with respect to $t$
gives the speed of source relative to observer expressed as
\begin{align}\label{eq:DEAM-71a}
  v_{s-o}(t)=&\frac{dr(t)}{dt}\nonumber\\
  =&-\frac{r_0v\sin(\omega t)}{R\sqrt{\left(1-\frac{r_0}{R}\right)^2+\frac{4r_0}{R}\cos^2\left(\frac{\omega t}{2}\right)}}
\end{align}
Hence, we have an approximate formula for the Doppler frequency expressed as
\begin{align}\label{eq:DEAM-71b}
  f_D(t)= \frac{r_0v\sin(\omega t)}{\lambda R\sqrt{\left(1-\frac{r_0}{R}\right)^2+\frac{4r_0}{R}\cos^2\left(\frac{\omega t}{2}\right)}}
\end{align}

Before deriving the more accurate formulas for the Doppler effect,
several special cases can be obtained from \eqref{eq:DEAM-71b}. First,
when $r_0\rightarrow 0$, meaning that observer is near the center of
circle, \eqref{eq:DEAM-71} gives constant distance of $r(t)=R$ and
\eqref{eq:DEAM-71b} gives $f_D=0$. Second, if $r_0\rightarrow \infty$,
\eqref{eq:DEAM-71b} can be simplified to
\begin{align}\label{eq:DEAM-71c}
  f_D(t)= \frac{v\sin(\omega t)}{\lambda }
\end{align}
Straightforwardly, $f_D(t)=0$, when $wt=n\pi$ with $n$ being an
integer, and $|f_D(t)|=v/\lambda=fv/c$ is maximum, when $\omega t=n\pi\pm
\pi/2$. Furthermore, when $r_0=R$, i.e., when observer is at {\bf C}, 
\begin{align}\label{eq:DEAM-71d}
  f_D(t)= \frac{v\sin(\omega t)}{2\lambda \left|\cos\frac{\omega t}{2}\right|}=\pm\frac{v\sin\left(\frac{\omega t}{2}\right)}{\lambda }
\end{align}
where $+/-$ is determined by whether source is approaching or leaving
observer. Explicitly, the maximum Doppler effect is generated when
$\omega t=(2n+1)\pi$.

Now, let us derive the more accurate formula for the Doppler effect.
Assume that source emits two adjacent maxima at $t_1=t$ and
$t_2=t+T$. Then, observer receives the two maxima at $t'_1=t+r(t)/c$
and $t'_2=t+T+r(t+T)/c$. The period of received signal by observer is
given by
\begin{align}\label{eq:DEAM-73}
  T'(t)=t'_2-t'_1=T+\frac{r(t+T)- r(t)}{c}
\end{align}
After substituting \eqref{eq:DEAM-72} into \eqref{eq:DEAM-73}, we obtain
\begin{align}\label{eq:DEAM-74}
  T'(t)=T+\frac{r_0}{c}&\left[\sqrt{\left(1-\frac{R}{r_0}\right)^2+\frac{4R}{r_0}\cos^2\left(\frac{\omega (t+T)}{2}\right)}\right.\nonumber\\
    &\left.-\sqrt{\left(1-\frac{R}{r_0}\right)^2+\frac{4R}{r_0}\cos^2\left(\frac{\omega t}{2}\right)}\right],~t\geq 0
\end{align}
This is a general formula that can be used for predicting the Doppler
effect, when an observer lying on the same plane of the source's
moving circle receives signals from the source, after necessary
rotation to make $x$-axis align with the observer. For example, if
{\bf S} is a satellite, a ground station may be located at {\bf D}.
{\bf S} at {\bf C} is the situation that the satellite is directly
overhead of the ground station.

Specifically, if observer is at {\bf C} as seen in
Fig.~\ref{figure-Circular-Acceleration} , we have $r_0=R$. Applying
this into \eqref{eq:DEAM-74} and after some simplification, we obtain
\begin{align}\label{eq:DEAM-75}
T'=T+\frac{2R}{c}\left[\left|\cos\left(\frac{\omega (t+T)}{2}\right)\right|-\left|\cos\left(\frac{\omega t}{2}\right)\right|\right],~t\geq 0
\end{align}
Referring to Fig.~\ref{figure-Circular-Acceleration}, if $(\omega t)_{
  \mod 2\pi}\geq 0$ and $(\omega (t+T))_{ \mod 2\pi}\leq \pi$, where
`$\mod$' is modulo-operation, \eqref{eq:DEAM-75} becomes
\begin{align}\label{eq:DEAM-76}
  T'=&T+\frac{2R}{c}\left[\cos\left(\frac{\omega (t+T)}{2}\right)-\cos\left(\frac{\omega t}{2}\right)\right]\nonumber\\
  =&T-\frac{4R}{c}\sin\left(\frac{\omega T}{4}\right)\sin\left(\frac{\omega (2t+T)}{4}\right),~t\geq 0
\end{align}
If $(\omega t)_{ \mod 2\pi}\geq \pi$ and $(\omega (t+T))_{ \mod
  2\pi}\leq 2\pi$, \eqref{eq:DEAM-75} becomes
\begin{align}\label{eq:DEAM-77}
  T'=&T+\frac{2R}{c}\left[-\cos\left(\frac{\omega (t+T)}{2}\right)+\cos\left(\frac{\omega t}{2}\right)\right]\nonumber\\
  =&T+\frac{4R}{c}\sin\left(\frac{\omega T}{4}\right)\sin\left(\frac{\omega (2t+T)}{4}\right),~t\geq 0
\end{align}
Furthermore, if $\omega T\approx 0$, \eqref{eq:DEAM-76} and
\eqref{eq:DEAM-77} can be approximated as
\begin{align}\label{eq:DEAM-78}
T'\approx T\left[1\mp \frac{\omega R}{c}\sin\left(\frac{\omega t}{2}\right)\right],~t\geq 0
\end{align}
when they are represented in one compact form. Accordingly, the
Doppler frequency is
\begin{align}\label{eq:DEAM-78a}
  f_D(t)\approx& \frac{\pm\frac{v}{\lambda}\sin\left(\frac{\omega t}{2}\right)}{1\mp \frac{v}{c}\sin\left(\frac{\omega t}{2}\right)}\nonumber\\
  \approx &\pm\frac{v}{\lambda}\sin\left(\frac{\omega t}{2}\right)
\end{align}
where $v/c\rightarrow 0$ is applied for obtained the second equation
from the first one. Notice that \eqref{eq:DEAM-78a} is the same as
\eqref{eq:DEAM-71d}.

Now, let us consider the case that {\bf S}({\bf O}) is the observer,
hence using {\bf O} to avoid confusion, while source is at {\bf A},
{\bf B}, {\bf C} or {\bf D}. Again, it is easy to analyze that the
Doppler effect is zero, provided that source is on the $z$-axis
including {\bf A} and the centre of the circle. For the other cases
that source is at {\bf B}, {\bf C} or {\bf D}, assume that source
emits two adjacent crests at $t_1=t$ and $t_2=t+T$. Assume that {\bf
  O} is on the positive $x$-axis at $t=0$. Then, {\bf O} receives the
first crest at $t_1'=t+r(t_1')/c$ and the second crest at
$t_2'=t+T+r(t_1'+T'(t_1'))/c$. Denoting $t'=t_1'$, we can obtain
a relation between $T'(t')$ and $T$ expressed as
\begin{align}\label{eq:DEAM-109}
  T'(t')=T+\frac{r_0}{c}&\left[\sqrt{\left(1-\frac{R}{r_0}\right)^2+\frac{4R}{r_0}\cos^2\left(\frac{\omega (t'+T'(t'))}{2}\right)}\right.\nonumber\\
    &\left.-\sqrt{\left(1-\frac{R}{r_0}\right)^2+\frac{4R}{r_0}\cos^2\left(\frac{\omega t'}{2}\right)}\right]
\end{align}
with $t'=t+r(t')/c$, $t\geq 0$.

As seen in \eqref{eq:DEAM-109}, $t'$ and $T'(t')$ are included in the
cosine term at the right-hand side, making it hard to simplify the
equation to obtain a straightforward relation between $T'(t')$ and
$T$. In this case, the Doppler frequency may only be obtained by
numerical search for a $T'(t')$ with $t'=t+r(t')/c$, to make.
\begin{align}\label{eq:DEAM-110}
\Bigg| T'(t')-T-\frac{r_0}{c}&\left[\sqrt{\left(1-\frac{R}{r_0}\right)^2+\frac{4R}{r_0}\cos^2\left(\frac{\omega (t'+T'(t'))}{2}\right)}\right.\nonumber\\
    &\left.\left.-\sqrt{\left(1-\frac{R}{r_0}\right)^2+\frac{4R}{r_0}\cos^2\left(\frac{\omega t'}{2}\right)}\right]\right|\rightarrow 0
\end{align}

In the evaluation of the Doppler effect in this case, two
approximation methods can be considered.

First, assume the case of large $R$, which is the case, for example,
of satellite communications. In this kind of applications,
approximations of $r(t')\approx r(t)$ and $r(t'+T'(t'))\approx r(t+T)$
can be very accurate. Accordingly, the same relation as
\eqref{eq:DEAM-74} can be obtained, except that $T'(t)$ needs to be
replaced by $T'(t')$. Since $t'\approx t+r(t)/c$, comparing to the
case where source moving observer stationary, the same Doppler
frequency in the case of source stationary observer moving comes
slightly later, by a time of about $r(t)/c$. Note that, due to the
same relation of \eqref{eq:DEAM-74}~\footnote{Eq.~\eqref{eq:DEAM-74}
can be easily computed, as $T'(t)$ is a simple function of $T$.}, all
the specific cases considered following \eqref{eq:DEAM-74} are also
satisfied. This explains that when $R$ is large, the same formulas can
be simultaneously used in both the case of source moving and observer
stationary, and the case of source stationary and observer moving.

Second, assume a small $R$ and also that the distance between source
and observer is relatively small, which is usually the case in indoor
and industrial applications. If $R$ is small, the motion spanning one
wavelength may be significant. However, as the result of a small
distance between source and observer, the wave propagation time from
source to observer can be ignored, yielding $t'\approx t$. Hence,
\eqref{eq:DEAM-109} can be approximated by a formula of
\begin{align}\label{eq:DEAM-111}
  T'(t)=T+\frac{r_0}{c}&\left[\sqrt{\left(1-\frac{R}{r_0}\right)^2+\frac{4R}{r_0}\cos^2\left(\frac{\omega (t+T'(t))}{2}\right)}\right.\nonumber\\
    &\left.-\sqrt{\left(1-\frac{R}{r_0}\right)^2+\frac{4R}{r_0}\cos^2\left(\frac{\omega t}{2}\right)}\right],~t\geq 0
\end{align}
Nonetheless, due to the term of $\cos^2\left(\frac{\omega
  (t+T'(t))}{2}\right)$, it is not straightforward to obtain a simple
relation between $T'(t)$ and $T$. Hence, searching methods may be
needed for finding $T'(t)$ for given $T$.

\begin{figure}[th]
  \begin{center}
 \includegraphics[angle=0,width=.65\linewidth]{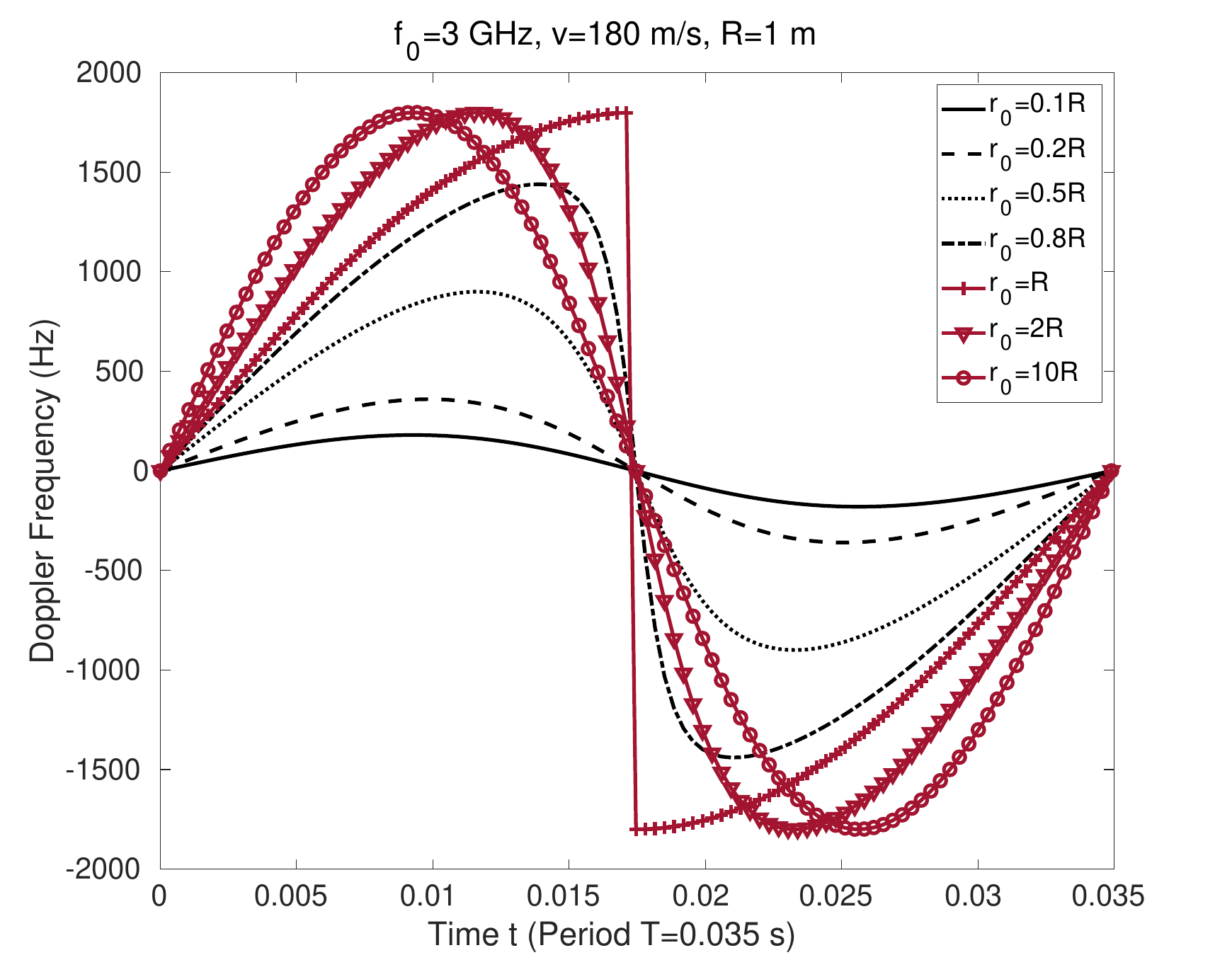}
  \end{center}
  \caption{Doppler effect of a circularly moving source on its
    transmitted radio signals received by a stationary observer.}
  \label{figure-Circular_Doppler_Effect1}
\end{figure}
\begin{figure}[th]
  \begin{center}
 \includegraphics[angle=0,width=.65\linewidth]{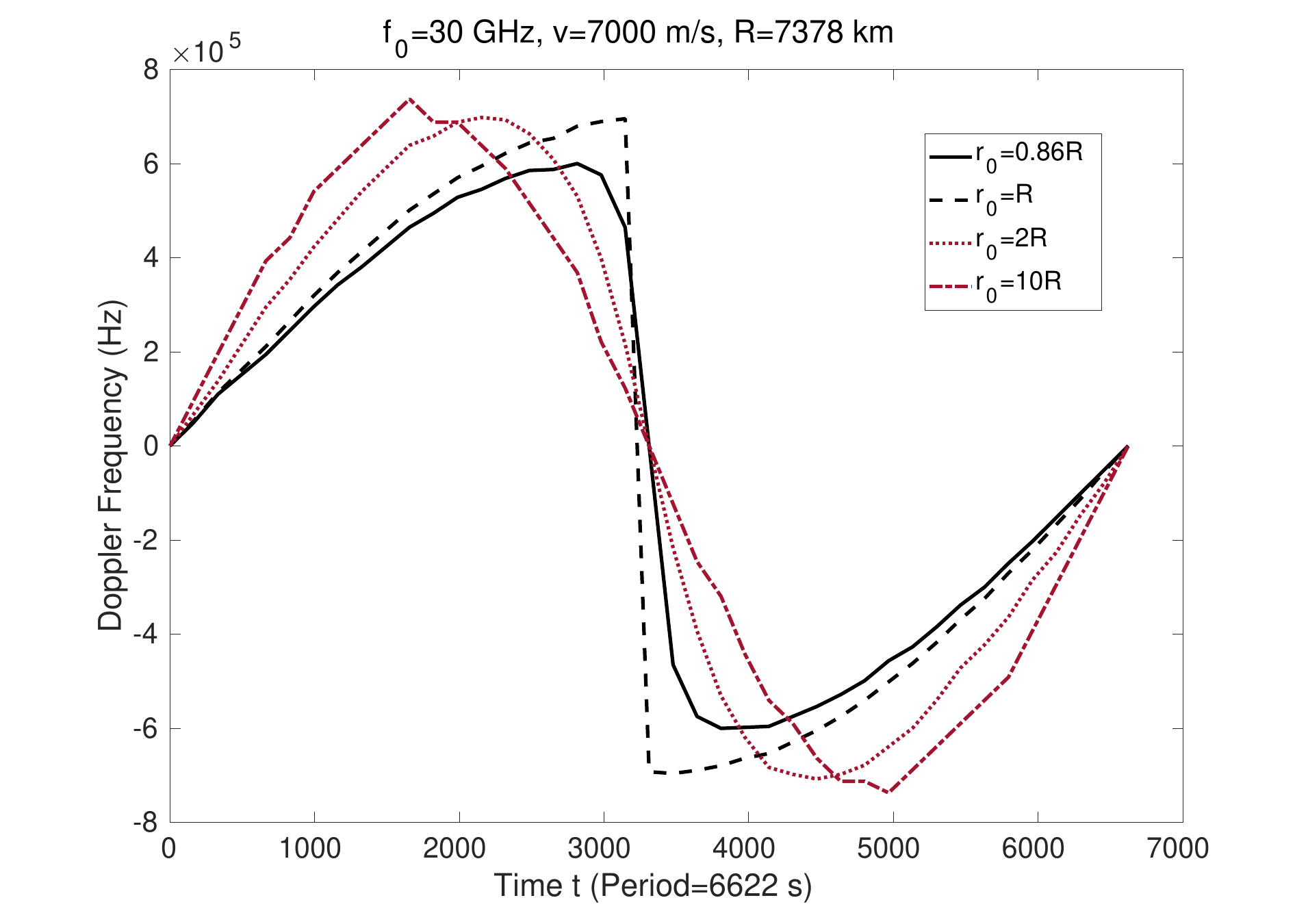}
  \end{center}
  \caption{Doppler effect of a circularly moving source on its
    transmitted radio signals received by a stationary observer.}
  \label{figure-Circular_Doppler_Effect2}
\end{figure}

\begin{example}\label{Example-DE-4x}
Fig.~\ref{figure-Circular_Doppler_Effect1} shows the Doppler frequency
generated by a radio emitter installed on, such as, a wheel of one
meter of radius, rotating at a speed of about 29 turns per second. The
observer is on the same plane of the wheel and has the different
distances from the wheel center as shown in the figure. It shows that,
when $r_0<R$, the Doppler frequency increases as $r_0$ increases, with
the phase of generating the maximum Doppler frequency shifting towards
$\omega t=2n+ \pi$ as $r_0$ increases towards $R$. At $r_0=R$, the
change of Doppler frequency is not continuous at $\omega t=2n+\pi$. In
contrast, when $r_0>R$, the maximum Doppler frequency is not much
affected by $r_0$, which can be implied by \eqref{eq:DEAM-71b} - where
the $\cos(\omega t)$ term is insignificant - and
\eqref{eq:DEAM-71c}. Furthermore, the phase giving the maximum Doppler
frequency converges to $\omega t=2n\pi\pm\pi/2$, which is also implied
by \eqref{eq:DEAM-71c}.
\end{example}

\begin{example}\label{Example-DE-5g}
Fig.~\ref{figure-Circular_Doppler_Effect2} shows the Doppler frequency
generated by a satellite travelling at a speed of about 7000 m/s
relative to the Earth's surface, when the radio signals emitted by the
satellite are detected by an observer at $r_0=0.86R$, on the Earth's
surface and at the opposite pole of satellite's orbit $r_0=R$, and at
two other altitudes of $r_0=2R$ and $10R$, respectively. Explicitly,
the Doppler effect appears similarly as that shown in
Fig.~\ref{figure-Circular_Doppler_Effect1}, with the actual Doppler
frequency depended on the radio frequency, in addition to the other
parameters as above-mentioned.
\end{example}

\subsection{Relativistic Doppler Effect in General Motion}\label{subsection-Doppler-FS.2.2}

To deal with the relativistic Dopper effect, the analysis in
Section~\ref{subsubsection-Doppler-FS.1.3} explained that \emph{the
relativistic Doppler effect is determined by two proper time intervals
relating to two different pairs of events occurring in two different
reference frames, i.e., the reference frame of source and the
reference frame of observer}~\ncite{Zanchini_2025}. Hence, given a
proper time interval, expressed as $T$, in the source's reference
frame, the proper time interval, referred to as $T'$, in an observer's
reference frame is not only contributed by the proper time interval
itself in the source's reference frame, but also by the effect from
the relative motion - uniform or accelerated motion - between source
and observer.  Section~\ref{subsubsection-Doppler-FS.1.3} considered
the relativistic Doppler effect in uniform motion, which yields
inertial reference frames. In this section, the relativistic Doppler
effect in the general motion scenarios where acceleration may exist is
analyzed. Hence, non-inertial frames instead of inertial frames must
be considered. Below, the relativistic Doppler effect in general
motion is first considered. Then, the relativistic Doppler effects in
several special application scenarios experiencing accelerated motions
are analyzed. Main references followed for the analysis are
\ncite{Zanchini_2025,Zanchini_2012,Chen_2014,KHOLMETSKII2020168191}.

Note that in the analysis below, we again assume that the relative
velocity between source and observer is positive when they move away
from each other, while negative when they move towards each other. We
assume that source and observer do not pass each other during the time
of considering the Doppler effect. Otherwise, the velocity within a
period has both positive and negative Doppler effect, making a part of
waveform contracted and the other part stretched, as that shown in
Fig.~\ref{figure-Doppler_distortion}.

In the analysis of the Doppler effect in systems with general motion,
one issue should be emphasised is - as pointed out in
\ncite{Zanchini_2025} - the time dilatation formula is correct only
when applied with an inertial frame of reference, but not correct when
applied with a non-inertial frame. In other words, the time dilatation
formula should not be mistakenly applied with the frames in
acceleration.

Additionally, the general analysis is focused, while formulas in
special cases are provided. Hence, the analysis is based on the
definitions of $\bm{v}(t)$, $\bm{a}(t)$, $v$, $r(t)$, $c$, $c'$,
etc., having similar meanings as those defined in
Section~\ref{subsubsection-Doppler-FS.1.3}.

\begin{figure}[th]
  \begin{center}
 \includegraphics[angle=0,width=.65\linewidth]{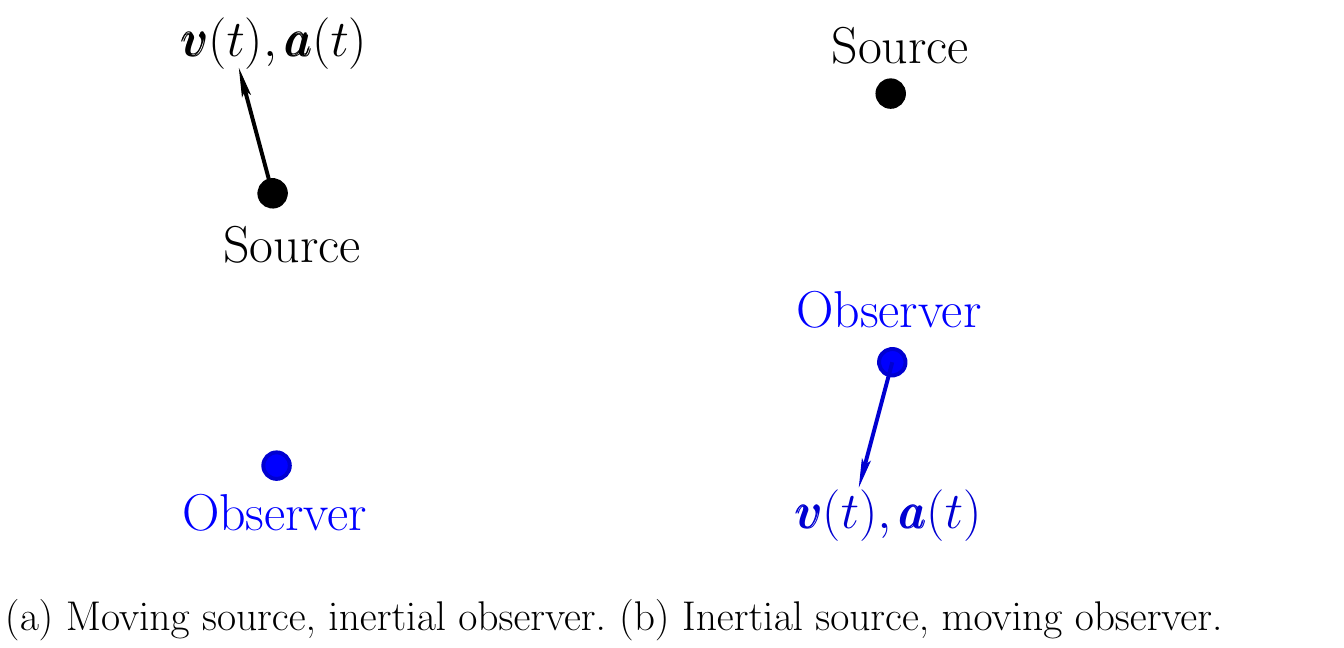}
  \end{center}
  \caption{Scenarios of (a) random moving source and inertial observer, and (b) inertial source and random moving observer.}
  \label{figure-Relativistic-Acceleration-DE}
\end{figure}
%

\subsubsection{General Moving Source and Inertial Observer}\label{subsection-Doppler-FS.2.2.1}

For the scenario of a moving source sending EM signals to a stationary
observer in a medium, as shown in
Fig.~\ref{figure-Relativistic-Acceleration-DE}(a), there are three
reference frames involved, the non-inertial reference frame of source,
the inertial reference frame of medium and the inertial reference
frame of observer, which is rest in medium and hence the same inertial
reference frame of medium. To analyse the Doppler effect, as in
Section~\ref{subsubsection-Doppler-FS.1.3}, three types of times,
periods and frequencies are defined, the proper time $t$, proper
period $T$ and the proper frequency $f$ of source in the reference
frame of source, the time $t''$, period $T''(t'')$ and the frequency
$f''(t'')$ of the source in the reference frame of medium, and the
proper time $t'$, proper period $T'(t)$ and the proper frequency
$f'(t)$ of observer in the reference frame of observer.  The distance
between source and observer in the reference frame of medium is
expressed as $r(t'')$ and all velocities are measured with respect to
the reference frame of medium. Since observer rests in medium's
reference frame, we have $t''=t'$, i.e., the clock holding by observer
runs synchronously as that in medium.

As explained in Section~\ref{section-Doppler-FS.01a}, Dopper effect
can be explained by the ratio of two periods or two frequencies of the
emitted signal at source and, correspondingly, received signal by
observer. When relativisitc Doppler effect is considered, an
refinement is the ratio of two proper periods or two proper
frequencies, expressed as
\begin{align}\label{eq:DEAM-87}
  \frac{T}{T'(t')}=\frac{f'(t')}{f}
\end{align}
From these ratios, the Doppler frequency can be expressed as
\begin{align}\label{eq:DEAM-88}
  f'_D(t')=f\left(\frac{f'(t')}{f}-1\right)=\frac{1}{T}\left(\frac{T}{T'(t')}-1\right)
\end{align}

Since the general moving non-inertial source and inertial observer are
considered, the time dilation should be derived with respect to the
inertial reference frame of observer (or medium). Assume that two
adjacent maxima are emitted by source at $t_1=t$ and $t_2=t+T$ in its
reference frame. Accordingly, when observed in the reference frame of
medium, these two maxima are emitted at $t_1''=t''$ and
$t_2''=t''+T''(t'')$\footnote{Note that, while $T$ is independent of
$t$, $T''(t'')$ is dependent on $t''$, since the transformation from
the non-inertial reference frame of source to the inertial reference
frame of observer is dependent on the velocity of source, which is a
function of $t''$. } at the distances of $r(t'')$ and
$r(t''+T''(t''))$. Hence, following
Section~\ref{subsection-Doppler-FS.2.1} (See
Fig.~\ref{figure-Doppler-effect-Acceleration}), $T'(t')$ and
$T''(t'')$ in the reference frame of medium follow the relationship of
\begin{align}\label{eq:DEAM-89}
 T'(t')=T''(t'')+\frac{r(t''+T''(t''))-r(t'')}{c'}
\end{align}
To find the Doppler effect of $T/T'(t')$, a relation between $T$ and
$T''(t)$ is needed.  In principle, this can be obtained from the
differential equation built on the time dilation relationship of
\begin{align}\label{eq:DEAM-90}
  dt''=\frac{dt}{\sqrt{1-\beta^2(t'')}}~\leftrightarrow~dt=\sqrt{1-\beta^2(t'')}dt''
\end{align}
{where $\beta(t'')=|\bm{v}(t'')|/c$ with $|\bm{v}(t'')|$ being the
  speed of source measured in the reference frame of medium. As above
  assumed, $t_1''=t''$ corresponds to $t_1=t$, and $t_2''=t''+T''(t)$
  corresponds to $t_2=t+T$. Hence, upon integrating \eqref{eq:DEAM-90}
  over these ranges, we can obtain a relationship between $T''(t)$ and
  $T$ as
\begin{align}\label{eq:DEAM-91}
  T=\int_{t''}^{t''+T''(t'')}\sqrt{1-\beta^2(\tau)}d\tau
\end{align}
Based on \eqref{eq:DEAM-91}, once a $\bm{v}(t)$ is given, we can
obtain a function $T=f(T''(t''),t'')$, from which $T''(t'')$ can be
obtained. Then, substituting it into \eqref{eq:DEAM-89}, we can obtain
the formula for the Doppler effect of $T/T'(t)$. }

However, solving \eqref{eq:DEAM-91} may be highly involved due to
$\bm{v}(t)$ and the square-root operation.  In most radio and
optical-based wireless communication scenarios,
$\beta(t'')=|\bm{v}(t'')|/c$ resulted from source's motion is nearly
the same across one period $T''(t'')$. Hence, in general, the
approximation of
\begin{align}\label{eq:DEAM-92}
  T''(t'')\approx \frac{T}{\sqrt{1-\bar{\beta}^2(t'')}}
\end{align}
can be used, where $\bar{\beta}(t'')=\overline{|\bm{v}(t'')|}/c$ with
$\overline{|\bm{v}(t'')|}$ the average speed of source over one
period of $T''(t'')$.  Substituting \eqref{eq:DEAM-92} into
\eqref{eq:DEAM-89} gives
\begin{align}\label{eq:DEAM-93}
 T'(t')=\frac{T}{\sqrt{1-\bar{\beta}^2(t'')}}+\frac{r(t''+T''(t''))-r(t'')}{c'}
\end{align}

Furthermore, if express
$r(t''+T''(t''))-r(t'')=\bar{v}(t'')T''(t'')=\bar{v}(t'')T/\sqrt{1-\bar{\beta}^2(t'')}$,
and apply it into \eqref{eq:DEAM-93}, a formula for the Doppler effect
can be found to be
\begin{subequations}
\begin{align}\label{eq:DEAM-94a}
  \frac{f'(t)}{f}=\frac{T}{T'(t')}=&\sqrt{1-\bar{\beta}^2(t'')}\left(1+\frac{\bar{v}(t'')}{c'}\right)^{-1}\\
  \label{eq:DEAM-94}
  \approx&\sqrt{1-\bar{\beta}^2(t'')}\left(1-\frac{\bar{v}(t'')}{c'}\right)\\
  \label{eq:DEAM-94b}
  =&\sqrt{1-\bar{\beta}^2(t')}\left(1-\frac{\bar{v}(t')}{c'}\right)
\end{align}
\end{subequations}
where the approximation is due to ignoring an insignificant term of
${\bar{v}^2(t)}/{c^2}$, and \eqref{eq:DEAM-94b} is due to $t'=t''$. In
\eqref{eq:DEAM-94}, $\bar{v}(t'')$ is the average speed of source
moving on the line connecting source and observer,
$\overline{|\bm{v}(t'')|}$ in $\bar{\beta}(t'')$ is the magnitude of
$\bm{v}(t'')$.

\begin{example}\label{Example-DE-6}
A Low Earth Orbit (LEO) satellite travels at a speed of about
$v=7.8$~km/s ($28,000$~km/h(our)), orbiting at an altitude of about
$h=200$~km above Earth's surface.  The communication signals sent by
this LEO satellite is operated on the frequency of $f=28$~GHz in the
Ka-band. Assume that $c'=c$.  a) Find the non-relativistic and
relativistic Doppler frequencies, when a receiver is located directly
below the LEO satellite or at the center of the spherical Earth.  b)
Find the maximum non-relativistic and relativistic Doppler
frequencies, when a receiver is fixed at a location on the satellite's
orbit.
\end{example}
\begin{solution}\label{Solution-DE-6}
  a) When the receiver is directly below the LEO satellite or at the
  center of the Earth, the velocity of satellite is approximately
  perpendicular with the line connecting the satellite and
  receiver. Hence, the non-relativistic Doppler frequency is zero.

  The relativistic Doppler frequency can be found from
  \eqref{eq:DEAM-94} by letting $\bar{v}(t'')=0$ due to the above, and
  $\overline{|\bm{v}(t'')|}=7.8$~km/s, giving
  \begin{align}
f_D'=f(\sqrt{1-\bar{\beta}^2(t)}-1)=-9.48~\textrm{Hz}\nonumber
    \end{align}
Hence, the frequency is slightly red-shifted.

b) When the receiver is fixed at a location on the orbit of LEO
satellite, the maximum Doppler shift is reached when the satellite
approaches the receiver with a speed of $\bar{v}(t'')=7.8$~km/s. The
non-relativistic Doppler frequency is given by
\begin{align}
f_D'=f\left(1+\frac{\bar{v}(t'')}{c}\right)^{-1}-f=728522.9~Hz\nonumber
\end{align}

From \eqref{eq:DEAM-94}, the relativistic Doppler frequency is given
by
\begin{align}
f_D'=f\sqrt{1-\bar{\beta}^2(t)}\left(1+\frac{\bar{v}(t'')}{c}\right)^{-1}-f\approx 728513.5~Hz\nonumber
\end{align}

The difference between the relativistic and non-relativistic Doppler
shifts is about $9.4$~Hz, approximately the value given in the case of
a).

\end{solution}

\subsubsection{Inertial Source and General Moving Observer}\label{subsection-Doppler-FS.2.2.2}

When the scenario of inertial source resting in medium and general
moving observer in Fig.~\ref{figure-Relativistic-Acceleration-DE}(b)
is considered, the frames of source and medium are inertial, but the
frame of observer is non-inertial, if observer's moving velocity is
not constant.  Hence, the time dilation should be derived with respect
to the reference frame of medium.

Assume that the variables defind in the previous
Section~\ref{subsection-Doppler-FS.2.2.1} are reused. Assume that two
adjacent maxima are emitted by source at $t_1=t$ and $t_2=t+T$ in its
reference frame. Accordingly, in medium's reference frame, these two
maxima are received by observer at $t_1''=t''$ and
$t_2''=t''+T''(t'')$\footnote{Note that, while $T$ is independent of
$t$, $T''(t'')$ is dependent on $t''$, as the result that observer is
moving relative to source in medium's reference frame. } at the
distances $r(t'')$ and $r(t''+T''(t))$ between observer and
source. Since source rests in the reference frame of medium, we have
$t=t''$, i.e., the clock holding by source runs synchronously as that
in medium.  Hence, following Section~\ref{subsection-Doppler-FS.2.1},
we have
\begin{align}\label{eq:DEAM-95}
 T''(t'')=T+\frac{r(t''+T''(t''))-r(t'')}{c'}
\end{align}

To find the Doppler effect of $T/T'(t)$, a relation between $T''(t'')$ and
$T'(t')$ is needed. This may be obtained with the aid of the
differential equation
\begin{align}\label{eq:DEAM-96}
 dt''=\frac{dt'}{\sqrt{1-\beta^2(t'')}}~\leftrightarrow~dt'=\sqrt{1-\beta^2(t'')}dt''
\end{align}
where $\beta(t'')=|\bm{v}(t'')|/c$ with $|\bm{v}(t'')|$ being the speed
of observer measured in medium's reference frame. However, solving
\eqref{eq:DEAM-96} may be highly involved.

In the applications where $\beta(t'')=|\bm{v}(t'')|/c$ resulted from
observer's motion is nearly the same across one period of $T''(t'')$,
the approximation of
\begin{align}\label{eq:DEAM-97}
  T''(t'')\approx \frac{T'(t')}{\sqrt{1-\bar{\beta}^2(t'')}}
\end{align}
can be used, where $\bar{\beta}(t'')=\overline{|\bm{v}(t'')|}/c$ and
$\overline{|\bm{v}(t'')|}$ is the average magnitude of $|\bm{v}(t'')|$
over one period of $T''(t'')$.  Substituting \eqref{eq:DEAM-97} into
\eqref{eq:DEAM-95} gives
\begin{align}\label{eq:DEAM-98}
 \frac{T'(t')}{\sqrt{1-\bar{\beta}^2(t'')}}=T+\frac{r(t''+T''(t''))-r(t'')}{c'}
\end{align}
Furthermore, if express $r(t''+T''(t''))-r(t'')=\bar{v}(t'')T''(t'')=\bar{v}(t''){T'(t')}/{\sqrt{1-\bar{\beta}^2(t'')}}$, and apply it
to \eqref{eq:DEAM-98}, a formula for the Doppler effect can be found
to be
\begin{align}\label{eq:DEAM-99}
  \frac{f'(t')}{f}=\frac{T}{T'(t')}=&\frac{1}{\sqrt{1-\bar{\beta}^2(t'')}}\times \left(1-\displaystyle\frac{\bar{v}(t'')}{c'}\right)\nonumber\\
  =&\frac{1}{\sqrt{1-\bar{\beta}^2(t)}}\times \left(1-\displaystyle\frac{\bar{v}(t)}{c'}\right)
\end{align}
where the equality in the second line is due to $t=t''$.

\begin{example}\label{Example-DE-6x}
  Assume that a LEO satellite system has the geometric and signaling
  parameters as that in Example~\ref{Example-DE-6}. Now the receiver
  is on the satellite.  a) Find the non-relativistic and relativistic
  Doppler frequencies, when a transmitter is located directly below
  the LEO satellite or at the center of the Earth.  b) Find the maximum
  non-relativistic and relativistic Doppler frequencies, when a
  transmitter is fixed at a location on the satellite's orbit.
  
\end{example}
\begin{solution}\label{Solution-DE-6x}
  a) When transmitter is directly below the LEO satellite, the
  velocity of LEO satellite is perpendicular with the line connecting
  the satellite and transmitter. Hence, the non-relativistic Doppler
  frequency is zero.

  The relativistic Doppler frequency can be found from
  \eqref{eq:DEAM-99} by letting $\bar{v}(t)=0$ due to the above
  mentioned, and $|\bm{v}(t)|=7.8$~km/s, giving
  \begin{align}
f_D=f\left[\left(\sqrt{1-\bar{\beta}^2(t)}\right)^{-1}-1\right]=9.48~\textrm{Hz}
    \end{align}
Hence, the frequency is slightly blue-shifted.

b) The maximum Doppler-shift is reached when satellite starts passing
the transmitter. Hence the speed for both (classic) Doppler effect and
time dilation is $-7.8$~km/s. Hence, the non-relativistic Doppler frequency is
\begin{align}
f_D'=f\left(1-\frac{\bar{v}(t)}{c}\right)-f=728504~Hz\nonumber
\end{align}

From \eqref{eq:DEAM-99}, the relativistic Doppler frequency is
calculated to be
\begin{align}
f_D'\approx  728513.5~Hz\nonumber
\end{align}

The difference between the relativistic and non-relativistic Doppler
shifts is about $9.5$~Hz, the approximated value given in the case of
a).

\end{solution}

\subsubsection{General Moving Source and  Observer}\label{subsection-Doppler-FS.2.2.3}

To analyze the relativistic Doppler effect in this more general
scenario where both source and observer are moving relative to
medium~\ncite{10.1119/1.14479,Zanchini_2025}, in addition to the
definitions used above, we associate the set of variables of $\{t'',
T_S''(t_S''), f_S''(t_S'')\}$ with source and the set of variables of
$\{t^*, T^*(t^*), f^*(t^*)\}$ with observer, both are relative to
medium. Furthermore, we assume that, in medium's reference frame,
there is an Ex-observer always resting on the line connecting source
and observer, so that the distance between source and observer is
equal to the distance $r(t'')$ between source and Ex-observer plus the
distance $r(t^*)$ between Ex-observer and observer. Then, assuming
that source emits two adjacent maxima at $t_1=t$ and $t_2=t+T$, in the
medium's frame, these two events occur at $t_1''=t''$ and
$t_2''=t''+T''(t'')$ with the distances of $r(t'')$ and
$r(t''+T''(t''))$ from the Ex-observer, respectively. These two maxima
are received by observer at $t_1^*=t^*$ and $t_2^*=t^*+T^*(t^*)$ with
the distances of $r(t^*)$ and $r(t^*+T^*(t^*))$ from the
Ex-observer. Hence, in the reference frame of medium, the relationship
of
\begin{align}\label{eq:DEAM-127}
T^*(t)=T''(t'')+\frac{r(t''+T''(t''))-r(t'')}{c'}+\frac{r(t^*+T^*(t^*))-r(t^*)}{c'}
\end{align}
`approximately' holds. Here, the use of `approximately' is because
during the events related to \eqref{eq:DEAM-127}, the Ex-observer may
not always be on the line connecting source and observer, making the
change of distance between source and observer is only approximately
the sum of the second and third terms in \eqref{eq:DEAM-127}.

Let us express $\bar{v}''(t'')={[r(t''+T''(t''))-r(t'')]}/{T''(t'')}$
and $\bar{v}^*(t^*)={[r(t^*+T^*(t^*))-r(t^*)]}/{T^*(t^*)}$,
respectively, the average velocities of source and observer move on
the line connecting source and observer. These velocities are relative
to the medium, with positive and negative mean moving away and towards
the Ex-observer, respectively.  Then, \eqref{eq:DEAM-127} can be
written as
\begin{align}\label{eq:DEAM-128}
T^*(t)\left(1-\frac{\bar{v}^*(t^*)}{c'}\right)=T''(t'')\left(1+\frac{\bar{v}''(t'')}{c'}\right)
\end{align}
Now considering the time dilation due to the motions of source and observer relative to medium, there are relations of
\begin{subequations}
\begin{align}\label{eq:DEAM-129a}
T^*(t^*)=\frac{T'(t')}{\sqrt{1-\bar{\beta}^2(t^*)}}\\
\label{eq:DEAM-129b}
T''(t'')=\frac{T}{\sqrt{1-\bar{\beta}^2(t'')}}
\end{align}
\end{subequations}
where $\bar{\beta}(t^*)=\overline{|\bm{v}^*(t^*)|}/c$ and $\bar{\beta}(t'')=\overline{|\bm{v}''(t'')|}/c$. Upon substituting \eqref{eq:DEAM-129a} and  \eqref{eq:DEAM-129b} into \eqref{eq:DEAM-128} and some simplification, the Doppler effect can be expressed as 
\begin{align}\label{eq:DEAM-130}
\frac{f'(t')}{f}=\left(\frac{1-{\bar{v}^*(t^*)}/{c'}}{1+{\bar{v}''(t'')}/{c'}}\right)\times\left(\frac{\sqrt{1-\bar{\beta}^2(t'')}}{\sqrt{1-\bar{\beta}^2(t^*)}}\right)
\end{align}
Explicitly, \eqref{eq:DEAM-94} and \eqref{eq:DEAM-99} are the special
cases of \eqref{eq:DEAM-130}. Furthermore, if
$\overline{|\bm{v}^*(t^*)|}=\bar{v}^*(t^*)$ and
$\overline{|\bm{v}''(t'')|}=\bar{v}''(t'')$, meaning that both source
and observer are moving on a line, and if also $c'=c$,
\eqref{eq:DEAM-130} is reduced to
\begin{subequations}\label{eq:DEAM-130a}
\begin{align}\label{eq:DEAM-130b}
  \frac{f'(t')}{f}=&\sqrt{\frac{1-{\bar{v}''(t'')}/c}{1+{\bar{v}''(t'')}/c}}\times\sqrt{\frac{1-{\bar{v}^*(t^*)}/c}{1+{\bar{v}^*(t^*)}/c}}\\
  \label{eq:DEAM-130c}
  =&\sqrt{\frac{1-{\bar{v}''(t'')}/c-{\bar{v}^*(t^*)}/c+{\bar{v}''(t'')}{\bar{v}^*(t^*)}/c^2}{1+{\bar{v}''(t'')}/c+{\bar{v}^*(t^*)}/c+{\bar{v}''(t'')}{\bar{v}^*(t^*)}/c^2}}\\
  \label{eq:DEAM-130d}
  = &\sqrt{\frac{1-\bar{u}(t,t')/c}{1+\bar{u}(t,t')/c}}
\end{align}
\end{subequations}
where, by definition,
\begin{align}\label{eq:DEAM-130e}
 \bar{u}(t,t')=\frac{{\bar{v}''(t'')}+{\bar{v}^*(t^*)}}{1+{\bar{v}''(t'')}{\bar{v}^*(t^*)}/c^2}
\end{align}
which is the relative moving velocity between source and observer,
measured at $t$ in terms of source or at $t'$ in terms of observer in
free space. Note that in free space, there exists no medium reference,
i.e., there are no $t''$ and $t^*$. Hence, on the lefthand side of
\eqref{eq:DEAM-130e}, $ \bar{u}(t,t')$ is directly represented with
respect to the reference frames of source and observer.

Specifically, if both source and observer move along the same line at
a uniform velocity $v$, then $\bar v''=v$ and $\bar v^*=-v$ (or vice
versa). In this case, \eqref{eq:DEAM-130a} yields ${f'(t')}/{f}=1$,
indicating that no relativistic Doppler effect occurs regardless of
the value of $v$. We will return the formulas in \eqref{eq:DEAM-130}
and \eqref{eq:DEAM-130e} in Section~\ref{subsection-Doppler-FS.4.1},
when considering the scenario where source and observer are in the
same uniform accelerated frame.

Above in Sections~\ref{subsection-Doppler-FS.2.2.1} - \ref{subsection-Doppler-FS.2.2.3}, the relativistic Doppler effect in general motion has been analyzed. It is shown that the Doppler effect is contributed by two components, one component accounts for the (classic) Doppler effect, and the other one is due to the time dilation resulted from relative motion. Furthermore, it is shown that the components for (classic) Doppler effect have the similar expressions obtained in Section~\ref{subsubsection-Doppler-FS.1.3} for uniform motion, with the uniform velocities in uniform motion replaced by the time-variant average velocities in general motion. Based on these observations, the relativistic Doppler effects in linear acceleration and circularly uniform motion are briefly analyzed below in Section~\ref{subsection-Doppler-FS.2.2.4} and \ref{subsection-Doppler-FS.2.2.5}, by following the analysis in Section~\ref{subsubsection-Doppler-FS.2.1.1} and \ref{subsubsection-Doppler-FS.2.1.2}.

\subsubsection{Linearly Uniform Acceleration}\label{subsection-Doppler-FS.2.2.4}

Assume a linear acceleration system between source and observer, which
has an acceleration $a$ and initial velocity $v_0$. Positive $a$ and
$v_0$ represent departing, while negative $a$ and $v_0$ mean
approaching, but the total effect is relied on their resulted
$v_0+at$. Then, according to the special
relativity~\ncite{Rafael-Ferraro-book}, the resultant velocity is
given by
\begin{align}\label{eq:DEAM-131a}
 v(t)=\frac{v_0+at}{\sqrt{1+\frac{(v_0+at)^2}{c^2}}}
\end{align}
Hence, the time-variant average velocity over one period
is
\begin{align}\label{eq:DEAM-13b}
\bar{v}(t)=&\frac{1}{T}\int_{t}^{t+T}\frac{v_0+a\tau}{\sqrt{1+\displaystyle\frac{(v_0+a\tau)^2}{c^2}}}d\tau
\end{align}
Completing the integration gives
\begin{align}\label{eq:DEAM-131c}
\bar{v}(t)=&\frac{c^2}{aT}\left[\sqrt{1+\displaystyle\frac{(v_0+a(t+T))^2}{c^2}}-\sqrt{1+\displaystyle\frac{(v_0+at)^2}{c^2}}\right]
\end{align}
Assume that $a$ is small enough for using the formula $H=aT^2/2$ - the
distance travelled in $[0,T]$ for an initial velocity $v_0=0$. Then,
we have $T=\sqrt{2H/a}$. Substituting it into \eqref{eq:DEAM-131c}
gives a $T$-independent formula of
\begin{align}\label{eq:DEAM-131}
\bar{v}(t)=&\frac{c^2}{\sqrt{2aH}}\left[\sqrt{1+\displaystyle\frac{(v_0+a(t+\sqrt{2H/a}))^2}{c^2}}-\sqrt{1+\displaystyle\frac{(v_0+at)^2}{c^2}}\right]
\end{align}

It can be shown that
\begin{align}\label{eq:DEAM-131d}
\lim_{t\rightarrow\infty}\bar{v}(t)=c
\end{align}
Hence, no matter what $a$ is, the resultant speed will never goes
beyond $c$ of the speed of light.

Hence, when source is moving and observer is stationary, following \eqref{eq:DEAM-94a}, we have
\begin{align}\label{eq:DEAM-132}
  \frac{f'(t)}{f}=&\sqrt{1-\bar{\beta}^2(t)}\left(1+\frac{\bar{v}(t)}{c'}\right)^{-1}
  \end{align}
where $\bar{\beta}(t)=\bar{v}(t)/c$. Similarly, when source is stationary and observer is moving, following \eqref{eq:DEAM-99}, we have
\begin{align}\label{eq:DEAM-133}
  \frac{f'(t')}{f}=&\frac{1}{\sqrt{1-\bar{\beta}^2(t')}}\times \left(1-\displaystyle\frac{\bar{v}(t')}{c'}\right)
  \end{align}
where $\bar{v}(t')$ is the same as \eqref{eq:DEAM-131} with $H$
replaced by $H'=a(T'(t'))^2/2$, denoting the distance of observer
travelling in $[0,T'(t)]$ for an initial velocity of $v_0=0$, while
$\bar{\beta}(t')=\bar{v}(t)/c$.

An inspection of equations \eqref{eq:DEAM-132} and \eqref{eq:DEAM-133}
reveals that the relativistic Doppler frequency will never approach
infinity as $t$ increases. This outcome sharply contrasts with the
non-relativistic Doppler frequency discussed in
Section~\ref{subsubsection-Doppler-FS.2.1.1}, where the frequency
increases linearly with $t$ and can, therefore, potentially exceed the
source frequency $f$ and eventually go to infinity.

\subsubsection{Circularly Uniform Motion}\label{subsection-Doppler-FS.2.2.5}

\begin{figure}[tb]
  \begin{center}
 \includegraphics[angle=0,width=.65\linewidth]{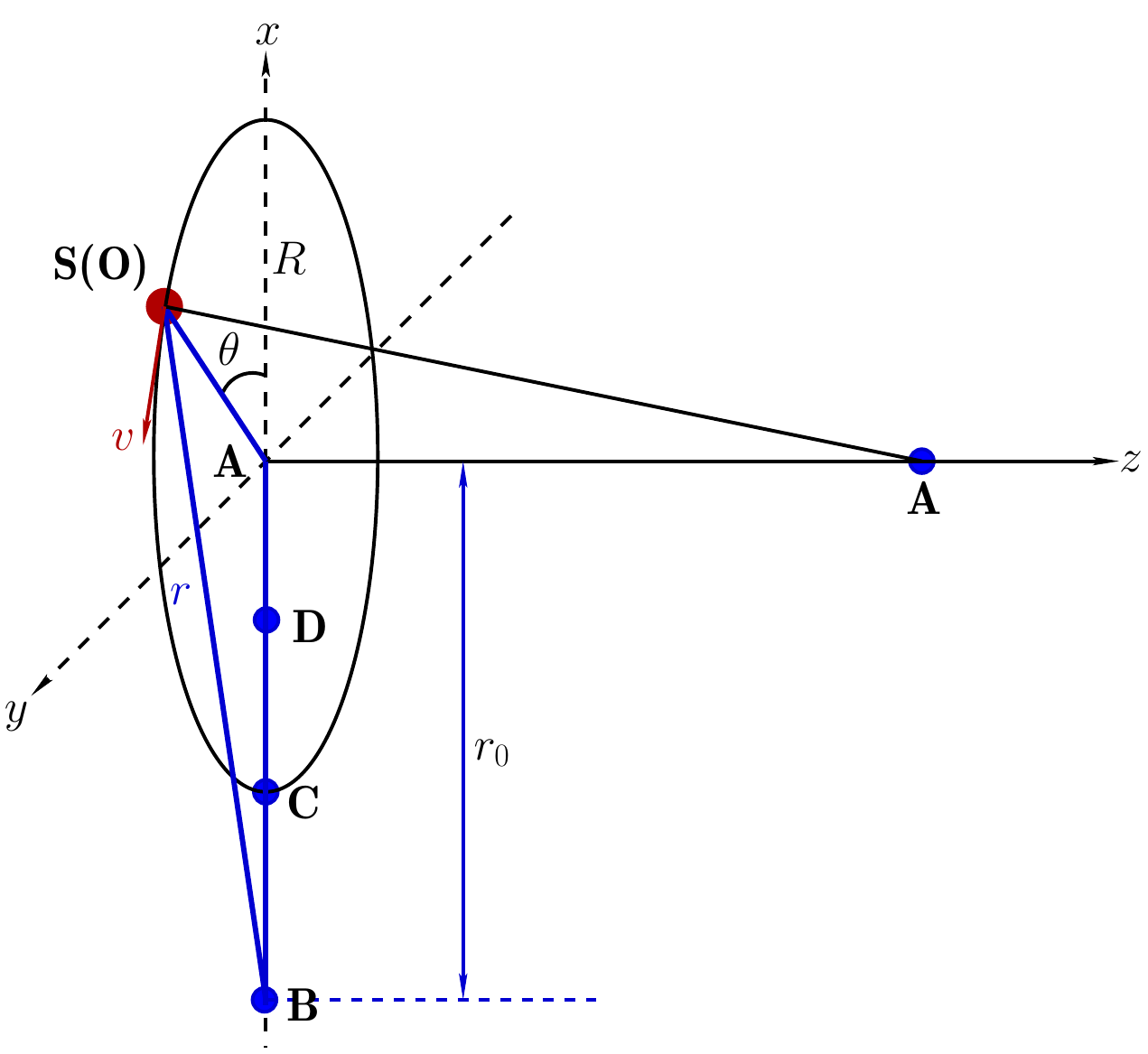}
  \end{center}
  \caption{Illustration of circularly uniform acceleration, where {\bf
      A}, {\bf B}, {\bf C} and {\bf D} are source/observer locations,
    and {\bf S} ({\bf O}) is for moving source (observer).}
  \label{figure-Circular-Acceleration-copy}
\end{figure}

For circularly uniform motion, the same system as that in
Section~\ref{subsubsection-Doppler-FS.2.1.2} is considered. All the
assumptions and settings considered there are also applied here,
except the relativistic Doppler effect is analyzed now. For
convenience of reading, Fig.~\ref{figure-Circular-Acceleration} is
repeated here as Fig.~\ref{figure-Circular-Acceleration-copy}.

Following the scenarios considered in
Fig.~\ref{figure-Circular-Acceleration}, if an observer (or source) is
installed on the $z$-axis, the average velocity projected on the line
connecting source and observer is zero. Hence, the (classic) Doppler
frequency shift is zero. However, source (or observer) has a speed of
$v=R\omega$, which generate time dilation, resulting in frequency
shift. Specifically, if source is moving on the circle and observer is
fixed on $z$-axis, there is
\begin{align}\label{eq:DEAM-138}
  \frac{f'}{f}=&\sqrt{1-v^2/c^2}\nonumber\\
  =&\sqrt{1-R^2\omega^2/c^2}
\end{align}
Hence, $f'<f$, generating red-shift. In contrast, if source is fixed
on $z$-axis while observer is rotating around the circle, the relation
is
\begin{align}\label{eq:DEAM-139}
  \frac{f'}{f}=&\frac{1}{\sqrt{1-R^2\omega^2/c^2}}
\end{align}
Therefore, there is blue-shift, as $f'>f$.

In the case that observer is on the $x$-axis, in
Section~\ref{figure-Circular-Acceleration}, it has given the general
formula of \eqref{eq:DEAM-74} for the Doppler effect, when an observer
is located, such as, at {\bf B}, {\bf C} or {\bf D}, which has a
distance $r_0$ from the center of the circle. When relativistic
Doppler effect is considered, this formula is only correct when all
variables are defined in the reference frame of medium.  Accordingly,
the formula needs to be presented as
\begin{align}\label{eq:DEAM-140}
  T'(t')=T''+\frac{r_0}{c'}&\left[\sqrt{\left(1-\frac{R}{r_0}\right)^2+\frac{4R}{r_0}\cos^2\left(\frac{\omega (t''+T'')}{2}\right)}\right.\nonumber\\
    &\left.-\sqrt{\left(1-\frac{R}{r_0}\right)^2+\frac{4R}{r_0}\cos^2\left(\frac{\omega t''}{2}\right)}\right],~t''\geq 0
\end{align}
where $t''$ and $T''$ are the variables associated with the moving
source in the reference frame of medium. With the aid of the Lorentz
transformation,
\begin{align}\label{eq:DEAM-141}
  t''=\frac{t}{\sqrt{1-R^2\omega^2/c^2}},~T''=\frac{T}{\sqrt{1-R^2\omega^2/c^2}}
\end{align}
Substituting them into \eqref{eq:DEAM-140}, a relationship between
$T'(t)$ and $T$, i.e., the relativistic Doppler effect can be
obtained. 

Note that \eqref{eq:DEAM-140} is not mathematically accurate, as some
minor effects are ignored. For example, the change of distance in one
period, i.e., the amount given by the bracket of \eqref{eq:DEAM-140},
should be small relative to the speed of light. Otherwise, special
relativity in terms of the change needs to be involved. Additionally,
in the Lorentz transformation of time, the effect of source motion is
also ignored.

Since the Lorentz factor is a constant, it can be expected that the
relativistic Doppler frequency behaves similarly as the (classic)
Doppler frequency, for example, as shown in
Example~\ref{Example-DE-5g}.  There might be slight corrections on time
and frequency, depending on the velocity of circular motion.

Similarly, when source is on the $x$-axis while an observer is
rotating on the circle, in the reference frame of medium, there is a
relation of
\begin{align}\label{eq:DEAM-142}
  T''(t'')=T+\frac{r_0}{c'}&\left[\sqrt{\left(1-\frac{R}{r_0}\right)^2+\frac{4R}{r_0}\cos^2\left(\frac{\omega (t''+T''(t''))}{2}\right)}\right.\nonumber\\
    &\left.-\sqrt{\left(1-\frac{R}{r_0}\right)^2+\frac{4R}{r_0}\cos^2\left(\frac{\omega t''}{2}\right)}\right],~t''\geq 0
\end{align}
Accordingly, the Lorentz transformations are
\begin{align}\label{eq:DEAM-143}
  t''=\frac{t'}{\sqrt{1-R^2\omega^2/c^2}},~T''(t'')=\frac{T'(t')}{\sqrt{1-R^2\omega^2/c^2}}
\end{align}
Substituting them into \eqref{eq:DEAM-142}, the relation for
estimating the relativistic Doppler effect can be obtained.

So far, the analysis in the previous sections has covered the Doppler
effect in various scenarios, ranging from uniform motion to general
motion, and from the (classic) Doppler effect to the relativistic
Doppler effect.  As a final remark, whenever a wave propagation medium
was assumed in the analysis, only a static and uniform medium was considered in the
analysis. In other words, medium was assumed to be a static reference
frame, while the states of source and observer were defined relative
to the medium. In practice, both non-uniform medium and moving medium
affect the characteristics of wave propagation within them.  A
static, non-uniform medium does not introduce a Doppler effect; it
only distorts the propagation speed and wavelength. However, a moving
medium may impact the Doppler effect. An example is provided below to
illustrate this impact.

Assume that a wave source is flying away at a speed $v_s$ relative to
a fixed receiver, and is sending EM signals of frequency $f$. We
further assume that the wave propagation medium between source and
receiver has a refractive index of $n$, and that this medium moves in
the direction from receiver to source at a speed $v_m$. The wave
propagation speed in the medium is $c'=c/n$. Accordingly, in the
principles of special relativity, the wave propagation speed from the
source to the receiver (as observed by the receiver)
is~\ncite{Book-Kenneth-Modern-Physics}
\begin{align}\label{eq:DEAM-144}
  c^*=\frac{c'-v_m}{1-c'v_m/c^2}=\frac{c/n-v_m}{1-v_m/(nc)}
\end{align}
Then, following the previous analysis in
Section~\ref{subsubsection-Doppler-FS.1.3.2}, the Doppler frequency is
expressed as
\begin{align}\label{eq:Doppler-FS-3.48a}
f'=&f\times\left(\frac{\sqrt{1-v_s^2/c^2}}{1+v_s/{c^*}}\right)
\end{align}
Explicitly, the moving medium affects the observed Doppler frequency
specifically through the value of $c^*$, which represents the wave
propagation speed relative to the receiver's frame.

\section{Doppler Effect in Accelerated Motion and Gravitational Fields}\label{section-Doppler-FS.4}

In Section~\ref{subsection-Doppler-FS.2.2}, the relativistic Doppler
effect in general motion has been analyzed, which can be specialized
to the following scenarios containing non-inertial acceleration
frame(s):
\begin{itemize}
\item source is in acceleration and observer is in an inertial frame,

\item source is in an inertial frame and observer is in acceleration,
  and

  \item both source and observer are in acceleration.
\end{itemize}

In this section, we first specifically analyze the Doppler effect in
uniform acceleration and, then, extend it to the Doppler effect (or
red/blue shift) by gravity, which is the focus of this
section. Accordingly, we assume that EM waves propagate in free space
and that the direction of wave propagation is collinear with the
motion of source and observer, meaning that wave propagates either
along or opposite to the direction of acceleration.

\subsection{Relativistic Doppler Effect in Uniform Acceleration}\label{subsection-Doppler-FS.4.1}

%
\begin{figure}[tb]
  \begin{center}
 \includegraphics[angle=0,width=.3\linewidth]{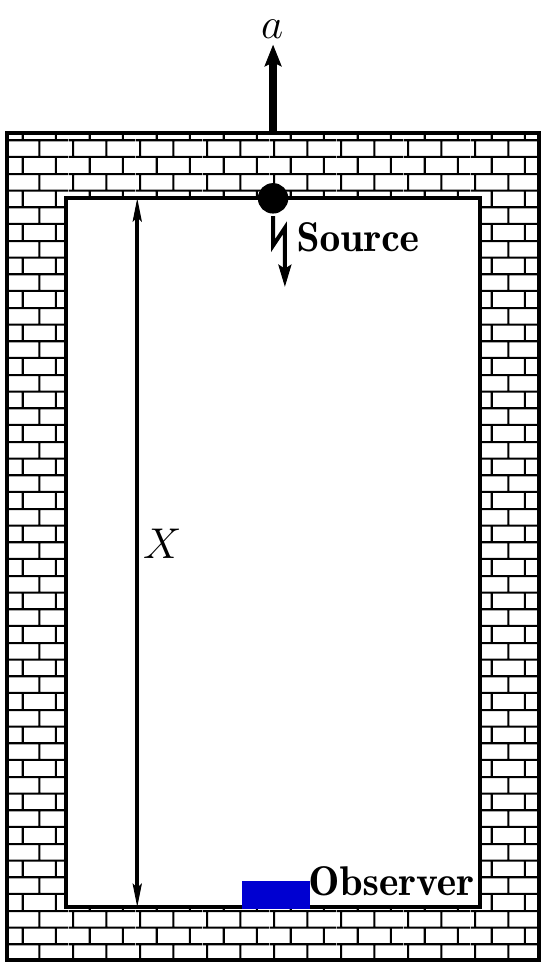}
  \end{center}
  \caption{An accelerating frame in which source emits EM signals to an observer of distance $X$ away.}
  \label{figure-Acceleration-frame}
\end{figure}

In Feynman's lecture notes~\ncite{Feynman-lecture-notes} and other
modern physics textbooks~\ncite{Book-Kenneth-Modern-Physics}, the
Doppler effect by acceleration has been derived directly from the
formula of the relativistic Doppler frequency, as seen in
\eqref{eq:Doppler-FS-XW-3.56-b}. Specifically, when free space and the
above-mentioned assumptions are considered, and when the relative
speed between source and observer is expressed as $\Delta v$,
\eqref{eq:Doppler-FS-XW-3.56-b} can be re-written as
\begin{align}\label{eq:Doppler-FS-147}
  f' =&f\times\sqrt{\frac{1-\Delta v/c}{1+\Delta v/c}}
\end{align}
where positive and negative values of $\Delta v$ indicate that source
and observer are moving away from and toward one another,
respectively. However, in some cases, such as, in acceleration frame,
the sign of $\Delta v/c$ should be determined based on `tendency'. If
source and observer tend to move away from each other, $\Delta v/c$ is
positive. Otherwise, when source and observer tend to move towards
each other, $\Delta v/c$ is negative.

Following \ncite{Feynman-lecture-notes,Book-Kenneth-Modern-Physics},
let us consider Fig.~\ref{figure-Acceleration-frame}, where source and
observer separated by a distance $X$ are in the same chamber (frame)
that is accelerated with an acceleration of $a$. Assume that source
emits an EM wave of frequency $f$ at $t$. Accordingly, the velocity of
source at $t$ is expressed as $v$. Assume that $v<<c$. Then, the time
required for the wave to propagate to observer is $\Delta t=X/c$. Hence, when
observer receives the wave, it has a velocity $v'=v+a\Delta
t=v+aX/c$. These make $\Delta v=-aX/c$, where the negative sign is
because source and observer are tend to move toward each other based
on the settings shown in the figure. Moreover, the sign of $\Delta
v/c$ can be determined as follows. When wave propagates in the same
direction of acceleration $a$, the sign is positive. Otherwise, when
wave propagates in the opposite direction of acceleration $a$, the
sign is negative. Substituting $\Delta v=-aX/c$ into
\eqref{eq:Doppler-FS-147} yields
\begin{subequations}
\begin{align}\label{eq:Doppler-FS-148a}
  f' =&f\times\sqrt{\frac{1+aX/c^2}{1-aX/c^2}}\\
  \label{eq:Doppler-FS-148}
  \approx& f\left(1+\frac{aX}{c^2}\right)
\end{align}
\end{subequations}
where the approximation is obtained by ignoring a term of
$a^2X^2/c^4$. From \eqref{eq:Doppler-FS-148}, the Doppler shift
satisfies
\begin{align}\label{eq:Doppler-FS-149}
  \frac{f_D(=f'-f)}{f} \approx&\frac{aX}{c^2}
\end{align}
Hence, when the wave propagates in the opposite direction to the
acceleration, the Doppler frequency is positive, resulting in a
blueshift. Conversely, when the wave propagates in the same direction
as the acceleration (making $a$ negative in
\eqref{eq:Doppler-FS-149}), the Doppler frequency is negative,
resulting in a redshift. In other words, if in the chamber of
Fig.~\ref{figure-Acceleration-frame}, when light is emitted in the
direction from observer to source, the light will experience redshift.

The formula of \eqref{eq:Doppler-FS-148} or \eqref{eq:Doppler-FS-149}
is derived as a first-order approximation from the relativistic
Doppler effect of \eqref{eq:Doppler-FS-147}, under the assumption of
$v<<c$, and the assumptions that the EM signal's period is small, the
acceleration $a$ is small, or their joint effect is ignorable. Under
these assumptions but except $v<<c$, in
\ncite{Stefano-Quattrini-2024}, the relativistic Doppler effect for
source/observer in any uniform acceleration was accurately analyzed,
showing that the formula of \eqref{eq:Doppler-FS-148} or
\eqref{eq:Doppler-FS-149} is in fact accurate. In other words, no
assumption of $v<<c$ is needed and the approximation in
\eqref{eq:Doppler-FS-148} or \eqref{eq:Doppler-FS-149} is in fact
equality. This is achieved because all the other effects are ideally
cancelled by themselves, leaving the Doppler effect exactly the same
as that shown in \eqref{eq:Doppler-FS-148} or
\eqref{eq:Doppler-FS-149}.

This can also be derived from \eqref{eq:DEAM-130a}, which is the
result from a rigorous analysis of relativistic Doppler effects. For
the current accelerating system shown in
Fig.~\ref{figure-Acceleration-frame}, let us assume a parallel
reference frame moving at the same speed as the source's moving speed
at time $t$, which can be set to $t=0$ for convenience. Hence, viewing
from this parallel reference frame, the speed of source and observer
at $t=0$ is $v=v'=0$. Since an infinitesimal signal period is
considered.  We can assume that source emits an EM impulse at $t=0$.
Then, we have $\bar{v}=0$ and $\bar{v}'=v'(t')$, where $v'(t')$ is the
velocity of observer at $t'$ when it receives the impulse. Assume that
within $t'$, the distance that observer moves is $\Delta X$. Then, we
have the following relations~\ncite{Semay_2006,Stefano-Quattrini-2024}
\begin{subequations}\label{eq:Doppler-FS-155}
  \begin{align}\label{eq:Doppler-FS-155a}
    \Delta X=&\frac{c^2}{a}\left[\sqrt{1+\left(\frac{at'}{c}\right)^2}-1\right]\\
    \label{eq:Doppler-FS-155b}
    t'=&\frac{X-\Delta X}{c}\\
       \label{eq:Doppler-FS-155c}
 \bar{v}'=&-\frac{at'}{\sqrt{1+\left(\displaystyle\frac{at'}{c}\right)^2}}
  \end{align}
\end{subequations}
where, again, the negative sign associated with $\bar{v}'$ is because
observer moves towards source.  From \eqref{eq:Doppler-FS-155a} and
\eqref{eq:Doppler-FS-155b}, $t'$ can be obtained
as~\ncite{Stefano-Quattrini-2024}
\begin{align}\label{eq:Doppler-FS-156}
t'=\frac{X}{c}\left(\displaystyle\frac{1+\displaystyle\frac{aX}{2c^2}}{1+\displaystyle\frac{aX}{c^2}}\right)
\end{align}
Substituting it into \eqref{eq:Doppler-FS-155c} gives
\begin{align}\label{eq:Doppler-FS-157}
\bar{v}'=&-\frac{\displaystyle\frac{aX}{c}\left(1+\frac{aX}{2c^2}\right)}{\displaystyle\frac{1}{2}\left(\frac{aX}{c^2}\right)^2+\frac{aX}{c^2}+1}
\end{align}
Finally, substituting $\bar{v}''(t'')=0$ and $\bar{v}^*(t^*)=\bar{v}'$
into \eqref{eq:DEAM-130b} or \eqref{eq:DEAM-130d}, it can be
simplified to obtain
\begin{align}\label{eq:Doppler-FS-158}
\frac{f'}{f}=1+\frac{aX}{c^2}
\end{align}
where $f'$ is independent of $t'$ as the result that both source and
observer move in the same acceleration frame. Hence, the formula of
\eqref{eq:Doppler-FS-148} or \eqref{eq:Doppler-FS-149} derived as a
first-order approximation from the relativistic Doppler effect of
\eqref{eq:Doppler-FS-147} is accurate, on the condition that the
period of EM wave is infinitesimal. Below we analyze the Doppler effect without
imposing this condition.

\begin{figure}[tb]
  \begin{center}
 \includegraphics[angle=0,width=.5\linewidth]{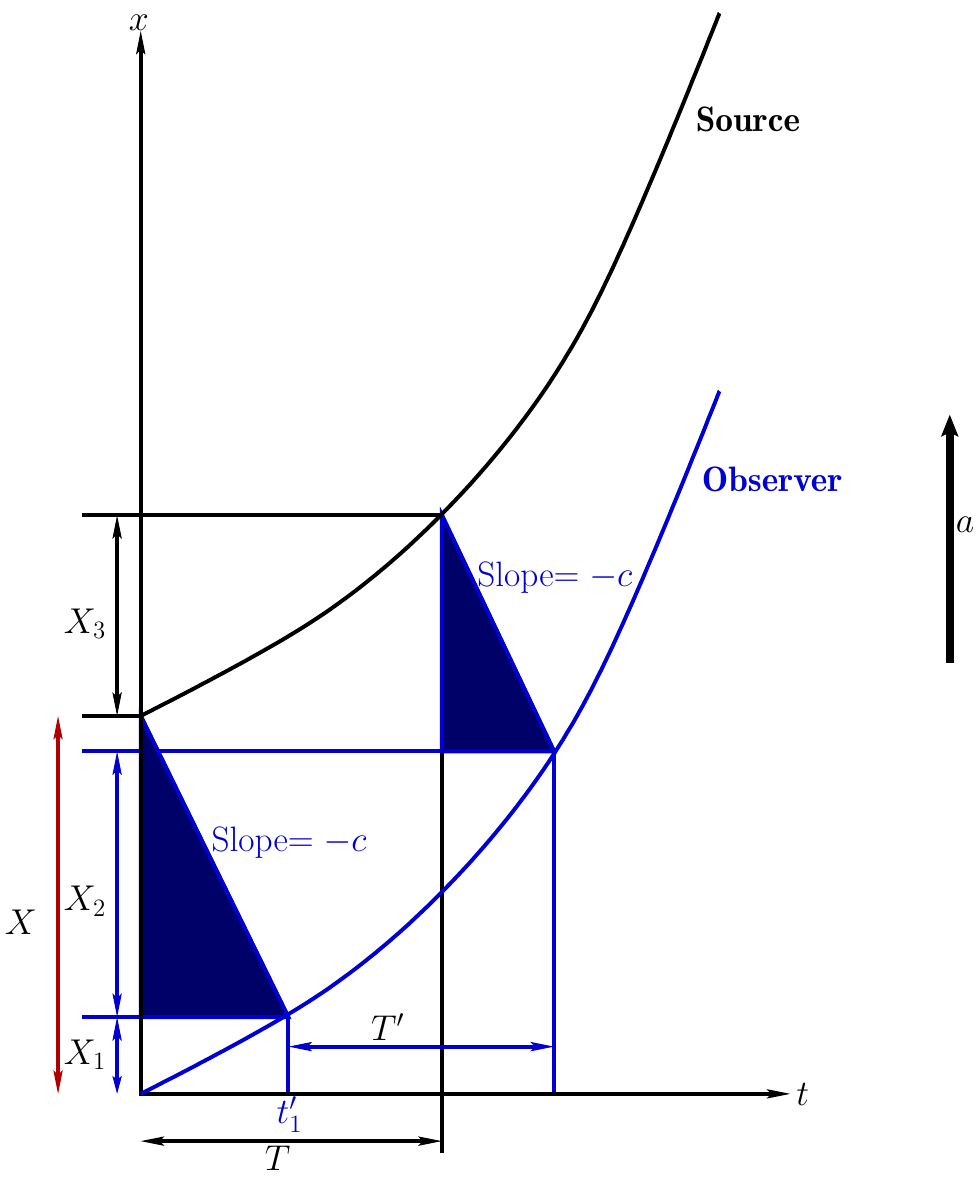}
  \end{center}
  \caption{Graphic representation of EM-wave propgation from source to observer within an acceleration frame.}
  \label{figure-Acceleration-Doppler-graphy}
\end{figure}
The system considered is graphically represented in
Fig.~\ref{figure-Acceleration-Doppler-graphy}, where the first and
second triangles explain the transmision states of the first and
second crests of an EM-wave. In the graph, it is assumed that at $t=0$,
in the parallel frame, source is at $X$ and observer is at $x=0$. From
Fig.~\ref{figure-Acceleration-Doppler-graphy}, the following geometric
relations can be identified:
\begin{subequations}\label{eq:Doppler-FS-159}
  \begin{align}\label{eq:Doppler-FS-159a}
    X_1=&\frac{c^2}{a}\left[\sqrt{1+\left(\frac{at_1'}{c}\right)^2}-1\right]\\\label{eq:Doppler-FS-159b}
    X_1+X_2=&\frac{c^2}{a}\left[\sqrt{1+\left(\frac{a(t_1'+T')}{c}\right)^2}-1\right]\\\label{eq:Doppler-FS-159c}
    X_3=&\frac{c^2}{a}\left[\sqrt{1+\left(\frac{aT}{c}\right)^2}-1\right]\\\label{eq:Doppler-FS-159d}
    t_1'=&\frac{X-X_1}{c}\\\label{eq:Doppler-FS-159e}
    T'+t_1'-T=&\frac{X+X_3-X_1-X_2}{c}
  \end{align}
\end{subequations}
Applying \eqref{eq:Doppler-FS-159d} to \eqref{eq:Doppler-FS-159e} yields
\begin{align}\label{eq:Doppler-FS-160}
  T'-T=\frac{X_3-X_2}{c}
\end{align}
From \eqref{eq:Doppler-FS-159a} and \eqref{eq:Doppler-FS-159d}, $t_1'$
can be derived to be
\begin{align}\label{eq:Doppler-FS-161}
t_1'=\frac{X}{c}\left(\displaystyle\frac{1+\displaystyle\frac{aX}{2c^2}}{1+\displaystyle\frac{aX}{c^2}}\right)
\end{align}
which is a fixed value. Hence, based on \eqref{eq:Doppler-FS-159a} -
\eqref{eq:Doppler-FS-159c} and \eqref{eq:Doppler-FS-160}, it can be
shown that $T'$ and $T$ follow the relation of 
\begin{align}\label{eq:Doppler-FS-162}
T'-T=\frac{c}{a}\left[\sqrt{1+\left(\frac{aT}{c}\right)^2}-\sqrt{1+\left(\frac{a(t_1'+T')}{c}\right)^2}+\sqrt{1+\left(\frac{at_1'}{c}\right)^2} -1\right]
\end{align}
From \eqref{eq:Doppler-FS-162}, although it is cumbersome to obtain an
explicit formula for $T/T'=f'/f$, for a given $T$, an explicit formula
for $f'=1/T'$ can be obtained, which can be expressed as
\begin{align}\label{eq:Doppler-FS-163}
  f'=\frac{2k(A+kt_1')}{A^2-k^2(t_1')^2-1}
\end{align}
associated with the definitions of
\begin{subequations}\label{eq:Doppler-FS-164}
\begin{align}\label{eq:Doppler-FS-164a}
  k=&\frac{a}{c}\\
 A=&kT+\left[\sqrt{1+\left(\frac{aT}{c}\right)^2}+\sqrt{1+\left(\frac{at_1'}{c}\right)^2} -1\right]
\end{align}
\end{subequations}

\begin{example}
Consider that in an accelerating system with an acceleration
$a=1000~m/s^2,~ 10^9~m/s^2$ and $9.18\times 10^9~m/s^2$, respectively,
a source sends radio signals of $f=1$~GHz to an observer having a
distance $X=1000$~m behind the source. Find the frequencies of the
signals received by the observer, when the formulas
\eqref{eq:Doppler-FS-158} and \eqref{eq:Doppler-FS-163} are
respectively applied.

  \begin{solution}

    First, when \eqref{eq:Doppler-FS-158} is used, the frequencies are
    \begin{align}
      f'=&1,000,000,000.011~\text{Hz}~(\text{for}~a=1000~m/s^2)\nonumber\\
      =&1,011,126,500.56~\text{Hz}~(\text{for}~a=10^9~m/s^2)\nonumber\\
      =&1,102,131,014.23~\text{Hz}~(\text{for}~a=9.18\times 10^9~m/s^2)\nonumber
    \end{align}

    Second, when \eqref{eq:Doppler-FS-163} is used, the frequencies
    are
    \begin{align}
      f'=&1,000,000,000.011~\text{Hz}~(\text{for}~a=1000~m/s^2)\nonumber\\
      =&1,011,126,500.55~\text{Hz}~(\text{for}~a=10^9~m/s^2)\nonumber\\
      =&1,102,131,013.23~\text{Hz}~(\text{for}~a=9.18\times 10^9~m/s^2)\nonumber
    \end{align}

    The differences between the frequencies calculated from the two
    formulas are respectively about $10^{-8}$~\text{Hz} $(\text{for}~a=1000~m/s^2)$,
    $0.01$~\text{Hz} $(\text{for}~a=10^9~m/s^2)$, and $1$~\text{Hz} $(\text{for}~a=9.18\times 10^9~m/s^2)$.
    
    \end{solution}
\end{example}

The above examples demonstrate that: a) the Doppler effect by
acceleration is typically very small and can be ignored for most
events; b) the approximated formula \eqref{eq:Doppler-FS-158} is safe
to use even when the acceleration is as high as $10^{10}~m/s^2$.  Note
that the accelerations on such order can only be found in extreme
environments like the vicinity of neutron stars or within high-energy
particle accelerators. Nevertheless, nowadays, it is possible for
electrons to be accelerated with an acceleration up to $10^{22}~m/s^2$
by the Laser-Plasma Wakefield Accelerators (LPFA). In the situation of
this kind, more accurate evaluation than that by
\eqref{eq:Doppler-FS-158} be necessary.

\subsection{Clock Postulate and K$\ddot{\textrm{u}}$ndig's Experiment Induced Researches}\label{subsection-Doppler-FS.4.2}

In the above subsection, although the acceleration on Doppler
effect is addressed, the Doppler effect is in fact by the
acceleration resulted velocity, not by the acceleration itself, as seen
from \eqref{eq:Doppler-FS-147} to \eqref{eq:Doppler-FS-148}, for
example. This means that the {\em clock
  postulate (hypothesis)}~\ncite{Don-Koks-Clock-Postulate,Book-Albert-Einstein-Relativity,Book-General-Relativity-1,Book-General-Relativity-2} has been implicitly applied in the analyses. The clock postulate can be stated as that the rate of an ideal clock depends only on its instantaneous velocity, not on its acceleration or other higher-order derivatives of motion~\ncite{Don-Koks-Clock-Postulate}. Considering specifically the Doppler effect, the clock postulate can be explained as follows. 
  
Assume a rest source and a general accelerating observer, which has an
acceleration $a(t)$ relative to the source. Furthermore, assume that
the velocity of observer at $t$ is $v(t)$. In this system, the
following reference frames and corresponding time intervals can be
defined:
\begin{align}\label{eq:Doppler-FS-16x5}
  \mathcal{K}_s~(dt,dt_v)\rightarrow \mathcal{K}_v~(dt_a)\rightarrow \mathcal{K}_a~(dt')
\end{align}
where $\mathcal{K}_s$ and $\mathcal{K}_a$ are the reference frames of source and the accelerating observer, respectively, while $\mathcal{K}_v$ is an inertial frame moving at velocity $v(t)$ in parallel with the observer's reference frame, which is referred to as a {\em momentarily comoving inertial frame (MCIF)}. Accordingly,  $dt$ is time interval at source and $dt_v$ is time interval at observer, both of which are in $\mathcal{K}_s$, while $dt_a$ and $dt'$ are the time intervals in $\mathcal{K}_v$ and $\mathcal{K}_a$, respectively. With these definitions, the Doppler effect can be represented as
\begin{align}\label{eq:Doppler-FS-166}
  \frac{f'(t)}{f}=&\frac{dt}{dt'}\nonumber\\
  =&\frac{dt}{dt_v}\times\frac{dt_v}{dt_a}\times\frac{dt_a}{dt'}
\end{align}
The clock postulate explains that, in the above equation,   
\begin{align}\label{eq:Doppler-FS-167}
\frac{dt_v}{dt_a}\times\frac{dt_a}{dt'}=\frac{dt_v}{dt'}
\end{align}
is satisfied, meaning that observer's accelerating frame $\mathcal{K}_a$ does not need to be considered. In other words, the transformation from $\mathcal{K}_v$ to $\mathcal{K}_a$ is identity, the Doppler effect is only affected by the instantaneous velocity resulted by the acceleration, and the acceleration itself  does not generate Doppler effect. Therefore, \eqref{eq:Doppler-FS-166} becomes
\begin{align}\label{eq:Doppler-FS-168}
  \frac{f'(t)}{f}=&\frac{dt}{dt_v}\times\frac{dt_v}{dt'}\nonumber\\
  =&\left(\frac{f'(t)}{f}\right)_{\textrm{classic}}\times \frac{1}{\sqrt{1-v^2(t)/c^2}}
\end{align}
where the first term is the classic Doppler effect in frame $\mathcal{K}_s$, and the second term is due to the Lorentz transformation between $\mathcal{K}_s$ and $\mathcal{K}_v$. But what if the clock postulate is not true? Here comes K$\ddot{\textrm{u}}$ndig's experiment and some researches induced by the results obtained from the experiment.

The experiment by W. K$\ddot{\textrm{u}}$ndig in 1963 measured the transverse Doppler effect in an accelerated system by means of the M$\ddot{\textrm{o}}$ssbauer effect~\ncite{PhysRev.129.2371}.  In the experiment, the transverse Doppler effect was measured via a rotating disk, with the source located at the disk center and the observer (absorber) on the rim of the disk, at a distance $R=9.3$~cm away from the center. Note that the classic Doppler effect in the experiment is $\left({f'(t)}/{f}\right)_{\textrm{classic}}=1$. The experimental objective was to test the relativistic dilation of time based on the formula
\begin{align}\label{eq:Doppler-FS-169}
\frac{E_s-E_o}{E_s}=\frac{\Delta E}{E_s}=\sqrt{1-\frac{v^2}{c^2}}-1\approx -\frac{v^2}{2c^2}
\end{align}     
where $E_s$ and $E_o$ represent the particle (Co$^{57}$) energy emitted at source and measured at observer, respectively, and $v=R\omega$ with $\omega$ the angular velocity.

The experiments with $\omega/2\pi$ between $300$ and $35,000$ revolutions per minutes (RPM) were carried out, obtained the results satisfying~\ncite{PhysRev.129.2371,Kholmetskii_2008}
\begin{align}\label{eq:Doppler-FS-170}
\frac{\Delta E}{E_s}=-(1.0065\pm0.011)\frac{v^2}{2c^2}
\end{align}     
which agreed closely with \eqref{eq:Doppler-FS-169}. Due to this, M$\ddot{\textrm{o}}$ssbauer's rotor experiments were not repeated for about half a century, until 2008 when Kholmetskii et al.~\ncite{Kholmetskii_2008} found an error in the data processing of the results obtained from K$\ddot{\textrm{u}}$ndig's experiment. They re-analyzed the results, yielding the relationship of
\begin{align}\label{eq:Doppler-FS-171}
\frac{\Delta E}{E_s}=-(1.192\pm0.011)\frac{v^2}{2c^2}
\end{align}     
which is about $20\%$ higher than K$\ddot{\textrm{u}}$ndig's result or the theoretical result of ${v^2}/{2c^2}$. 

The finding by Kholmetskii et al. has inspired a lot of the followed researches, as shown in \ncite{KHOLMETSKII2020168191} and the references therein. Specifically, relating to the Doppler effect, in light of the Maximal Acceleration theory by Caianiello~\ncite{Caianiello-Maximal-Acceleration}\footnote{In \ncite{Caianiello-Maximal-Acceleration}, Caianiello analyzed the fundamental upper limit of a particle's acceleration based on the Heisenberg uncertainty principles in quantum mechanics, showing that $a_{\max}=2mc^3/\hbar$, where $m$ is particle's rest mass and $\hbar$ is the normalized Planck's constant.}, in \ncite{Friedman_2011}, Friedman suggested to introduce a time dilation by acceleration to the Doppler effect. Reviewing \eqref{eq:Doppler-FS-166}, this means that the transformation from frame $\mathcal{K}_v$ to frame $\mathcal{K}_a$ is not identity, but introduces a time dilation of $dt_a/dt'={1}/{\sqrt{1-a^2/a_{\max}}}$, in addition to the time dilation $dt_v/dt_a={1}/{\sqrt{1-v^2(t)/c^2}}$ yielded by the transformation from frame $\mathcal{K}_s$ to frame $\mathcal{K}_v$. Consequently, the Doppler effect follows a formula of
\begin{align}\label{eq:Doppler-FS-172}
  \frac{f'(t)}{f}=&\frac{1}{\sqrt{1-v^2(t)/c^2}}\times \frac{1}{\sqrt{1-a^2/a_{\max}}}\times\left(\frac{f'(t)}{f}\right)_{\textrm{classic}}
\end{align}
   
One of the objectives of \ncite{Friedman_2011} was to find the maximum acceleration using the results obtained from K$\ddot{\textrm{u}}$ndig's  experiment, which was given as $a_{\max}=(1.006\pm0.063)\times 10^{19}~m/s^2$. However, as mentioned in \ncite{KHOLMETSKII2020168191}, the analysis in \ncite{Potzel-MA} shows that the lowest limit of maximum acceleration, if exists, should be near $10^{21}~m/s^2$. 

Some of other researches carried out experiments trying to find a constant $k$ in
\begin{align}\label{eq:Doppler-FS-173}
\frac{\Delta E}{E_s}=-k\frac{v^2}{c^2}
\end{align}     
These include \ncite{Kholmetskii_2009,Kholmetskii-2011,10.1063/1.4912716,2016CaJPh94780Y}, and their results demonstrated a value of $k=2/3$. 

\subsection{Relativistic Doppler Effect by Gravity}\label{subsection-Doppler-FS.4.3}

In a uniform gravitational field with field strength $g$, its Doppler effect can be directly obtained by Einstein's {\em principle of equivalence}, which states~\ncite{Book-Kenneth-Modern-Physics}: ``{There is no local experiment that can be done to distinguish between the effects of a uniform gravitational field in a non-accelerating inertial frame and the effects of a uniformly accelerating (non-inertial) reference frame.}'' Here by local it means that the experiment is carried out within a sufficiently small space in which the gravitational field is uniform. Relating to Doppler effect, the principle of equivalence means that the Doppler effect generated by a uniform gravitational field of strength $g$ is the same as the Doppler effect generated by an accelerating reference frame with the acceleration $g$. 

Hence, following our analysis in Section~\ref{subsection-Doppler-FS.4.3}, all the formulas can be directly applied for estimating the Doppler effect by the gravitational field, by letting $a=g$ and $X=\Delta h$, where $\Delta h$ denotes the difference in height between source and observer. Specifically, for simplicity, corresponding to \eqref{eq:Doppler-FS-148}      
\begin{subequations}\label{eq:Doppler-FS-174}
\begin{align}\label{eq:Doppler-FS-174a}
  f'\approx& f\left(1+\frac{g\Delta h}{c^2}\right)\\
  \label{eq:Doppler-FS-174b}
  =&f\left(1+\frac{\Delta \Phi}{c^2}\right)
\end{align}
\end{subequations}
where $\Delta \Phi$ is the difference between the gravitational potential at source and that at observer, expressed as
\begin{align}\label{eq:Doppler-FS-175}
\Delta \Phi=\Phi_s-\Phi_o
\end{align}
The gravitational potential by a body of mass $M$ is defined as~\ncite{Book-University-Physics-Young,Book-Carroll-Spacetime-Geometry}
\begin{align}\label{eq:Doppler-FS-176}
\Phi_a=-\frac{GM}{r_a}
\end{align}
with ``$a$'' for ``$s$'' or ``$o$'', and $G=6.674\times
10^{-11}~m^3\cdot kg^{-1}\cdot s^{-2}$ is the gravitational
constant. The formula of \eqref{eq:Doppler-FS-176} explains that the
gravitational potential at $r_a=\infty$ is zero. However, we note that
the location of $r_a$ should be an external point of mass $M$, meaning
that both source and observer are outside mass $M$. Otherwise, if mass
$M$ is spherical, the gravitational potential at any point inside the
mass is the same~\ncite{Book-University-Physics-Young}.

Substituting \eqref{eq:Doppler-FS-175} and \eqref{eq:Doppler-FS-176} into \eqref{eq:Doppler-FS-174b}
\begin{subequations}\label{eq:Doppler-FS-177}
\begin{align}\label{eq:Doppler-FS-177a}
  f' =&f\left(1+\frac{\Phi_s-\Phi_o}{c^2}\right)\\
  \label{eq:Doppler-FS-177b}
  =&f\left(1+\frac{GM}{c^2}\left[\frac{1}{r_o}-\frac{1}{r_s}\right]\right)
\end{align}
\end{subequations}
Hence, when source is above observer, i.e., when the altitude of source is higher than that of observer, \eqref{eq:Doppler-FS-177b} gives $f'>f$, yielding blueshift. By contrast, when source is below observer, \eqref{eq:Doppler-FS-177b} gives $f'<f$, yielding redshift.

Let $f'=1/dt_o$ and $f=1/dt_s$. The time dilation between source and observer in a gravitational field of strength $g$ can be approximately expressed as
\begin{align}\label{eq:Doppler-FS-178}
  dt_o=&dt_s\left(1+\frac{GM}{c^2}\left[\frac{1}{r_o}-\frac{1}{r_s}\right]\right)^{-1}
\end{align}

Above, while the analysis starts with a constant acceleration $g$, the ended formulas in \eqref{eq:Doppler-FS-177} and \eqref{eq:Doppler-FS-178} are general, where no condition on constant $g$ is required. One of its applications is in satellite communication systems. When a ground station with a smaller $r_s$, and larger $g$, sends radio signals to a satellite with a larger $r_o$, and smaller $g$, gravitational field will generate redshift on the radio signals. On the opposite way, gravitational field will generate blueshift on the radio signals. Also, a clock on satellite ticks faster than a clock on Earth's surface.

Moreover, in \eqref{eq:Doppler-FS-177a}, the two gravitational potentials are not required to be generated by a same body of mass, but can be separately by two bodies, meaning that $\Phi_s=-GM_s/r_s$ and  $\Phi_o=-GM_o/r_0$, giving
\begin{align}\label{eq:Doppler-FS-179}
  f' =&f\left(1+\frac{G}{c^2}\left[\frac{M_o}{r_o}-\frac{M_s}{r_s}\right]\right)
\end{align}

For example, when a photon leaving a star's surface with the potential of about $-1.9\times 10^{11}$~Joules (J)/kg is observed on the Earth, where the gravitational potential is about $-6.2\times 10^7$~J/kg, the relative redshift is about $\Delta f/f=(1.9\times 10^{11}-6.2\times 10^7)/c^2=2.11\times 10^{-6}$, when assuming that the star and the Earth do not impact each other, and there are also no other impacts.       

\begin{example}\label{Example-DE-11}
A ground navigation device, at $r_{o}=6370$~km, receives the positioning signals on $f=1.5$~GHz band from a GPS satellite that has an altitude of $26,600$~km. a) Calculate the frequency shift generated by the gravitational differences. b) Calculate the total time shift of the clock on the GPS satellite with respect to the time on Earth over one day. 
\begin{solution}

a) Substituting the respective values into \eqref{eq:Doppler-FS-177b} gives
\begin{align}
  f_D=&f'-f=f\left(\frac{GM}{c^2}\left[\frac{1}{r_o}-\frac{1}{r_s}\right]\right)\nonumber\\
  =&1.5\times 10^9\left(\frac{6.674\times 10^{-11}\times 5.972\times 10^{24}}{9\times 10^{16}}\left[\frac{1}{6.37\times 10^6}-\frac{1}{2.66\times 10^{7}}\right]\right)\nonumber\\
  =&0.7931~\textrm{Hz}\nonumber
\end{align}

b) Following the above calculations, the total time shift over one (Earth) day is
\begin{align}
  dt_s=&24\times 3600\times 0.7931/1.5\times 10^9=45.68~\textrm{microseconds ($\mu$s)}\nonumber
\end{align}
 
\end{solution}
\end{example}   
This example demonstrates that although the Doppler shift is insignificant, the accumulation of time shifts due to gravity is significant.

Above the Doppler effect and time dilation yielded by gravitational field are purely obtained from the principle of equivalence and the extension, no general relativity theory is needed. However, the principle of equivalence is only applicable for uniform gravitational fields, but not for non-uniform, typically, spherical, gravitational fields. Therefore, there is no theoretical basis for \eqref{eq:Doppler-FS-177b} -  \eqref{eq:Doppler-FS-179}.  To fill this gap, let us consider the spherically symmetric gravitational field. In this field, Einstein's field equation (EFE) has a unique solution given by the {\em Schwarzschild metric}~\ncite{Book-Carroll-Spacetime-Geometry,Book-General-Relativity-1}, which, after re-introducing the constant $c$, can be expressed as\footnote{Note that in \ncite{Book-Carroll-Spacetime-Geometry,Book-General-Relativity-1}, the Schwarzschild metric is represented with $c=1$.}
\begin{align}\label{eq:Doppler-FS-180}
ds^2=-c^2\left(1-\frac{2GM}{rc^2}\right)dt^2+\left(1-\frac{2GM}{rc^2}\right)^{-1}dr^2+r^2\left(d\theta^2+\sin^2\theta d\phi^2\right)
\end{align}
where $r,~\theta$ and $\phi$ are the usual spherical polar coordinates, and $ds$ represents the infinitesimal distance. From the Schwarzschild metric we understand that a stationary observer ($dr=d\theta=d\phi=0$) located at a radial distance $r$ from the body of mass $M$ measures a time interval of
\begin{align}\label{eq:Doppler-FS-181}
dt_r^2=-\frac{ds^2}{c^2}=\left(1-\frac{2GM}{rc^2}\right)dt^2
\end{align}
i.e., 
\begin{align}\label{eq:Doppler-FS-182}
dt_r=dt\sqrt{1-\frac{2GM}{rc^2}}=dt\sqrt{1+\frac{2\Phi(r)}{c^2}}
\end{align}
where $dt$ is the time interval that a stationary observer measures at
a distance of $r\rightarrow\infty$, and $\Phi(r)=-GM/r$ is the
gravitational potential at radius $r$. Note that the potential
$\Phi(r)=-GM/r$ is obtained from assuming that Earth is spherical. If
non-spherical Earth is required for high accuracy measurement in, such
as, Section~\ref{section-Doppler-FS.6.1}, $\Phi(r)$ should include the
oblateness term, having the formula of~\ncite{Ashby2003}
\begin{align}\label{eq:Doppler-FS-182b}
  \Phi(r)=-\frac{GM}{r}\left[1-J_2\left(\frac{a_1}{r}\right)^2L_2(\cos\phi)\right]
\end{align}
where $J_2=1.0826300\times 10^{-3}$ is Earth's quadrupole moment
coefficient, $a_1=6.3781370\times 10^{6}$ is Earth's equatorial
radius in meter, $\phi$ is the polar angle measured downward from the axis of
rotational symmetry, and $L_2$ is the Legendre polynomial of degree
$2$.

According to \ncite{Ashby2003}, $dt_r$ in \eqref{eq:Doppler-FS-182}
can also be formulated relative to $dt_0$ of a clock on the geoid (or
surface of the Earth) as
\begin{align}\label{eq:Doppler-FS-182a}
dt_r=dt_0\sqrt{1+\frac{2(\Phi(r)-\Phi_0)}{c^2}}
\end{align}
where $\Phi_0$ is the gravity potential at the geoid.

Now assume two observers, located at the radial distances of $r_1$ and
$r_2(>r_1)$, respectively. Then, we have
\begin{align}\label{eq:Doppler-FS-183}
  \frac{f_{r_1}}{f_{r_2}}=\frac{dt_{r_2}}{dt_{r_1}}=&\frac{dt_{r_2}}{dt_{0}}\times \left(\displaystyle\frac{dt_{r_1}}{dt_{0}}\right)^{-1}\nonumber\\
  =&\sqrt{\frac{1+{2(\Phi(r_2)-\Phi_0)}/{c^2}}{1+{2(\Phi(r_1)-\Phi_0)}/{c^2}}}
\end{align}
which gives both the time relationship and the frequency relationship between the two observers. 

The formulas in \eqref{eq:Doppler-FS-183} are exact results. When both $r_1c^2>>2GM$ and $r_2c^2>>2GM$, the following approximations hold:
\begin{align}\label{eq:Doppler-FS-184}
\frac{f_{r_1}}{f_{r_2}}=&\frac{dt_{r_2}}{dt_{r_1}}\nonumber\\
\approx&\sqrt{({1+{2(\Phi(r_2)-\Phi_0)}/{c^2}})({1-{2(\Phi(r_1)-\Phi_0)}/{c^2}})}\nonumber\\
\approx &\sqrt{1+{2\Phi(r_2)}/{c^2}-{2\Phi(r_1)}/{c^2}}\nonumber\\
\approx &1+\frac{\Phi(r_2)}{c^2}-\frac{\Phi(r_1)}{c^2}
\end{align}
which is \eqref{eq:Doppler-FS-177}, obtained via the extension of the result given by the principle of equivalence.   

\begin{remark}
From \eqref{eq:Doppler-FS-181}  or \eqref{eq:Doppler-FS-182} we find that when $r=R_s=2GM/c^2$, $dt_r=0$. This $R_s$ is the Schwarzschild radius, defines the event horizon of a black hole~\ncite{Book-Carroll-Spacetime-Geometry,Book-General-Relativity-1}. 
\end{remark}

\section{Doppler Effect by Atmosphere}\label{section-Doppler-FS.5}
  
The Earth's atmosphere significantly affects satellite communications and navigation systems~\ncite{11018358}. It diminishes radio signals through processes of scattering and absorption. The non-uniform distribution of air in the atmosphere also obstructs direct LoS signal transmission, introducing timing delays. Furthermore, dynamic atmospheric phenomena, such as rapid fluctuations of electrons in the ionosphere, and variations in temperature, pressure, and humidity within the troposphere, introduce serious randomness to the signals propagating through it. Crucially, this randomness is often difficult to accurately model and predict.

The dynamics in Earth's atmosphere also introduces Doppler effect. In principle, static media do not generate Doppler effect on the signals propagating through them, while dynamic media do. Before analyzing the Doppler effect by the dynamics of Earth's atmosphere, let us consider two simple scenarios to explain the principles.

First, assume that an EM-signal is emitted by a stationary source in
vacuum. Then, this signal is propagated through a first layer of
medium with a refractive index $n_1$, and a second layer of medium
with a refractive index $n_2$, to an observer that is rest in the
second medium. Assume that the distances that the signal travelled in
the first and second media are $r_1$ and $r_2$, respectively. Now
assume that source emits two adjacent wave crests at $t_1=0$ and
$t_2=T$. Ignoring the common distance travelled in the vacuum,
observer receives the first crest at
$t_1'=\frac{r_1}{c/n_1}+\frac{r_2}{c/n_2}$, and the second at
$t_2'=T+\frac{r_1}{c/n_1}+\frac{r_2}{c/n_2}$. Hence, the period of
received signal is $T'=t_2'-t_1'=T$. There is no Doppler effect.

Second, following the above example, now assume that the first medium is not on top of the second medium. Instead, the first wave crest is propagated through the first medium directly to observer, and the second wave crest is propagated through the second medium directly to observer. In both cases, the geometric distances that signal travels in media are the same, expressed as $r$. Accordingly, the first crest is received by observer at $t_1'=\frac{r}{c/n_1}$, and  the second crest at $t_2'=T+\frac{r}{c/n_2}$. Consequently, the period of received signal is $T'=t_2'-t_1'=T+\frac{r}{c/n_2}-\frac{r}{c/n_1}$, yielding the Doppler effect expressed as
\begin{align}\label{eq:Doppler-FS-185}
f'=f\left(1+\frac{r[n_2-n_1]}{\lambda}\right)^{-1}
\end{align}
Hence, there is Doppler effect if $n_2\neq n_1$, $n_2>n_1$ yields redshift, while $n_2<n_1$ yields blueshift.  

The Doppler effect, and also the other effects to be addressed in
Chapter~\ref{Chapter-SGCM}, by Earth's atmosphere on radio
propagations are primarily driven by the troposphere and
ionosphere. The troposphere extends from the Earth's surface to an
altitude of several tens of kilometers. It is the non-ionized portion
of the atmosphere and contains the bulk of its air and water
vapor. Above it lies the ionosphere, which extends upward for several
hundred kilometers. Because of their distinct physical properties,
these layers affect radio propagation differently. Consequently, the
following sections analyze their respective Doppler effects
separately, beginning with the ionosphere and followed by the
troposphere.

For simplicity, the following Doppler effect analysis is based on
\eqref{eq:Doppler-FS-3.16u}, repeated as
\begin{subequations}\label{eq:Doppler-FS-190}
\begin{align}\label{eq:Doppler-FS-190a}
  f_D=&-\frac{f}{c}\frac{dP(n,t)}{dt}\\
  \label{eq:Doppler-FS-190b}
  =&-\frac{f}{c}\frac{d}{dt}\int_{\mathcal{P}(t)}n(p,t)dp
\end{align}
\end{subequations}
where, to clarify, $P(n,t)$ is the phase path from wave source
(satellite or ground station) to observer (ground station or
satellite, correspondinly), $\mathcal{P}(t)$ is the geometric path
from source to observer, which is not necessary a straight line, and
$n(p,t)$ is the refractive index at a point $p$ on $\mathcal{P}(t)$.

Assume that source and observer are at $p_s(t)$ and $p_o(t)$,
respectively, which are the starting and ending points of
$\mathcal{P}(t)$ at $t$. Then, with the aid of the Leibniz integral
rule, \eqref{eq:Doppler-FS-190b} can be written as
\begin{align}\label{eq:Doppler-FS-191}
  f_D=&-\frac{f}{c}\left[\int_{\mathcal{P}(t)}\frac{\partial n(p,t)}{\partial t}dp+n(p_o(t),t)\frac{p_{o}(t)}{dt} - n(p_s(t),t)\frac{p_{s}(t)}{dt}\right]\nonumber\\
  =&f_{DA}+f_{DM}
\end{align}
where, by definition,
\begin{subequations}\label{eq:Doppler-FS-192}
\begin{align}\label{eq:Doppler-FS-192a}
  f_{DA}=&-\frac{f}{c}\int_{\mathcal{P}(t)}\frac{\partial n(p,t)}{\partial t}dp\\
  \label{eq:Doppler-FS-192b}
  f_{DM}=&-\frac{f}{c}\left[n(p_o(t),t)\frac{p_{o}(t)}{dt} - n(p_s(t),t)\frac{p_{s}(t)}{dt}\right]
\end{align}
\end{subequations}
Explicitly, $f_{DM}$ is the Doppler effect because of the relative
motion between source and observer, which has been analyzed in the
previous sections. In contrast, $f_{DA}$ is the Doppler effect by the
medium along the propagation path.  Hence, $f_{DA}$ captures the
frequency shifts caused by time-varying atmospheric conditions (e.g.,
passing weather fronts, pressure changes, or ionospheric
fluctuations), even if both source and observer are stationary. Hence,
to focus on the atmospheric Doppler effect, contributions from other
phenomena, such as motion and gravity, are excluded from the analysis,
unless their inclusion is essential. In other words, the following
analyses focus mainly on $f_{DA}$.

In the expression for $f_{DA}$ in \eqref{eq:Doppler-FS-192a}, the time
derivative of the refractive index, ${\partial n}/{\partial t}$, is a
scalar field. Consequently, its line integral with respect to the arc
length $dp$ is independent of the direction of integration. This is
expressed as
\begin{align}\label{eq:Doppler-FS-193}
  f_{DA}=&-\frac{f}{c}\int_{\mathcal{P}(t)}\frac{\partial n(p,t)}{\partial t}dp=-\frac{f}{c}\int_{-\mathcal{P}(t)}\frac{\partial n(p,t)}{\partial t}dp
\end{align}
In the context of satellite communications, this mathematical identity
implies that the atmospheric Doppler contribution for the
ground-to-satellite link (uplink) is identical to that of the
satellite-to-ground link (downlink) for a given path at time $t$.

Furthermore, because line integration satisfies the additive property,
the total path $\mathcal{P}(t)$ can be partitioned into component
segments. Specifically, by dividing the path into $\mathcal{P}_T(t)$,
traversing the troposphere, and $\mathcal{P}_I(t)$, passing through
the ionosphere, we obtain
\begin{align}\label{eq:Doppler-FS-194}
  f_{DA}=&\underbrace{-\frac{f}{c}\int_{\mathcal{P}_T(t)}\frac{\partial n(p,t)}{\partial t}dp}_{f_{DA,T}}\underbrace{-\frac{f}{c}\int_{\mathcal{P}_I(t)}\frac{\partial n(p,t)}{\partial t}dp}_{f_{DA,I}}
\end{align}
This demonstrates that the Doppler effects induced by the troposphere
and ionosphere can be analyzed as independent, additive terms.

Below we first analyze the Doppler effect by the ionosphere, i.e.,
$f_{DA,I}$ in \eqref{eq:Doppler-FS-194}.

\subsection{Doppler Effect by Ionosphere}\label{section-Doppler-FS.5.1}

As shown in the expression \eqref{eq:Doppler-FS-194}, the analysis of
the ionospheric Doppler effect depends on two issues: a) the
mathematical modeling of the propagation path $\mathcal{P}(t)$, and b)
the mathematical modeling of the local refractive index $n$ associated
with $\mathcal{P}(t)$. The refractive index $n$ is primarily
determined by the local electron density and distribution, which, in
cold plasma, can be calculated from the Appleton-Hartree
equation~\ncite{Budden_1988,SEZEN201972,SCOTTO20141642}
\begin{align}\label{eq:Doppler-FS-195}
n^2=1-\frac{X}{1-jZ-\displaystyle\frac{Y^2\sin^2\theta}{2(1-X-jZ)}\pm\sqrt{\frac{Y^4\sin^4\theta}{4(1-X-jZ)^2}+Y^2\cos^2\theta}}
\end{align}
where
\begin{itemize}

\item $j=\sqrt{-1}$;

\item $X=f_N^2/f^2$: the ratio of the square of the plasma frequency $f_{N}$ to the wave frequency $f$; 

\item $Y=f_H/f$: the ratio of the electron gyro-frequency $f_{H}$ to the wave frequency;

\item $Z=f_v/f$: the ratio of the electron collision frequency $f_v$ to the wave frequency;

\item $\theta$: the angle between the wave vector and the magnetic field vector;

\item $Y\cos\theta$ and $Y\sin\theta$: Longitudinal and transverse components of $Y$ relative to the wave propagation direction; 

\item $\pm$ signs: `+' is for the refractive index of ordinary wave, while `-' is for the refractive index of extraordinary wave.

\end{itemize}
Furthermore, $f_N=(2\pi)^{-1}\sqrt{N_eq^2/\varepsilon_0 m_e}$, where $N_e$ is the electron density, $\varepsilon_0$ is the free space dielectric permittivity, $q$ is the electron charge and $m_e$ is the mass of electron. $f_H=(2\pi)^{-1}qB_0/m_e$,  with $B_0$ the strength of geomagnetic field. 

Explicitly, it is not easy to analytically derive an expression for
$n$ that can be applied to \eqref{eq:Doppler-FS-194} for further
Doppler effect analysis. The refractive index in various ionospheric
scenarios has been analyzed in \ncite{Budden_1988}. In our case of
focusing on the principles and considering satellite communications,
which typically use frequencies $>1$~GHz, the refractive index can be
approximately expressed as~\ncite{Range-rate-Doppler,4066048}
\begin{align}\label{eq:Doppler-FS-196}
n\approx\sqrt{1-X}=\sqrt{1-\frac{\kappa N_e}{f^2}}
\end{align}
where $\kappa=q^2/4\pi^2\varepsilon_0 m_e=80.61~m^3/s^2$. When $\kappa<<f^2$, $n$ can be further approximated as
\begin{align}\label{eq:Doppler-FS-197}
n\approx 1-\frac{\kappa N_e}{2f^2}
\end{align}
which is a function of the electron density distribution. Substituting this into $f_{DA,I}$ in \eqref{eq:Doppler-FS-194} gives 
\begin{align}\label{eq:Doppler-FS-198}
  f_{DA,I}=&\frac{\kappa}{2fc}\int_{\mathcal{P}_I(t)}\frac{\partial N_e(p,t)}{\partial t}dp
\end{align}

\begin{example}\label{Example-DE-12}

Assume that signals are transmitted from ground station to satellite. Assume that Earth's ionosphere is modeled as a simple flat ionosphere layer~\ncite{Range-rate-Doppler} in which the electron density increases linearly with altitude above a minimum height $h_0$ and changes with time, expressed as
\begin{align}\label{eq:Doppler-FS-199}
N_e(z,t)=\alpha[z-h_0],~z\geq h_0 
\end{align}
where both $\alpha$ and $h_0$ may be functions of time, depending on the models.

In relation with this simple ionosphere model, a ray model for propagation path is proposed, which is a straight line from ground to $h_0$, and then traces a parabola above $h_0$, represented as~\ncite{Budden_1988,Range-rate-Doppler}
\begin{align}\label{eq:Doppler-FS-200}
x=h_0\cot\theta_E+\frac{2f^2}{\alpha \kappa}\cos\theta_E\left(\sin\theta_E-\left[\sin^2\theta_E-\frac{\alpha\kappa(z-h_0)}{f^2}\right]^{1/2}\right),~z\geq h_0
\end{align}
where $x$ is the horizontal distance, measured from the wave launching
point, and $\theta_E$ is the ray's launch elevation angle, which is
the angle between the local horizontal plane and the ray's direction
towards a considered satellite.

In \eqref{eq:Doppler-FS-199}, when assuming that $\alpha$ is a constant while $h_0$ is time-varying, according to \ncite{Range-rate-Doppler}, the Doppler frequency can be found to be
\begin{align}\label{eq:Doppler-FS-201}
f_{DA,I}=-\frac{fV_z}{c}\left(\sin\theta_E+\cos^2\theta_E\times\log\left[\sqrt{\frac{1+\sin\theta_E}{1-\sin\theta_E}}\right]\right)
\end{align}
where $V_z=dh_0/dt$ is the time-varying rate of $h_0$.

Note that, in \ncite{Range-rate-Doppler}, the case of time-varying $\alpha$ and constant $h_0$ in \eqref{eq:Doppler-FS-199} has also been considered, giving
\begin{align}\label{eq:Doppler-FS-202}
f_{DA,I}=-\frac{f}{c}\sin\theta_E\left(\frac{D}{3\cos^2\theta_E}-\frac{8f^2\sin\theta_E\cos\theta_E}{3\alpha\kappa}\right)\frac{d\theta_E}{dt}
\end{align}
where $D$ is a constant given by
\begin{align}\label{eq:Doppler-FS-203}
D=2\left(h_0\cot\theta_E+\frac{2f^2}{\alpha\kappa}\cos\theta_E\sin\theta_E\right)
\end{align}

\end{example}

Other models of electron density includes the exponential model, which can in general be expressed as~\ncite{4065735}
\begin{align}\label{eq:Doppler-FS-204}
N_e(h,t)=N_0\exp(-\alpha[h-h_0]),~h\geq h_0
\end{align}
where $h$ is altitude and the other parameters can be set according to scenarios, as seen, for example, in \ncite{4065735}.

Another fundamental model for the electron density in the ionosphere is the Chapman model~\ncite{S_Chapman_1931-a,S_Chapman_1931-b,Hargreaves_1992,Budden_1988}. Specifically, the widely used model for the vertical profile of electron density in the E and F layers is represented by the $\alpha$-Chapman function as
\begin{align}\label{eq:Doppler-FS-205}
N_e(h)=N_{\max}\exp\left(0.5[1-z-\exp(-z)]\right)
\end{align}
where $N_{\max}$ is the peak electron density and $z=(h-h_{\max})/H$. Here $h_{\max}$ is the altitude of peak density, typically $250$-$400$~km for the F2 layer, and $H=\kappa_B T_K/mg$ is the ionospheric scale height, where $\kappa_{B}$ is the Boltzmann constant, $T_K$ is the absolute plasma temperature, $m$ is the mean ion mass and $g$ is the acceleration by gravity.    

Example~\ref{Example-DE-12} demonstrates that, even for such a simple
ionospheric scenario, deriving the analytical solution for the Doppler
shift is highly involved. Hence, for convenience of ionospheric
calculations, the total electron content (TEC) is introduced, which is
defined as the integral of the free electron density along the direct
path between the satellite (A) and the receiver (B), expressed as
\begin{align}\label{eq:Doppler-FS-206}
\textrm{TEC}=\int_A^BN_e(s)ds
\end{align}
Accordingly, for a high-frequency satellite link, when applying \eqref{eq:Doppler-FS-206}
into \eqref{eq:Doppler-FS-198}, the Doppler shift introduced specifically by the ionosphere can be written as
\begin{align}\label{eq:Doppler-FS-207}
f_{DA,I}=\frac{\kappa}{2fc}\frac{d(\textrm{TEC})}{dt}
\end{align}

Note that, at the NASA CDDIS Global Ionosphere Maps and NOAA's GloTEC
Data Center\footnote{https://www.swpc.noaa.gov/products/us-total-electron-content, https://www.earthdata.nasa.gov/data/space-geodesy-techniques/gnss/data-products-holdings},
real-time and historical vertical TEC (VTEC) values for specific grid
coordinates on Earth are regularly published in IONEX format. Once
VTEC data are known, they can be converted to slant TEC (STEC) data
for path-specific ionospheric calculations using a geometric mapping
function~\ncite{TEC-mapping}.

\subsection{Doppler Effect by Troposphere}\label{section-Doppler-FS.5.2}

To analyze the Doppler effect by the troposphere, let us return
\eqref{eq:Doppler-FS-190b}, which includes the effect from both medium
and relative motion. To focus on the effect by medium in troposphere,
its Doppler effect can be written as
\begin{align}\label{eq:Doppler-FS-208}
  f_{DA,T}=&-\frac{f}{c}\frac{d}{dt}\left[\int_{\mathcal{P}_T(t)}n(p,t)dp-\int_{\mathcal{G}_T}dp\right]\nonumber\\
  =&-\frac{f}{c}\frac{d}{dt}(\Delta p(n,t))
\end{align}
with $\Delta p$ defined as
\begin{align}\label{eq:Doppler-FS-209}
  \Delta p(n,t)=&\int_{\mathcal{P}_T(t)}n(p,t)dp-\int_{\mathcal{G}_T}dp
\end{align}
Consider the case that signals are transmitted from a ground station
to a satellite. Then, in \eqref{eq:Doppler-FS-209}, the first integral
in the bracket is taken along the bended signal path through the
troposphere, while the second integral is along the tropospheric
potion of the slant path\footnote{The slant path between a ground
station and a satellite is the LoS path from the ground station to the
satellite.}.  It is shown
that~\ncite{Hopfield-Troposphere-Refractivity-1963}, when $n$ along
path $\mathcal{P}_T(t)$ and that along path $\mathcal{G}_T$ are
sufficiently similar, $\int_{\mathcal{P}_T(t)}n(p,t)dp\approx
\int_{\mathcal{G}_T}n(p,t)dp$, giving
\begin{align}\label{eq:Doppler-FS-210}
 \Delta p(n,t)\approx &\int_{\mathcal{G}_T}n(p,t)dp-\int_{\mathcal{G}_T}dp\nonumber\\
  =&\int_{\mathcal{G}_T}[n(p,t)-1)]dp
\end{align}
As noted in \ncite{Hopfield-Troposphere-Refractivity-1963}, the error
generated by the replacement of $\mathcal{P}_T(t)$ using
$\mathcal{G}_T$ is a second-order effect, which may reach about
$10\%$ of $\Delta p$ at the horizon, but is typically negligible
provided that the elevation angle is above $3^o$ or $4^o$.

Since the refractive index $n$ of air in troposphere is very small,
the refractivity defined as $N\equiv 10^6(n-1)$ is usually instead
used in studies. Applying this into \eqref{eq:Doppler-FS-210} yields
\begin{align}\label{eq:Doppler-FS-211}
 \Delta p(n,t)
  =&10^{-6}\int_{\mathcal{G}_T}N(p,t)dp
\end{align}

In contrast to the ionosphere, where refractive index is depended on frequency and ions, as shown in \eqref{eq:Doppler-FS-195} and \eqref{eq:Doppler-FS-197}, typically, the refractivity in the troposphere is a function of
temperature, pressure and humidity. Considering these factors, for example, an
expression that is good to $0.5\%$ in $N$ for frequencies upto
$30$~GHz has been obtained from a set of meteorological ballon data, which
is expressed as~\ncite{4051437}
\begin{align}\label{eq:Doppler-FS-212}
N=\frac{77.6}{T_K}\left(P+\frac{4810 e}{T_K}\right)
\end{align}
where $P$ is the total pressure and $e$ is the partial presure of
water vapor, both in millibars, and $T_K$ is absolute temperature.

\begin{figure}[tb]
  \begin{center}
 \includegraphics[angle=0,width=.9\linewidth]{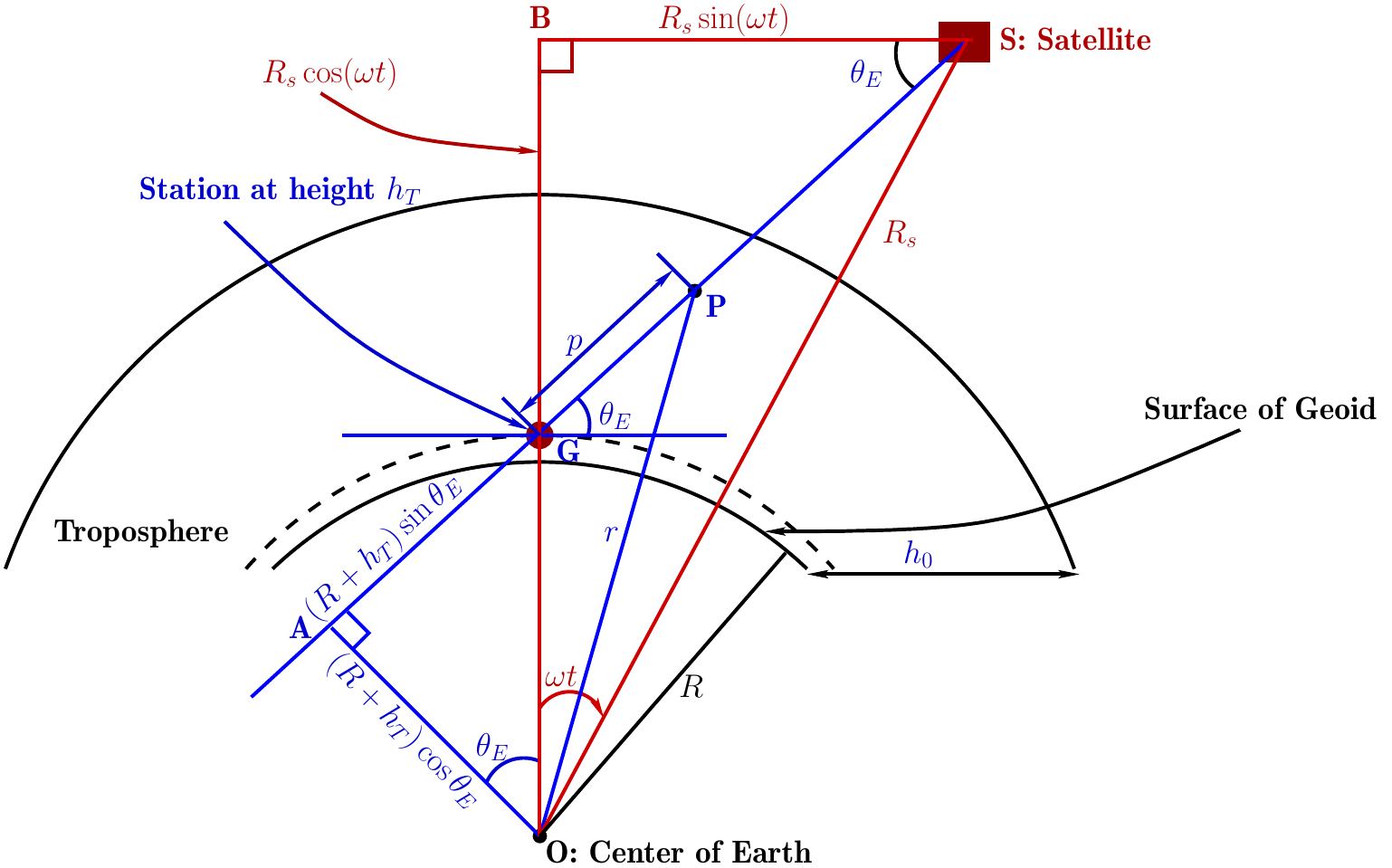}
  \end{center}
  \caption{Illustration of geometries for tropospheric analysis.}
  \label{figure-Troposphere-illustraction}
\end{figure}

Furthermore, during the passing interval of a satellite, the
refractivity $N$ of air in the troposphere is approximately a function
of height above the Earth, but is independent of both the horizontal
locations and time. Therefore, \eqref{eq:Doppler-FS-211} can be
further approximated as
\begin{align}\label{eq:Doppler-FS-213}
 \Delta p(n,t)\approx &10^{-6}\int_{R+h_T}^{R+h_0}N(r)\left(\frac{dp}{dr}\right)dr
\end{align}
where $r$ is the radial distance from the Earth's center, $R$ is the radius of the Earth (or of the geoid at the ground station), $h_T$ is the height of the ground station, and $h_0$ is the height above which the tropospheric refraction becomes negligible, as shown in Fig.~\ref{figure-Troposphere-illustraction}. 

Substituting \eqref{eq:Doppler-FS-213} into \eqref{eq:Doppler-FS-208} gives the Doppler effect by the troposphere expressed as
\begin{align}\label{eq:Doppler-FS-214}
  f_{DA,T}\approx &-\frac{10^{-6}f}{c}\frac{d}{dt}\int_{R+h_T}^{R+h_0}N(r)\left(\frac{dp}{dr}\right)dr
\end{align}
 
From Fig.~\ref{figure-Troposphere-illustraction} we can know that, in
\eqref{eq:Doppler-FS-214}, $p$ is a function of $r$ and the elevation
angle $\theta_E$. For a given $r$, the elevation angle $\theta_E$ is
time-varying during the passing of a satellite. Considering these
relationships, \eqref{eq:Doppler-FS-214} can be then represented as
\begin{align}\label{eq:Doppler-FS-215}
  f_{DA,T}=&-\frac{10^{-6}f}{c}\int_{R+h_T}^{R+h_0}N(r)\frac{\partial}{\partial \theta_E}\left(\frac{dp}{dr}\right)\frac{d\theta_E(t)}{dt}dr\nonumber\\
  =&-\frac{10^{-6}f}{c}\left[\int_{R+h_T}^{R+h_0}N(r)\frac{\partial}{\partial \theta_E}\left(\frac{dp}{dr}\right)dr\right]\frac{d\theta_E(t)}{dt}\nonumber\\
  =&\frac{10^{-6}f}{c}\times f(\theta_E)\times \frac{d\theta_E(t)}{dt}
\end{align}
where $f(\theta_E)$ is defined as
\begin{align}\label{eq:Doppler-FS-216}
 f(\theta_E)=-\int_{R+h_T}^{R+h_0}N(r)\frac{\partial}{\partial \theta_E}\left(\frac{dp}{dr}\right)dr
  \end{align}

Based on the geometries shown in
Fig.~\ref{figure-Troposphere-illustraction} (using $\bigtriangleup$OAP and $\bigtriangleup$OAG), it can be found that
\begin{align}\label{eq:Doppler-FS-217}
p(r,\theta_E)=\sqrt{r^2-(R+h_T)^2\cos^2\theta_E}-(R+h_T)\sin\theta_E
\end{align}
Then, the derivative of $p$ first with respect to $r$ and then with
respect to $\theta_E$ can be calculated to be
\begin{align}\label{eq:Doppler-FS-218}
\frac{\partial}{\partial \theta_E}\left(\frac{dp}{dr}\right)=-\frac{r(R+h_T)^2\sin(2\theta_E)}{{2[r^2-(R+h_T)^2\cos^2\theta_E]^{3/2}}}
\end{align}
Substituting it into \eqref{eq:Doppler-FS-216} and completing the
integration for a given $N(r)$, $f(\theta_E)$ can be obtained.

Next, to derive ${d\theta_E(t)}/{dt}$ in \eqref{eq:Doppler-FS-215},
for simplicity, the Earth is assumed to be a sphere with radius $R$,
the orbit of satellite is assumed to be circular with a radiun of
$R_s$, and the angular speed of satellite is expressed as
$\omega$. Then, from the geometries shown in
Fig.~\ref{figure-Troposphere-illustraction} (using $\bigtriangleup$SBO and  $\bigtriangleup$SBG), we have
\begin{align}\label{eq:Doppler-FS-219}
\tan\theta_E(t)=\frac{R_s\cos(\omega t)-(R+h_T)}{R_s\sin(\omega t)}
\end{align}
when it is assumed that $t=0$ occurs at the point when the satellite
is right over the ground station. Then, it can be shown that
\begin{align}\label{eq:Doppler-FS-220}
  \frac{d\theta_E}{dt}=&\omega\left[\frac{R_s(R+h_T)\cos(\omega t)-R_s^2}{R_s^2+(R+h_T)^2-2R_s(R+h_T)\cos(\omega t)}\right]\nonumber\\
  \approx & \omega\left[\frac{R_sR\cos(\omega t)-R_s^2}{R_s^2+R^2-2R_sR\cos(\omega t)}\right]
\end{align}
where the approximation is due to $h_T<<R$.

\begin{figure}[tb]
  \begin{center}
 \includegraphics[angle=0,width=.65\linewidth]{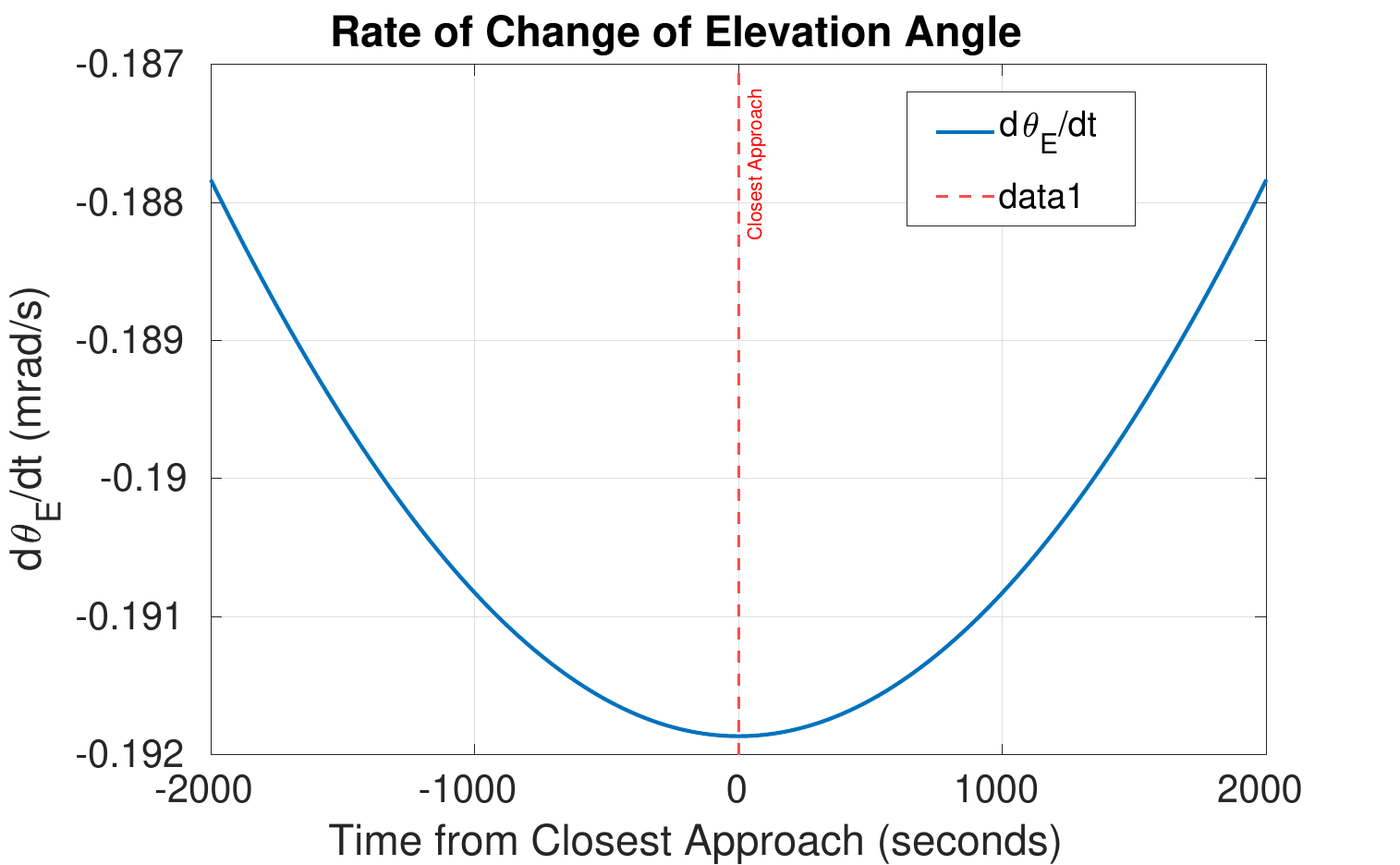}
  \end{center}
  \caption{Illustration of $d\theta_E/dt$ for a GPS satellite with the
    parameters of $R= 6378.137$~km, $R_s=26560$~km and the orbital
    period of $T_s=43082$~s (12 h).}
  \label{figure-Compute1_dThetaE_dt}
\end{figure}

Consider a GPS satellite, which has the orbital period of about
$T_s=12$ hours, Fig.~\ref{figure-Compute1_dThetaE_dt} shows the rate
of change of elevation angle during the passing time of a satellite,
where $t=0$ in the middle aligns with the point when the satellite is
overhead of the ground station.  As the absolute value of $d\theta
_{E}/dt$ reaches its maximum at $t=0$, the angular sensitivity of the
signal path through the atmosphere is greatest when the satellite is
directly overhead. Consequently, the dynamics of the
elevation-dependent Doppler corrections are most pronounced at this
point of closest approach.

\begin{remark}
  
In \ncite{Hopfield-Troposphere-Refractivity-1963}, the reference of
$t=0$ was defined at the point when satellite is closest to the center
of Earth. Accordingly, the expression for ${d\theta_E}/{dt}$ was
derived, which is different from \eqref{eq:Doppler-FS-220}.
\end{remark}

In literature, there are various formulas proposed for modeling
$N(r)$~\ncite{4065735,Hopfield-Troposphere-Refractivity-1963,Hopfield-Troposphere-Refractivity,4051437}. Below
two examples from
\ncite{Hopfield-Troposphere-Refractivity-1963,Hopfield-Troposphere-Refractivity}
as well as their resulted $f(\theta_E)$ are provided.

\begin{example}\label{Example-DE-13}
In \ncite{Hopfield-Troposphere-Refractivity-1963}, an quadratically
approximated model of $N(r)$ to the three-part model~\ncite{4065735}
was proposed, which is expressed as
\begin{align}\label{eq:Doppler-FS-221}
  N(r)=a\left[r-(R+h_0)\right]^2
\end{align}
where $a$ can be obtained from boundary conditions. For example, given
$N_T$ at $r=R+h_T$ (location of ground station), $a$ can be found to
be $a=N_T/(h_0-h_T)$.  Upon substituting this and
\eqref{eq:Doppler-FS-218} into \eqref{eq:Doppler-FS-216} and
completing the integration, it can be shown that
\begin{align}\label{eq:Doppler-FS-222}
  f(\theta_E)=&N_Tr_T\left[\cos\theta_E+\frac{r_T\sin(2\theta_E)}{h^2_{tro}}\left[\sqrt{r_T^2\sin^2\theta_E+2r_Th_{tro}+h_{tro}^2}\right.\right.\nonumber\\
    &\left.\left. -r_T\sin\theta_E+r_{tro}\ln\left(\frac{r_T(1+\sin\theta_E)}{r_{tro}+\sqrt{r_T^2\sin^2\theta_E+2r_Th_{tro}+h_{tro}^2}}\right)\right]\right]
\end{align}
where, by definition, $r_T=R+h_T$, $h_{tro}=R-h_T$ and
$r_{tro}=R+h_0$.

The results provided in \ncite{Hopfield-Troposphere-Refractivity-1963}
demonstrated that, with the aid of this quartic tropospheric
refractivity model, approximately $90\%$ of the tropospheric effect on
the Doppler effect could be removed, and hence allowing to achieve the
satellite-relied navigation of higher accuracy.
\end{example}

\begin{example}\label{Example-DE-14} 

In \ncite{Hopfield-Troposphere-Refractivity}, a quartic tropospheric
refractivity model was developed for evaluation of Doppler effect in
troposphere. In this model, the refractivity profile in
\eqref{eq:Doppler-FS-212} is divided into two components, expressed as
\begin{align}\label{eq:Doppler-FS-223}
N(r)=N_d(r)+N_w(r)
\end{align}
where $N_d$ is the `dry' component for dry air, and $N_w$ is the `wet'
component for water vapour. The models for these components are
expressed as
\begin{align}\label{eq:Doppler-FS-224}
  N_x(r)=\left\{\begin{array}{ll}
  \frac{N_{T_x}}{(h_{0x}-h_T)^4}(h_{0x}-r)^4, & r\leq h_{0x}\\
  0, & r> h_{0x}
\end{array}  \right.
\end{align}
where `$x$' stands for `$d$' or `$w$'. In \eqref{eq:Doppler-FS-224},
$N_{T_x}$ is the refractivity at ground station, and $h_{0d}$ and
$h_{0w}$ are the heights above which the corresponding tropospheric
refraction becomes negligible\footnote{Note that, as shown in
\ncite{Hopfield-Troposphere-Refractivity}, the height $h_{0d}$ may
reach about $40$~km, while $h_{0w}$ only reaches $12$~km of
maximum.}. All heights are measured with respect to the geoid. Then,
according to \ncite{Hopfield-Troposphere-Refractivity}, it can be derived that
\begin{align}\label{eq:Doppler-FS-225}
 f(\theta_E)=r_T\times\sum_{x\in\{d,w\}}N_{T_x}F_{4x}(\theta_E)
\end{align}
where for ${x\in\{d,w\}}$, $F_{4x}(\theta_E)$ is expressed as
\begin{align}\label{eq:Doppler-FS-226}
  F_{4x}(\theta_E)=&\cos\theta_E\left(1+\frac{4l_1}{h_{tro_x}^4}\left[\frac{l_{3x}^3-l_1^2}{3}+l_{3x}\left(l_2^2+\frac{3r_{tro_x}^2}{2}\right) \right.\right.\nonumber\\
    & -l_1\left(l_2^2-\frac{3r_Tr_{tro_x}}{2}+3r_{tro_x}^2\right)\nonumber\\
    &\left.\left. +\left(\frac{3r_{tro_x}l_2^2}{2}+r_{tro_x}^2\right)\ln\left(\frac{r_T+l_1}{r_{tro_x}+l_{3x}}\right)    \right]\right)
\end{align}
In \eqref{eq:Doppler-FS-226}, $r_T=R+h_T$, $h_{tro_x}=h_{0x}-h_T$, $r_{tro_x}=r_T+h_{tro_x}$, $l_1=r_T\sin\theta_E$, $l_2=r_T\cos\theta_E$, and $l_{3x}=\sqrt{r^2_{tro_x}-l_2^2}$. 

The results provided in \ncite{Hopfield-Troposphere-Refractivity} show
that the Doppler effects by both `dry' and `wet' can be closely
modelled by the proposed models for $N_d(r)$ and $N_w(r)$.  Their
effectiveness is evidenced by their close matching to any local
average profile of refractivity, in terms of the reduction of Doppler
residuals and navigation errors.

\begin{figure}[tb]
  \begin{center}
 \includegraphics[angle=0,width=.65\linewidth]{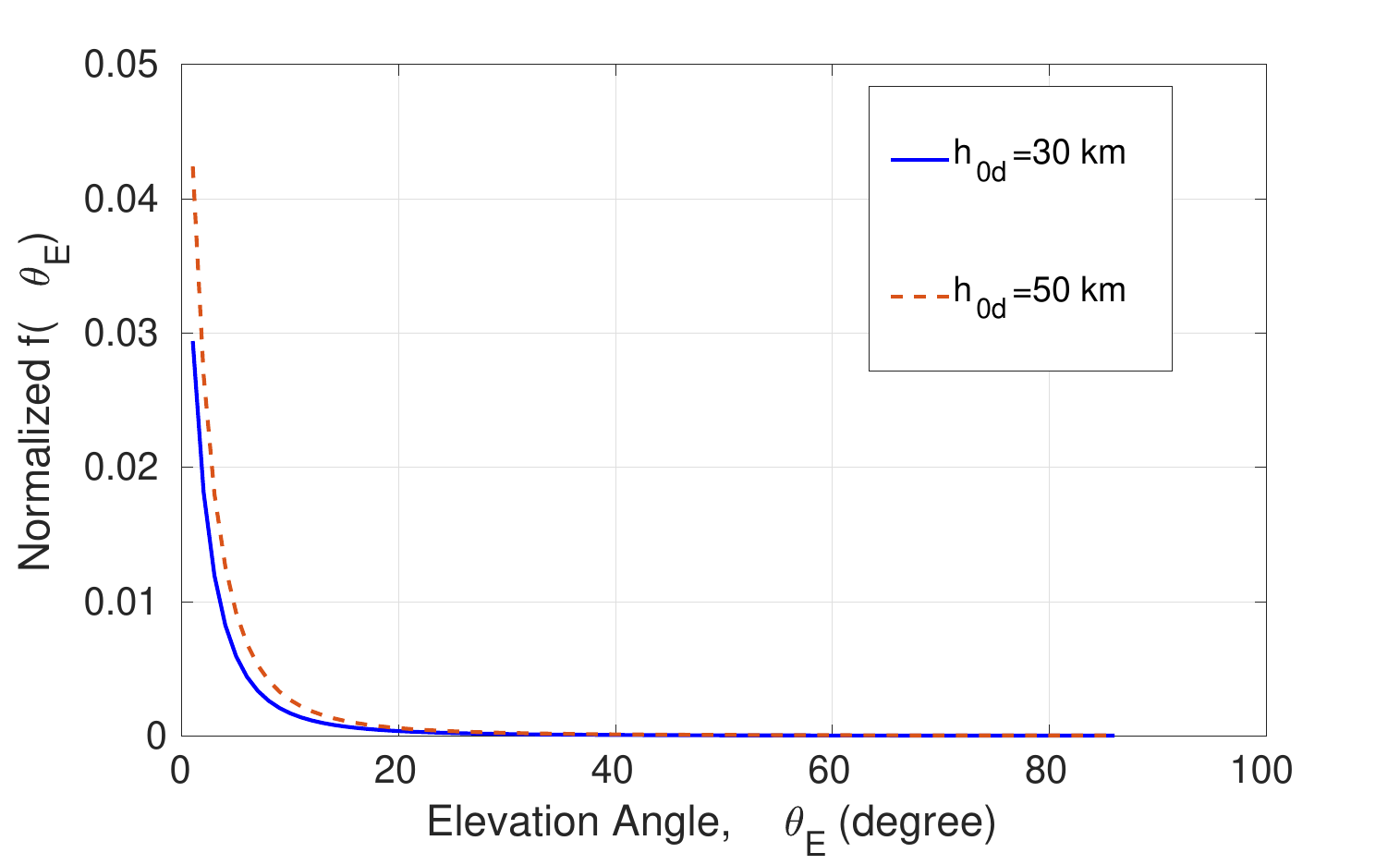}
  \end{center}
  \caption{Illustration of $f(\theta_E)$ for $h_{0d}=30$ and $50$
    kilometers, when other parameters are $R= 6378.137$~km, $h_T=0$,  $N_{T_d} = 315$,  $N_{T_w} = 50$ and $h_{0w} = 12$~km.}
  \label{figure-Two_Quartic_1}
\end{figure}

Fig.~\ref{figure-Two_Quartic_1} illustrates the profile of
$f(\theta_E)$ with $\theta_E$, showing that the impact reduces as the
elevation angle increases. Note that the model also becomes less
accurate as the elevation angle is small, such as, below $10$
degree. Note additionally that $f(\theta_E)$, i.e.,
\eqref{eq:Doppler-FS-222}, in Example~\ref{Example-DE-13} appears a
similar shape as the curves in Fig.~\ref{figure-Two_Quartic_1}, as
shown in \ncite{Hopfield-Troposphere-Refractivity-1963}.

\end{example}

\section{Doppler Effect in Applications}\label{section-Doppler-FS.6}

Having analyzed the Doppler effect by individual phenomena in
previous sections, in this section, we examine two representative
applications where multiple Doppler-generating factors may occur
simultaneously.

\subsection{Land Target Sensing}\label{section-Doppler-FS.6.1}

\begin{figure}[tb]
  \begin{center}
 \includegraphics[angle=0,width=.6\linewidth]{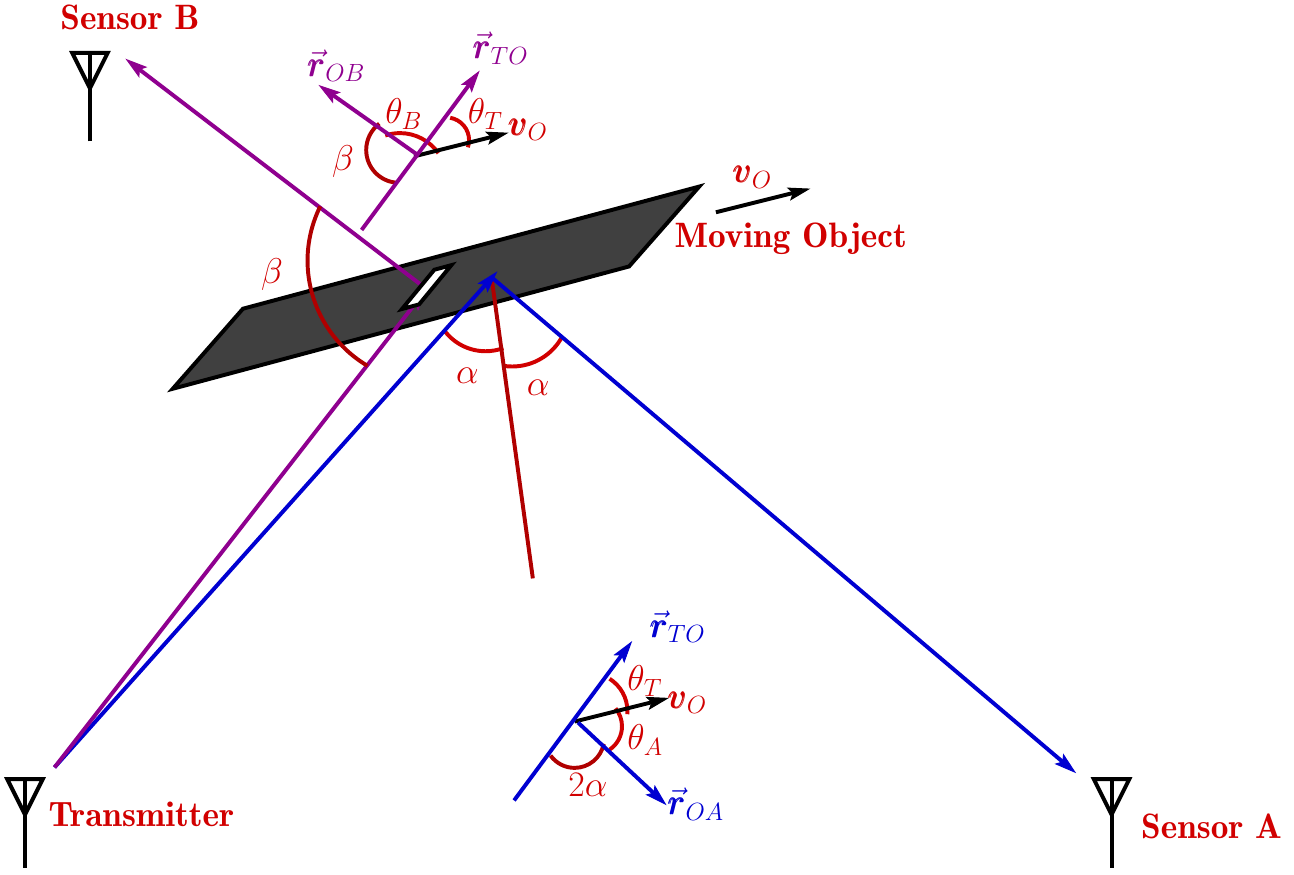}
  \end{center}
  \caption{Illustration for analyzing the Doppler shift by a moving
    object being sensed.}
  \label{figure-Application-1}
\end{figure}

Fig.~\ref{figure-Application-1} shows a land wireless sensing scenario
where a transmitter sends signals of frequency $f$ to probe objects in
the vicinity. Transmitter, sensors and object are all assumed on the same, such as $xy$, plane. Although the analyses are similar, two sensing scenarios
are considered. In Scenario A, signals sent by transmitter are
reflected to Sensor A, where the Doppler shift is estimated. In this
case, the angle of incidence is equal to the angle of reflection,
both expressed as $\alpha$ in Fig.~\ref{figure-Application-1}. By contrast,
in Scenario B, signals sent by transmitter is diffracted by the moving
object to Sensor B, where the Doppler shift is estimated.

Both transmitter and sensors are assumed to be fixed, while the object
being sensed is moving with a velocity vector $\bm{v}_o$, and
$\hat{\bm{r}}_{XY}$ is a unit vector in the direction from position
$X$ to position $Y$. Let $\arg(\bm{z})$ be the angle of $\bm{z}$
defined referred to $x$-axis. The angles in the figure are defined as
follows: $\theta_A=\arg(\bm{v}_o)-\arg(\hat{\bm{r}}_{OA})$,
$\theta_T=\arg(\hat{\bm{r}}_{TO})-\arg(\bm{v}_o)$,
$\theta_B=\arg(\hat{\bm{r}}_{OB})-\arg(\bm{v}_o)$, and
$\beta=180-(\arg(\hat{\bm{r}}_{OB})-\arg(\hat{\bm{r}}_{TO}))$. As in
previous sections, the relative velocity of moving away is positive
and that moving towards each other is negative. To embrace generality,
relativistic Doppler effect is analyzed, with motion is relative to
the Earth's reference frame. Also assumed is signal transmission in
vacuum.

Let us first analyze Scenario A. According to
Fig.~\ref{figure-Application-1}, the velocity projected on the line
connecting transmitter and object is
\begin{align}\label{eq:Doppler-FS-227}
  v_{TO}=|\bm{v}_{o}|\cos\theta_T
\end{align}
and the velocity projected on the line
connecting object and sensor A is
\begin{align}\label{eq:Doppler-FS-228}
  v_{OA}=-|\bm{v}_{o}|\cos\theta_A
\end{align}

Assume that the proper times at transmitter, object and Sensor A are expressed as $T$, $T_o$ and $T'$, respectively. Signal transmission from transmitter to object follows the case of rest source moving observer, which has been analyzed in Section~\ref{subsubsection-Doppler-FS.1.3}. Then, by combining \eqref{eq:Doppler-FS-XW-3.64} and \eqref{eq:Doppler-FS-XW-3.65}, $T$ and $T_o$ has the relation of 
\begin{align}\label{eq:Doppler-FS-229}
T=T_o\frac{1-{v_{TO}}/{c}}{\sqrt{1-|\bm{v}_o|^2/c^2}}
\end{align}
By contrast, signal transmission from object to Sensor A is in the case of moving source rest observer, which has also been analyzed in Section~\ref{subsubsection-Doppler-FS.1.3}. Correspondingly, based on \eqref{eq:Doppler-FS-3.48}, the relationship between $T_o$ and $T'$ can be represented as 
\begin{align}\label{eq:Doppler-FS-230}
T_o=T'\times \frac{\sqrt{1-|\bm{v}_o|^2/c^2}}{1+{v}_{OA}/{c}}
\end{align}
Substituting $T_o$ of \eqref{eq:Doppler-FS-230} into \eqref{eq:Doppler-FS-229} gives 
\begin{align}\label{eq:Doppler-FS-231}
T=T'\frac{1-{v_{TO}}/{c}}{1+{v}_{OA}/{c}}
\end{align}
Eq.\eqref{eq:Doppler-FS-231} shows that the relativistic terms in \eqref{eq:Doppler-FS-229} and \eqref{eq:Doppler-FS-230} cancel each other, making the formulas in both relativistic and non-relativistic cases the same. This result is expected, as both transmitter and Sensor A are in the same reference frame, and the motion of object is also relative to this reference frame. 

From \eqref{eq:Doppler-FS-231} the Doppler shift $f_D=f'-f=1/T'-1/T$ can be derived to have the formula of
\begin{align}\label{eq:Doppler-FS-232}
f_D=&-\frac{f({v_{TO}}/{c}+{v}_{OA}/{c})}{1+{v}_{OA}/{c}}\nonumber\\
\approx &-\frac{f(v_{TO}+v_{OA})}{c}
\end{align}
Substituting \eqref{eq:Doppler-FS-227} and \eqref{eq:Doppler-FS-228} into \eqref{eq:Doppler-FS-232} yields
\begin{align}\label{eq:Doppler-FS-233}
f_D(\theta_T,\theta_A)=&-\frac{f|\bm{v}_{o}|(\cos\theta_T-\cos\theta_A)}{c}
\end{align}
Furthermore, according to the geometry in Fig.~\ref{figure-Application-1}, we have $\theta_A=180-2\alpha-\theta_T$. Applying this into \eqref{eq:Doppler-FS-233} results in
\begin{align}\label{eq:Doppler-FS-234}
f_D(\alpha,\theta_T)=&-\frac{f|\bm{v}_{o}|[\cos\theta_T+\cos(2\alpha+\theta_T)]}{c}
\end{align}

When $\alpha\neq 0$, the Doppler shift of \eqref{eq:Doppler-FS-234} is that in the bi-static sensing scenario~\ncite{Lie-Liang-Optimization-book,10769985}, where signal transmitter and sensor are at different locations, hence unable to share a common clock. By contrast, in the mono-static sensing scenario where transmitter and sensor are co-located~\ncite{5776640}, allowing to share the same clock, $\alpha= 0$. Accordingly, \eqref{eq:Doppler-FS-234} becomes  
\begin{align}\label{eq:Doppler-FS-235}
f_D(\alpha,\theta_T)=&-\frac{2f|\bm{v}_{o}|\cos\theta_T}{c}
\end{align}

In the context of Scenario B, following the similar steps as above, a Doppler shift formula similar to \eqref{eq:Doppler-FS-233} is
\begin{align}\label{eq:Doppler-FS-236}
f_D(\theta_T,\theta_B)=&-\frac{f|\bm{v}_{o}|(\cos\theta_T-\cos\theta_B)}{c}
\end{align}
According to the geometry in Fig.~\ref{figure-Application-1}, the relationship of $\theta_B=180-\beta+\theta_T$ holds. Hence, 
\begin{align}\label{eq:Doppler-FS-237}
f_D(\beta,\theta_T)=&-\frac{f|\bm{v}_{o}|[\cos\theta_T+\cos(\beta-\theta_T)]}{c}
\end{align}

In practice, when the locations of transmitter, Sensor A and Sensor B are known, and when $\alpha$ and $\beta$ are already estimated, the velocity $\bm{v}_o$ of the object may be calculated after Sensors A and B estimated the respective Doppler shifts.  This can be achieved via solving the simultaneous equations of \eqref{eq:Doppler-FS-234} and \eqref{eq:Doppler-FS-237} for $|\bm{v}_{o}|$ and $\theta_T$.     

\subsection{Satellite Navigation}\label{section-Doppler-FS.6.2}

%
\begin{figure}[tb]
  \begin{center}
 \includegraphics[angle=0,width=.6\linewidth]{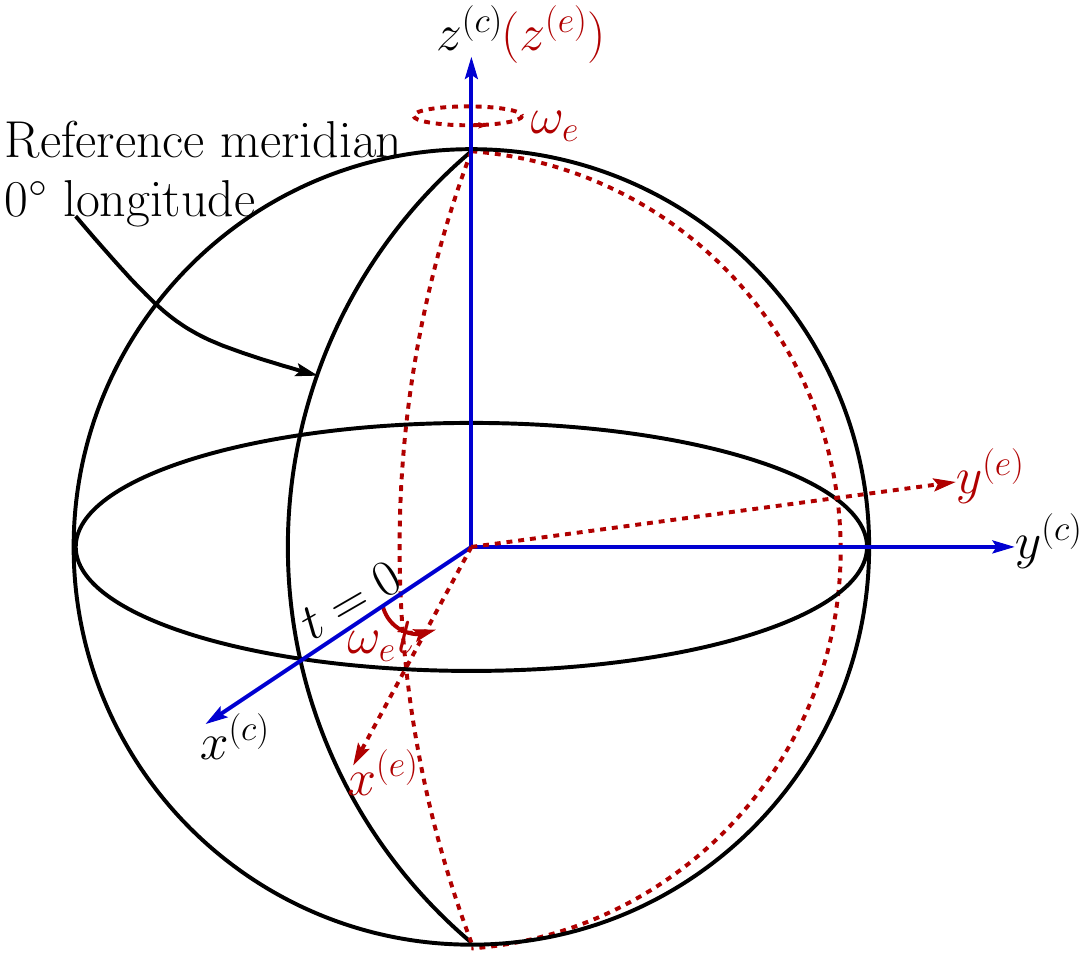}
  \end{center}
  \caption{Illustration of ECI and ECEF frames.}
  \label{figure-Earth-frame-transform}
\end{figure}

In satellite communication and navigation systems, the motion states and positions are all given with respect to the Earth-Centered Inertial (ECI) frame. Referring to Fig.~\ref{figure-Earth-frame-transform}, the coordinates in the ECI frame are defined as follows~\ncite{noureldin2013fundamentals,grewal2007global}:
\begin{itemize}

\item Origin: Located at the Earth's center of mass (geocenter).

\item $z^{(c)}$-axis: Aligned with the Earth's rotation axis, pointing
  towards the North Pole.

\item $x^{(c)}$-axis: Points towards the vernal equinox, defined as
  the intersection of the Earth's equatorial plane and the ecliptic
  plane (the plane of the Earth's orbit around the sun).

\item $y^{(c)}$-axis: Completes a right-handed orthogonal system, lying in the equatorial plane and directed $90^{\circ }$ east of the $x^{(c)}$-axis.

\end{itemize}    

By contrast, in satellite-based navigation or sensing, the motion
states and positions of objects, such as land vehicles, airplanes,
unmanned aerial vehicles (UAVs), etc., on or near Earth's surface, are
referred to the Earth-Centered, Earth-Fixed (ECEF) frame, i.e.,
Earth's surface or ground. Referring to
Fig.~\ref{figure-Earth-frame-transform}, the coordinates in the ECEF
frame are defined as
follows~\ncite{noureldin2013fundamentals,grewal2007global}:
\begin{itemize}

\item Origin: Coincident with the Earth's center of mass, which is the same as the origin in ECI frame.

\item $z^{(e)}$-axis: Points toward the North Pole, specifically aligned with the Conventional Terrestrial Pole (CTP).

\item  $x^{(e)}$-axis: Formed by the intersection of the equatorial plane and the reference meridian (Greenwich meridian).

\item $y^{(e)}$-axis: Completes a right-handed orthogonal system, lying in the equatorial plane and directed $90^{\circ }$ east of the $x^{(e)}$-axis. 

\end{itemize}

Hence, to use the signals sent by satellites to carry out sensing/navigation, the motion states and position vectors need to be transformed from the ECEF frame to the ECI frame. Express the angular velocity of Earth's rotation relative to the ECI frame as $\bm{\omega}_{ce}^{(e)}=[0,0,\omega_e]^T$, which is the angular velocity of the ECEF frame rotating with respect to the ECI frame. Then, referring to Fig.~\ref{figure-Earth-frame-transform}, it can be shown that the transformation matrix from the ECI frame  to the ECEF frame is~\ncite{noureldin2013fundamentals}
\begin{align}\label{eq:Doppler-FS-238}
\bm{R}_{ec}=\begin{bmatrix}
\cos(\omega_e t) & \sin(\omega_e t) & 0\\
-\sin(\omega_e t) & \cos(\omega_e t) & 0\\
0 & 0 & 1
\end{bmatrix}
\end{align}
where $t=0$ is defined as the instant when the $x^{(e)}$-axis of the
ECEF frame aligns with the $x^{(c)}$-axis of the ECI frame.
$\bm{R}_{ec}$ is an orthonormal matrix for carrying out the
transformation from ECI frame to ECEF frame.

Accordingly, the transformation matrix from the ECEF frame to the ECI frame is
\begin{align}\label{eq:Doppler-FS-239}
\bm{R}_{ce}=\bm{R}_{ec}^{-1}=\bm{R}_{ec}^T
\end{align}
$\bm{R}_{ce}$ carries out the transformation from ECEF frame to ECI
frame.

With the above preparation, we can now go ahead to analyze the Doppler
effect experienced by a signal sent by a satellite and received by a
moving sensing station. Assume that the velocity of the satellite is
$\bm{v}_s$ in the ECI frame, the velocity of the sensing station is
$\bm{v}_o^{(e)}$ in the ECEF frame, and the position vector of the
sensing station is $\bm{r}_o^{(e)}$ in the ECEF frame. Then, the
linear velocity of the sensing station in the ECI frame is given by the
formula
\begin{align}\label{eq:Doppler-FS-240}
\bm{v}_o^{(c)}=\frac{\bm{R}_{ce}\bm{v}_o^{(e)}+\bm{\omega}_{ce}^{(e)}\times \left(\bm{R}_{ce}\bm{r}_o^{(e)}\right)}{1+\beta_{o,1}\beta_{o,2}}
\end{align}
where `$\times$' is the cross-product operation between two
vectors\footnote{The cross-produce of two 3D vectors $\bm{a}$ and
$\bm{b}$ can be calculated as $\bm{a}\times
\bm{b}=\left|\begin{array}{c c c} i & j & k\\ a_1 & a_2 & a_3\\ b_1 &
b_2 & b_3
\end{array}\right|$, or $\bm{a}\times \bm{b}=|\bm{a}||\bm{b}|\sin(\theta)\bm{n}$, where $\theta$ is the angle between $\bm{a}$ and $\bm{b}$ in the plane containing them and $\bm{n}$ is a unit vector perpendicular to the plane containing $\bm{a}$ and $\bm{b}$, with directions of $\bm{a}$, $\bm{b}$ and $\bm{n}$ following the righthand rule. } ,  $\beta_{o,1}=|\bm{R}_{ce}\bm{v}_o^{(e)}|/c$ and
$\beta_{o,1}=|\bm{\omega}_{ce}^{(e)}\times
\left(\bm{R}_{ce}\bm{r}_o^{(e)}\right)|/c$ account for the effect of
special relativity, $\bm{v}_o^{(c)}$ is a function of time $t$ due to
$\bm{R}_{ce}$'s dependence on time, $\bm{v}_o^{(e)}$ may also be
time-dependent.

In most cases, both Earth's rotating speed and the moving speeds of
objects on or near Earth's surface - even aircraft for example - are
significantly smaller than the speed of light $c$. Then,
\eqref{eq:Doppler-FS-240} can be closely approximated
by~\ncite{noureldin2013fundamentals}
\begin{align}\label{eq:Doppler-FS-241}
\bm{v}_o^{(c)}=\bm{R}_{ce}\bm{v}_o^{(e)}+\bm{\omega}_{ce}^{(e)}\times
\left(\bm{R}_{ce}\bm{r}_o^{(e)}\right)
\end{align}
However, in some other cases, such as, when the Doppler shift between
two satellites or between a satellite and a flying rocket is
estimated, ignoring the effect of the factor
$1/(1+\beta_{o,1}\beta_{o,2})$ may result in noticeable errors. Hence,
the formula of \eqref{eq:Doppler-FS-240} needs to be used. For
simplicity, the analysis below is based on \eqref{eq:Doppler-FS-241}. 

When only velocity is considered, based on \eqref{eq:Doppler-FS-241},
in satellite-aided sensing/navigation of a sensing station, the
objective is to derive $\bm{v}_o^{(e)}$ in the ECEF frame via,
firstly, estimating the velocity $\bm{v}_o^{(c)}$ in the ECI frame, by
making use of its relationship with the Doppler shift measured at the
sensing station. Once $\bm{v}_o^{(c)}$ in the ECI frame is obtained,
\eqref{eq:Doppler-FS-241} can be inverted to obtain $\bm{v}_o^{(e)}$
as
\begin{align}\label{eq:Doppler-FS-242}
\bm{v}_o^{(e)}=\bm{R}_{ce}^T\left[\bm{v}_o^{(c)}-\bm{\omega}_{ce}^{(e)}\times
\left(\bm{R}_{ce}\bm{r}_o^{(e)}\right)\right]
\end{align}
when assuming that $\bm{r}_o^{(e)}$ is a priori.

Let press the linear velocity of a satellite as $\bm{v}_s^{(c)}$,
which is known when the satellite's orbit is known. Express a unit
line vector in parallel with the slant path from the sensing station
to the satellite as $\hat{\bm{r}}_{o\rightarrow s}$, which can be
assumed to be known when the satellite is fixed at a point on its
orbit and the position vector of sensing station is known.  Then, it
would be desirable that an ideal formula as
\begin{align}\label{eq:Doppler-FS-243}
  f_D=-\frac{f(\bm{v}_s^{(c)}  \cdot \hat{\bm{r}}_{o\rightarrow s} -\bm{v}_o^{(c)}\cdot \hat{\bm{r}}_{o\rightarrow s})}{c}
\end{align}
can be used. In practice, however, obtaining a measurement that
strictly adheres to \eqref{eq:Doppler-FS-243} is challenging due to
several physical and technical limitations. Below the primary factors
that interfere with the measured Doppler shift are analyzed.

\begin{figure}[tb]
  \begin{center}
 \includegraphics[angle=0,width=.99\linewidth]{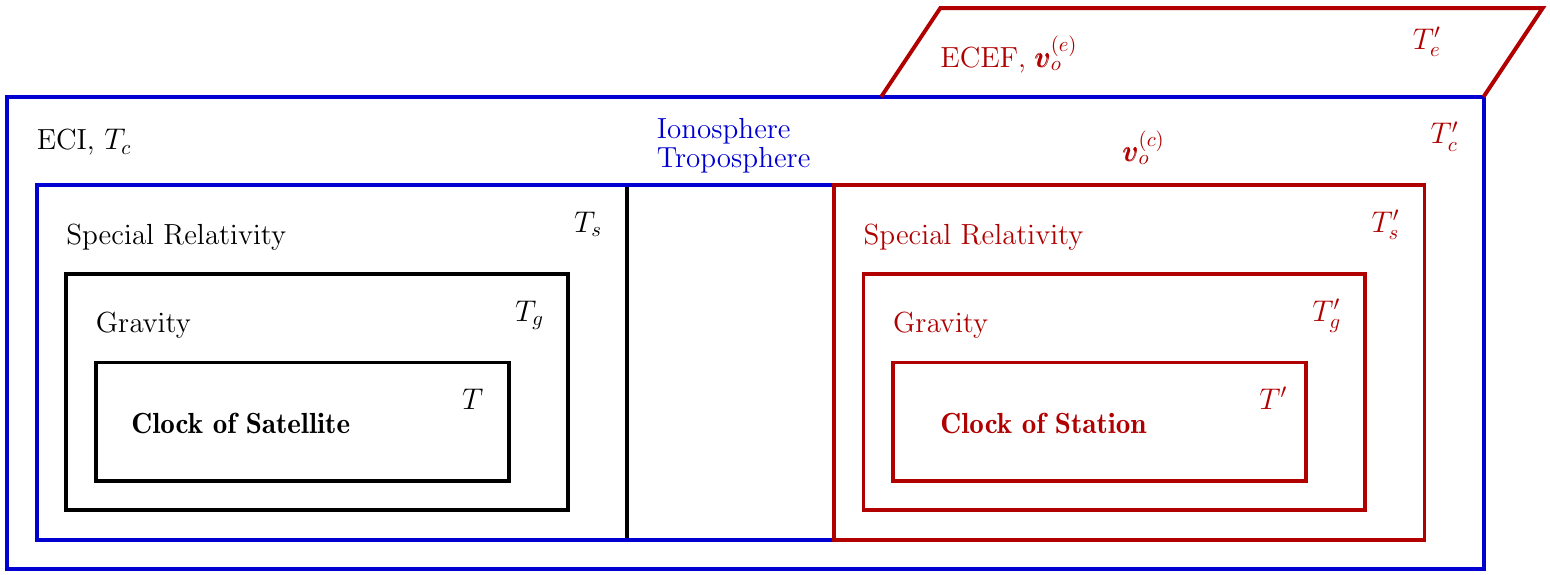}
  \end{center}
  \caption{Illustration of reference frames at satellite and sensing station, and physical factors affecting the measurement of Doppler effect at sensing station.}
  \label{figure-Frame-relation}
\end{figure}

Fig.~\ref{figure-Frame-relation} shows the reference frames involved
with the satellite and sensing station. As seen, the satellite and
sensing station only share the ECI frame. Let us analyze the physical
factors on the Dopper effect measured at the sensing station in the ECI
frame, one-by-one from the most inner square to the biggest square
representing the ECI frame.

First, assume that $T$ is the period of the signal sent by the satellite, which is
recorded by a clock carried by the satellite. Correspondingly, assume
that $T'$ is the period of the signal received by the sensing station,
which is recorded by a clock carried by the sensing station. Since both
satellite and sensing station lie in the Earth's gravity field, we can
write the concerned Doppler effect as
\begin{align}\label{eq:Doppler-FS-245}
  \frac{f'}{f}=&\frac{T}{T'}=\frac{T_g}{T'_g}
\end{align}
where ${T_g}$ and ${T'_g}$ are used to explicitly indicate that the
recorded periods are under the effect of Earth's gravity. 

Second, corresponding to $T_g$, assume that $T_s$ is the period
measured by a clock in the satellite's frame but free of
gravity. Similarly, corresponding to $T'_g$, assume that $T'_s$ is the
period measured by a clock in the sensing station's frame free of
gravity. Furthermore, for simplicity, assume that, relative to the
center of the Earth, $\bm{r}^{(s)}$ and $\bm{r}^{(e)}$ are the
position vectors of satellite and sensing station in all reference
frames\footnote{Otherwise, the Lorentz transformation can be applied
to transform $\bm{r}$ in one reference frame to $\bm{r}'$ in a new
reference as
$\bm{r}'=\bm{r}_{\perp}+\gamma(\bm{r}_{\parallel}-\bm{v}t)$, where
$\bm{r}_{\perp}$ and $\bm{r}_{\parallel}$ are the parallel and
perpendicular components to the velocity vector $\bm{v}$, and $\gamma$
is the Lorentz factor.}. Hence, $|\bm{r}^{(s)}|$ and
$|\bm{r}^{(e)}|$ are, respectively, the distances of satellite and
sensing station from the Earth's center. Then, according to
\eqref{eq:Doppler-FS-182a} in Section~\ref{subsection-Doppler-FS.4.3},
relative to the gravity potential $\Phi_0$ at the geoid, we have
\begin{align}\label{eq:Doppler-FS-246}
  \frac{f'}{f}=&\frac{T_g}{T'_g}=\frac{\sqrt{1+\frac{2(\Phi(|\bm{r}^{(s)}|)-\Phi_0)}{c^2}}}{\sqrt{1+\frac{2(\Phi(|\bm{r}^{(e)}|)-\Phi_0)}{c^2}}}\times\left(\frac{T_s}{T'_s}\right)
\end{align}

Third, corresponding to $T_s$, assume that $T_c$ is the period of
satellite's transmitted signal measured by a clock in the ECI
frame. Accordingly, corresponding to $T'_s$, assume that $T'_c$ is the
period of received signal measured by a clock also in the ECI
frame. Since in the ECI frame, the satellite moves at velocity
$\bm{v}_s^{(c)}$ and the sensing station moves at velocity
$\bm{v}_o^{(c)}$. Hence,  \eqref{eq:Doppler-FS-246} evolves to the ECI frame as
\begin{align}\label{eq:Doppler-FS-247}
  \frac{f'}{f}=&\frac{\sqrt{1+\frac{2(\Phi(|\bm{r}^{(s)}|)-\Phi_0)}{c^2}}}{\sqrt{1+\frac{2(\Phi(|\bm{r}^{(e)}|)-\Phi_0)}{c^2}}}\times\left(\frac{T_s}{T'_s}\right)\nonumber\\
  =&\frac{\sqrt{1+\frac{2(\Phi(|\bm{r}^{(s)}|)-\Phi_0)}{c^2}}}{\sqrt{1+\frac{2(\Phi(|\bm{r}^{(e)}|)-\Phi_0)}{c^2}}}\times\frac{\sqrt{1-|\bm{v}_s^{(c)}|^2/c^2}}{\sqrt{1-|\bm{v}_o^{(c)}|^2/c^2}}\times\left(\frac{T_c}{T'_c}\right)
\end{align}

Before going further to consider ${T_c}/{T'_c}$ in the ECI frame, let
us apply some approximation on the final equation in
\eqref{eq:Doppler-FS-247}. Assume that
${2(\Phi(|\bm{r}^{(e)}|)-\Phi_0)}/{c^2}<<1$ and
$|\bm{v}_o^{(c)}|^2/c^2<<1$. Then, \eqref{eq:Doppler-FS-247} can be
approximated as
\begin{align}\label{eq:Doppler-FS-248}
  \frac{f'}{f}\approx &{\sqrt{\left(1+\frac{2(\Phi(|\bm{r}^{(s)}|)-\Phi_0)}{c^2}\right)\left(1-\frac{2(\Phi(|\bm{r}^{(e)}|)-\Phi_0)}{c^2}\right)}}\nonumber\\
  &\times{\sqrt{\left(1-\frac{|\bm{v}_s^{(c)}|^2}{c^2}\right)\left(1+\frac{|\bm{v}_o^{(c)}|^2}{c^2}\right)}}\times\left(\frac{T_c}{T'_c}\right)\nonumber\\
  \approx  &\sqrt{1+\frac{2[\Phi(|\bm{r}^{(s)}|)-\Phi(|\bm{r}^{(e)}|)]}{c^2}}\times{\sqrt{1-\frac{|\bm{v}_s^{(c)}|^2}{c^2}+\frac{|\bm{v}_o^{(c)}|^2}{c^2}}}\times\left(\frac{T_c}{T'_c}\right)
\end{align}
where from the first to second equation, the $1/c^4$-related terms were
ignored. Now, expanding the two square-roots and keeping only the
linearly dependent terms, we  obtain
\begin{align}\label{eq:Doppler-FS-249}
  \frac{f'}{f}\approx&\left[1+\frac{\Phi(|\bm{r}^{(s)}|)-\Phi(|\bm{r}^{(e)}|)}{c^2}+\frac{|\bm{v}_o^{(c)}|^2-|\bm{v}_s^{(c)}|^2}{2c^2}\right]\times\left(\frac{T_c}{T'_c}\right)\nonumber\\
  =&\left[1+I_G+I_S\right]\times\left(\frac{T_c}{T'_c}\right)
\end{align}
where, by definition,
\begin{align}\label{eq:Doppler-FS-250}
  I_G=&\frac{\Phi(|\bm{r}^{(s)}|)-\Phi(|\bm{r}^{(e)}|)}{c^2}\nonumber\\
  I_S=&\frac{|\bm{v}_o^{(c)}|^2-|\bm{v}_s^{(c)}|^2}{2c^2}
\end{align}
through which the gravity and the special relativity related dynamics
generate interference on the measurement of the Dopper effect. As seen
in \eqref{eq:Doppler-FS-250}, once the orbit of the satellite and the
position of the sensing station are known, both $|\bm{r}^{(s)}|$ and
$|\bm{r}^{(e)}|$ are known and hence, the effect of $I_G$ can be
effectively removed. Similarly, once the orbit of the satellite is
known, $|\bm{v}_s^{(c)}|$ is known, whose interference on the Doppler
measurement can also removed. However, $|\bm{v}_o^{(c)}|$ is the
magnitude of the velocity to be measured in ECI frame, it may need to
stay for very accurate sensing/navigation. However, when the speed of
sensing station in the ECEF frame is small, $|\bm{v}_o^{(c)}|$ may be
approximated by the Earth's rotating speed.  In this case, both $I_G$
and $I_S$ are known, forming the systematic errors that can be removed
from the measurement of Doppler effect.

\begin{table}[t]
\caption{Basic data for GNSS, as well as MEO and LEO systems, where signal delay is one-way delay.}
\label{table-Satellite-data}
\begin{center}
\begin{tabular}{|l|l|l|l|}\hline
  \bf Type      & \bf Orbit Altitude & \bf Orbit Speed  & \bf Signal Delay  \\ \hline
  Global Navigation  & 20,000~km $\sim$ & 3.4~km/s $\sim$  & 60~ms $\sim$\\ 
  Satellite Systems (GNSS)   & 26,000~km  & 3.9~km/s & 90~ms \\ \hline
  Medium Earth   & 2,000~km $\sim$ & 3~km/s $\sim$  & 65~ms $\sim$\\ 
  Orbit (MEO)  System  & 35,786~km  & 7~km/s & 140~ms \\ \hline
   Low Earth & 160~km $\sim$ & 7.5~km/s $\sim$  & 1.5~ms $\sim$\\ 
   Orbit (LEO) System  & 2,000~km  & 7.8~km/s & 7~ms \\ \hline
  \end{tabular}
\end{center}
\end{table} 

Now let us forward to analyze the classic Doppler effect of
${T_c}/{T'_c}$ in the ECI frame. While focusing on this, let us leave
\eqref{eq:Doppler-FS-249} for the
moment. Table~\ref{table-Satellite-data} lists some basic data for the
GNSS as well as MEO and LEO systems.

\begin{figure}[tb]
  \begin{center}
 \includegraphics[angle=0,width=.6\linewidth]{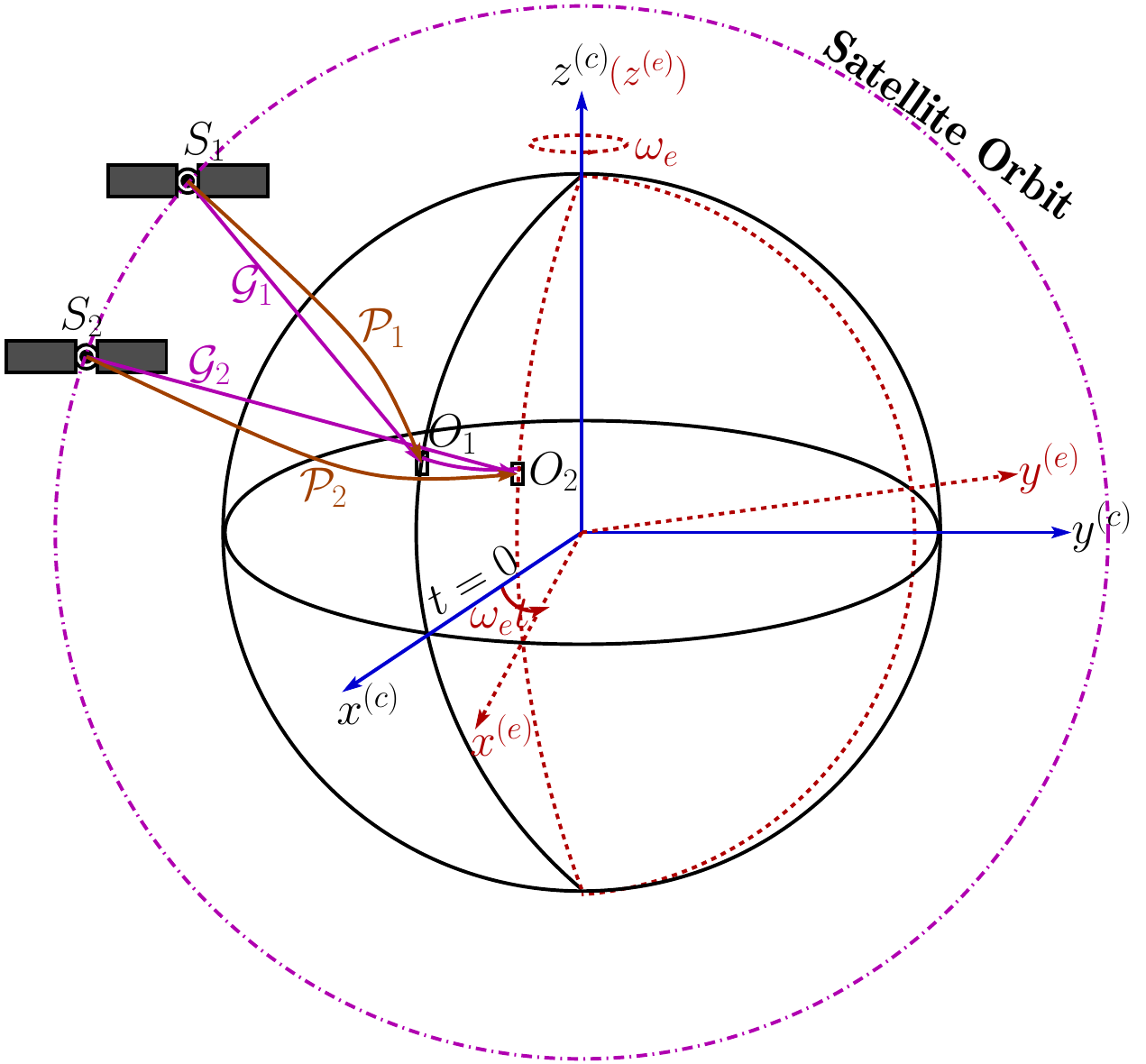}
  \end{center}
  \caption{Illustration shows the effect of rotating Earth and
    circularly moving satellite, where $\mathcal{G}_1$ and
    $\mathcal{G}_2$ are slant paths and $\mathcal{P}_1$ and
    $\mathcal{P}_2$ are bended optical (physical propagation)
    path. The drawings are not to scale, but for conceptual purposes
    only. }
  \label{figure-Moving-Earth-effect}
\end{figure}

Since both satellite rotates and sensing station moves circularly in
the ECI frame, it seems that this is the case of general moving source
and observer analyzed in
Section~\ref{subsection-Doppler-FS.2.1.3}. However, even such a
general model may not be accurate enough for analyzing the Doppler
effect experienced by a signal transmitted between satellite and
sensing station. In the model used in
Section~\ref{subsection-Doppler-FS.2.1.3}, it was assumed that there
is a fixed point that approximately lies on the lines between source
and observer, when they transmit/receive the first and second crests
of a wave.  However, in satellite systems, as seen in
Table~\ref{table-Satellite-data}, satellite moves fast, Earth is
rotating, and the signal delay from transmitter to receiver is
significant. Consequently, as shown in
Fig.~\ref{figure-Moving-Earth-effect}, even without considering the
effect of the dynamics of atmosphere, the geometric signal propagation
path $\mathcal{G}_1$, i.e., slant path, of the first wave crest may be
significantly different from the path $\mathcal{G}_2$ conveying the
second wave crest from satellite to sensing station, generating the
so-called Sagnac effect~\ncite{Ashby2004}. In satellite-based
sensing/navigation, the Sagnac effect should be taken into account to
improve accuracy.

In the ECI frame, the effect of atmosphere also needs to be addressed,
as physically, radio signals in atmosphere do not propagate along
slant paths, but along the bended optical paths, as shown by
$\mathcal{P}_1$ and $\mathcal{P}_2$ in
Fig.~\ref{figure-Moving-Earth-effect}.  We have analyzed this issue in
Section~\ref{section-Doppler-FS.5}. However, the analysis there is on
$d(Path)/dt$, considering one-path only. Therefore, to obtain the
analytical results of high accuracy, in satellite signal based
sensing/navigation, the two paths conveying two adjacent wave crests
need to be separately analyzed, to jointly address the Sagnac effect
and atmospheric effect. The analysis below abides this requirement.

Assume that satellite sends two adjacent wave crests at $t_1=t_0$ and
$t_2=t_0+T_c$. Correspondingly, the position vectors are
$\bm{r}^{(s)}(t_1)$ and $\bm{r}^{(s)}(t_2)$, and velocities are
$\bm{v}_s^{(c)}(t_1)=\bm{\omega}_s\times \bm{r}^{(s)}(t_1)$ and
$\bm{v}_s^{(c)}(t_2)=\bm{\omega}_s\times \bm{r}^{(s)}(t_2)$, where
$\bm{\omega}_s$ is satellite's orbital angular velocity. These two
crests follow two {\em physical paths}, $\mathcal{P}_1$ and
$\mathcal{P}_2$, to sensing station. Correspondingly, the slant paths
(or LoS paths) between sensing station and satellite are
$\mathcal{G}_1$ and $\mathcal{G}_2$. For sensing/navigation, $t_1$ and
$t_2$, and $\bm{r}^{(s)}(t_1)$ and $\bm{r}^{(s)}(t_2)$ (hence,
$\bm{v}_s^{(c)}(t_1)$ and $\bm{v}_s^{(c)}(t_2)$) are assumed a-priori
information.

Corresponding to the satellite, at sensing station, assume that the
time of receiving the first crest is $t'_1=t'_0$, and the time of
receiving the second crest is $t'_2=t'_0+T'_c$. Accordingly, the
position vectors are $\bm{r}_o^{(e)}(t'_1)$ and
$\bm{r}_o^{(e)}(t'_2)$. The corresponding velocities of sensing station
are $\bm{v}_o^{(c)}(t'_1)$ and $\bm{v}_o^{(c)}(t_2)$, whose relations
with the Earth's rotational velocity $\bm{\omega}_{ce}^{(e)}$ and the
sensing station's velocity $\bm{v}_o^{(e)}$ are given in
\eqref{eq:Doppler-FS-241}.

Based on the above settings, the following relationships can be built:
\begin{subequations}\label{eq:Doppler-FS-252}
\begin{align}\label{eq:Doppler-FS-252a}
  t'_1=&t_0+\frac{1}{c}\int_{\mathcal{P}_1}n(p,t)dp\\
\label{eq:Doppler-FS-252b}
t'_2=&t_0+T_c+\frac{1}{c}\int_{\mathcal{P}_2}n(p,t)dp
\end{align}
\end{subequations}
where $n(p,t)$ is the refraction index of atmosphere along an
integration path, and $t$ is explicitly used to indicate the
time-varying nature of refraction index in atmosphere. In general,
$n(p,t)=n(p,t_0)$ can be assumed over a short interval of such as
$T_c$. Hence, the period of received signal in the ECI frame is
\begin{align}\label{eq:Doppler-FS-253}
  T'_c=t'_2-t'_1=&T_c+\frac{1}{c}\int_{\mathcal{P}_2}n(p,t)dp-\frac{1}{c}\int_{\mathcal{P}_1}n(p,t)dp\nonumber\\
  =&T_c\left(1+\frac{1}{cT_c}\left[\int_{\mathcal{P}_2}n(p,t)dp-\int_{\mathcal{P}_1}n(p,t)dp\right]\right)
\end{align}
giving
\begin{align}\label{eq:Doppler-FS-254}
  \frac{T_c}{T'_c}=&\left(1+\frac{1}{cT_c}\left[\int_{\mathcal{P}_2}n(p,t)dp-\int_{\mathcal{P}_1}n(p,t)dp\right]\right)^{-1}
\end{align}

Let us introduce the slant paths $\mathcal{G}_1$ and $\mathcal{G}_2$
to the above formula. Then, \eqref{eq:Doppler-FS-254} can be modified
to the form of
\begin{align}\label{eq:Doppler-FS-255}
  \frac{T_c}{T'_c}=&\left(1+\frac{1}{cT_c}\left[\int_{\mathcal{G}_2}dp-\int_{\mathcal{G}_1}dp\right]+\frac{1}{cT_c}\left[\int_{\mathcal{P}_2}n(p,t)dp-\int_{\mathcal{G}_2}dp\right]\right.\nonumber\\
  &\left. -\frac{1}{cT_c}\left[\int_{\mathcal{P}_1}n(p,t)dp-\int_{\mathcal{G}_1}dp\right]\right)^{-1}\nonumber\\
  =&\left(1+\frac{d_2-d_1}{cT_c}\right.\nonumber\\
  &\left.+\frac{1}{cT_c}\left[\left(\int_{\mathcal{P}_2}n(p,t)dp-\int_{\mathcal{G}_2}dp\right) -\left(\int_{\mathcal{P}_1}n(p,t)dp-\int_{\mathcal{G}_1}dp\right)\right]\right)^{-1}\nonumber\\
  =&\left(1+\frac{d_2-d_1}{cT_c}-I_{Atmo}\right)^{-1}\nonumber\\
  \approx &1-\frac{d_2-d_1}{cT_c}+I_{Atmo}
\end{align}
where $d_1$ and $d_2$ are the distances of the first and second slant
paths. Hence, ${(d_2-d_1)}/{T_c}$ is the average relative speed
between satellite and sensing station during $t_0$ (satellite starts
transmission) and $t'_2$ (sensing station receives the second crest). In \eqref{eq:Doppler-FS-255},
\begin{align}\label{eq:Doppler-FS-256}
 I_{Atmo}=-\frac{1}{cT_c}\left[\left(\int_{\mathcal{P}_2}n(p,t)dp-\int_{\mathcal{G}_2}dp\right) -\left(\int_{\mathcal{P}_1}n(p,t)dp-\int_{\mathcal{G}_1}dp\right)\right]
\end{align}
is the interference generated by the atmosphere, including both the
ionosphere and troposphere, as analyzed in
Sections~\ref{section-Doppler-FS.5.1} and
\ref{section-Doppler-FS.5.1}, respectively. The last approximation in
\eqref{eq:Doppler-FS-255} is obtained by neglecting the
$1/c^2$-related terms. 

Now, we substitute \eqref{eq:Doppler-FS-255} back into
\eqref{eq:Doppler-FS-249}, obtaining
\begin{align}\label{eq:Doppler-FS-257}
  \frac{f'}{f}=&\left(1+I_G+I_S\right)\times\left(1-\frac{d_2-d_1}{cT_c}+I_{Atmo}\right)\nonumber\\
  \approx &1-\frac{d_2-d_1}{cT_c}+I_G+I_S+I_{Atmo}
\end{align}
where the approximation is due to the ignorance of the second-order
terms. Finally, from \eqref{eq:Doppler-FS-257} the Doppler shift
measured by the sensing station in the ECI frame can be obtained,
which is
\begin{align}\label{eq:Doppler-FS-258}
  f_D=&-\frac{f(d_2-d_1)}{cT_c}+I'_G+I'_S+I'_{Atmo}
\end{align}
where $I'_G=fI_G$, $I'_S=fI_S$ and $I'_{Atmo}=fI_{Atmo}$.

\begin{remark}
So far, all the approximations imposed in the Doppler effect analysis
are very accurate for the satellite sensing/navigation applications
considered. This can be convinced, as the approximation in
\eqref{eq:Doppler-FS-249} is ignoring the $1/c^4$-related terms, that
in \eqref{eq:Doppler-FS-255} is ignoring the $1/c^2$-related terms and
that in \eqref{eq:Doppler-FS-257} is ignoring the second-order terms,
which are proportional to $1/c^2$ at least.
  \end{remark}

Associated with \eqref{eq:Doppler-FS-249}, we have analyzed that the
interference of $I_G$ (and hence $I'_G$) can be near-ideally removed,
the interference of $I_S$ (and hence $I'_S$) can be near-fully removed
for most applications where the receiver's local velocity
$\bm{v}_o^{(e)}$ is well-characterized or negligible compared to the
tangential rotational velocity of the Earth. The interference by
atmosphere is more complicated, which includes both the predictable
component, generating systematic error, and the random component,
generating spread Doppler spectrum and random error. Hence, to attain
more accurate sensing/navigation, the predictable component should be
mitigated to the minimum level. Below we analyze the interference by
atmosphere.

Let express
\begin{align}\label{eq:Doppler-FS-259}
  I'_{Atmo}=I'_I+I'_T
\end{align}
where $I'_T$ and $I'_I$ are respectively the contributions from the
troposphere and ionosphere, which, based on \eqref{eq:Doppler-FS-256},
are expressed as
\begin{subequations}\label{eq:Doppler-FS-260}
\begin{align} \label{eq:Doppler-FS-260a}
 I'_{T}=&-\frac{f}{cT_c}\left[\left(\int_{\mathcal{P}_{2,T}}n(p,t)dp-\int_{\mathcal{G}_{2,T}}dp\right) -\left(\int_{\mathcal{P}_{1,T}}n(p,t)dp-\int_{\mathcal{G}_{1,T}}dp\right)\right]\\
 \label{eq:Doppler-FS-260b}
 I'_{I}=&-\frac{f}{cT_c}\left[\left(\int_{\mathcal{P}_{2,I}}n(p,t)dp-\int_{\mathcal{G}_{2,I}}dp\right) -\left(\int_{\mathcal{P}_{1,I}}n(p,t)dp-\int_{\mathcal{G}_{1,I}}dp\right)\right]
\end{align}
\end{subequations}
where $\mathcal{P}_{2,T}$, $\mathcal{P}_{1,T}$, $\mathcal{G}_{2,T}$,
$\mathcal{G}_{1,T}$ are the portions of paths in the troposphere,
while $\mathcal{P}_{2,I}$, $\mathcal{P}_{1,I}$, $\mathcal{G}_{2,I}$,
$\mathcal{G}_{1,I}$ are the corresponding potions of paths in the
ionosphere.

In \eqref{eq:Doppler-FS-260a}, $I'_{T}$ is the Doppler effect by the
medium in the troposphere. When assuming that the troposphere is
sufficiently stable during the passing interval of satellite,
following the analysis in Section~\ref{section-Doppler-FS.5.2},
$I'_{T}$ can be approximated as
\begin{align} \label{eq:Doppler-FS-261}
 \bar{I}'_{T}\approx &-\frac{f}{c}\frac{d}{dt}\int_{\mathcal{G}_{T}}(n(p,t)-1)dp
\end{align}
which, accordingly, has the expression of
\eqref{eq:Doppler-FS-214}. Then, with the modeling of $N(r)$ and
information about the satellite, $\bar{I}'_{T}$ in
\eqref{eq:Doppler-FS-261} can be evaluated, as detailed in
Section~\ref{section-Doppler-FS.5.2}. Consequently, the interference
$I'_{T}$ can be cancelled, leaving the randomly distributed residue expressed as
\begin{align} \label{eq:Doppler-FS-262}
  \Delta I'_T=I'_T-\bar{I}'_{T}
\end{align}

In the context of the interference $I'_{I}$ in
\eqref{eq:Doppler-FS-260b} by the ionosphere, using the fact that
signal from satellite to sensing station is not much bended by the
ionosphere, $\mathcal{P}_{2,I}=\mathcal{G}_{2,I}$ and
$\mathcal{P}_{1,I}=\mathcal{G}_{1,I}$ can be assumed. Hence,
\eqref{eq:Doppler-FS-260b} can be re-written as
\begin{align}\label{eq:Doppler-FS-263}
 \bar{I}'_{I}=&-\frac{f}{cT_c}\left[\int_{\mathcal{G}_{2,I}}(n(p,t)-1)dp -\int_{\mathcal{G}_{1,I}}(n(p,t)-1)dp\right]
\end{align}
As analyzed in Section~\ref{section-Doppler-FS.5.1}, the refractive
index $n(p,t)$ is primarily determined by the local electron density and
distribution, and is frequency dependent. Specifically, in
satellite-based systems, the refractive index can be closely
approximated as that in \eqref{eq:Doppler-FS-196}. Let in
\eqref{eq:Doppler-FS-196} $X=f_N^2/f^2$, where $f_N$ is the plasma
frequency.  Then, with the aid of the series approximation, $n(p,t)$
can be expressed as
\begin{align}\label{eq:Doppler-FS-264}
  n(p,t)  \approx\sqrt{1-\frac{f_N^2}{f^2}}=1-\frac{f_N^2}{2f^2}+\mathcal{O}\left(\frac{1}{f^4}\right)
\end{align}
Applying it into \eqref{eq:Doppler-FS-263} gives
\begin{align}\label{eq:Doppler-FS-265}
 \bar{I}'_{I}=&\frac{1}{c}\left[\frac{\alpha_2}{f}-\frac{\alpha_1}{f}+\mathcal{O}\left(\frac{1}{f^3}\right)\right]
\end{align}
where $\alpha_i=0.5T_c^{-1}f_N^2\int_{\mathcal{G}_{i,I}}dp$ for $i=1$ and $2$.

Based on \eqref{eq:Doppler-FS-265}, a dual-frequency
method~\ncite{4066048} can be introduced to mitigate the first-order
effect from the ionosphere. In detail, using
\eqref{eq:Doppler-FS-265},  \eqref{eq:Doppler-FS-258} can be re-written
as
\begin{align}\label{eq:Doppler-FS-266}
  f_D=&f'_D+\bar{I}'_{I}\nonumber\\
  =&f'_D+\frac{1}{c}\left[\frac{\alpha_2}{f}-\frac{\alpha_1}{f}+\mathcal{O}\left(\frac{1}{f^3}\right)\right]
\end{align}
where $f'_D$ contains the Doppler shift by all other factors, except
that by the ionosphere. Then, assume that two signals with frequencies
of $f_1$ and $f_2$ are simultaneously transmitted by
satellite. Correspondingly, the sensing station measures the Doppler
shifts of
\begin{align}\label{eq:Doppler-FS-267}
  f_{D,f_1}=&f'_{D,f_1}+\frac{1}{c}\left[\frac{\alpha_2}{f_1}-\frac{\alpha_1}{f_1}+\mathcal{O}\left(\frac{1}{f_1^3}\right)\right]\nonumber\\
  f_{D,f_2}=&f'_{D,f_2}+\frac{1}{c}\left[\frac{\alpha_2}{f_2}-\frac{\alpha_1}{f_2}+\mathcal{O}\left(\frac{1}{f_2^3}\right)\right]
\end{align}
where $f_{D,f_2}$ and $f'_{D,f_2}$ mean that they are
frequency-dependent.  Then, the sensing station calculates
\begin{align}\label{eq:Doppler-FS-268}
  f_2f_{D,f_2}-f_1f_{D,f_1}=(f_2f'_{D,f_2}-f_1f'_{D,f_1})+\mathcal{O}\left(\frac{1}{f^2_{1,2}}\right)
\end{align}
showing that the first-order interference by the ionosphere is ideally
cancelled, and the motion information of the sensing station is
embedded in $(f_2f'_{D,f_2}-f_1f'_{D,f_1})$. Hence, if the other
interferences have been cancelled before the operation in
\eqref{eq:Doppler-FS-268}, the velocity of the sensing station in ECI
frame can be directly derived from $f_2f_{D,f_2}-f_1f_{D,f_1}$ by
treating the remaining $\mathcal{O}\left({1}/{f^2_{1,2}}\right)$ and
the other possible residues as random interference.

To this point, let us assume that the systematic errors contributed by
gravity, special relativity, troposphere and ionosphere are all
corrected, leaving only the random errors. Then,
\eqref{eq:Doppler-FS-258} can be modified to
\begin{align}\label{eq:Doppler-FS-269}
  f_D=&-\frac{f(d_2-d_1)}{cT_c}+\varepsilon
\end{align}
where $\varepsilon$ accounts for the above-mentioned random errors,
and $f_D$ is kept using for simplicity. By making use of
\eqref{eq:Doppler-FS-243} but replacing the involved velocities by
their averaging ones, it becomes 
\begin{align}\label{eq:Doppler-FS-270}
  \bar{f}_D=-\frac{f(\bar{\bm{v}}_s^{(c)} \cdot \hat{\bm{r}}_{o\rightarrow
      s} -\bar{\bm{v}}_o^{(c)}\cdot \hat{\bm{r}}_{o\rightarrow s})}{c}
\end{align}
Based on this formula, once $\bar{f}_D$ is known, and since
$\bar{\bm{v}}_s^{(c)}$ and $\hat{\bm{r}}_{o\rightarrow s}$ can be
assumed the a-priori, there is only one variable
$\bar{\bm{v}}_o^{(c)}$ of the average velocity of the sensing station
in the ECI frame. Using \eqref{eq:Doppler-FS-270} to modify
\eqref{eq:Doppler-FS-269}, yielding
\begin{align}\label{eq:Doppler-FS-271}
  f_D=&\bar{f}_D-\left(\bar{f}_D+\frac{f(d_2-d_1)}{cT_c}\right)+\varepsilon\nonumber\\
  =&\bar{f}_D+I_{Sag}+\varepsilon
\end{align}
where $I_{Sag}=-\left(\bar{f}_D+{f(d_2-d_1)}/{cT_c}\right)$ accounts for the change in the geometry during the
signal's transmission time. In the ECI frame, this is the
manifestation of the Sagnac effect~\ncite{Ashby2004}, representing the
fact that the receiver moves while the signal is in flight. This can
be precisely estimated using the known Earth rotation $\bm{\omega
}_{ce}^{(e)}$ and the satellite ephemeris ~\ncite{HuFarrell2024}.

Finally, it is worth noting that while the primary factors
contributing to Doppler shift errors along the slant path have been
addressed, further improvements in sensing/navigation accuracy may
require the consideration of higher-order effects and subtle
relativistic corrections, such as those detailed in~\ncite{Ashby2003}.


\section{Doppler Effect of Acoustic Waves}\label{section-Doppler-FS.7}

In the previous sections, although EM-waves were mainly assumed, the
analyses having considered the effect of medium, including
Sections~\ref{subsubsection-Doppler-FS.1.3} and
\ref{subsection-Doppler-FS.2.2}, can be applied for the acoustic wave
scenarios, directly or after slight modification by, such as,
neglecting the relativistic effect. Hence, this section emphasizes the
difference of the Doppler effect with acoustic waves from that of the
Doppler effect with EM-waves.

To carry out the analysis, the following definitions are used:
\begin{itemize}

\item $\hat{\bm{r}}_{s\rightarrow o}$: Unit vector in parallel with the line connecting source and observer and pointing in the direction from source to observer.

\item $\bm{v}_s$: Velocity of moving source. Hence, the moving velocity of source in the direction of $\hat{\bm{r}}_{s\rightarrow o}$ is $v_s=\bm{v}_s\cdot \hat{\bm{r}}_{s\rightarrow o}$.

\item $\bm{v}'_m$:  Velocity of moving medium. Hence, the moving velocity of medium in the direction of $\hat{\bm{r}}_{s\rightarrow o}$ is $v'_m=\bm{v}'_m\cdot \hat{\bm{r}}_{s\rightarrow o}$.

\item $\bm{v}_o$: Velocity of moving observer. Hence, the moving velocity of observer in the direction of $\hat{\bm{r}}_{s\rightarrow o}$ is $v_o=\bm{v}_o\cdot \hat{\bm{r}}_{s\rightarrow o}$.

\item $v$: Speed of acoustic wave in stationary medium. Hence, the wave propagation speed in the direction of $\hat{\bm{r}}_{s\rightarrow o}$ is $v_w=v+v'_m$.

\item $f,~T,~\lambda$: Frequency, period and wavelength of source acoustic wave.

\item $f',~T',~\lambda'$: Frequency, period and wavelength of observed acoustic wave.

\end{itemize}
Based on the above definitions, the analysis below can simply assume motions along the line connecting source and observer. In the other cases, corresponding formulas can be obtained by replacing $v_s$, $v_o$, $v_w$ accordingly using the above definitions.

Assume that source emits two adjacent crests at time $t_1=t_0$ and $t_2=t_0+T$. Accordingly, the time instants of receiving these two crests are $t'_1=t'_0$ and $t'_2=t'_0+T'$. Based on the above definitions, the relation of
\begin{align}\label{eq:Doppler-FS7-274}
T'=T+\frac{-v_sT}{v_w}+\frac{v_oT'}{v_w}
\end{align}
holds, which gives
\begin{align}\label{eq:Doppler-FS7-275}
f'=f\left(\frac{v_w-v_o}{v_w-v_s}\right)
\end{align}

If $v_o=0$ and source is moving with speed $v_s$, there is
\begin{align}\label{eq:Doppler-FS7-276}
f_w=\frac{f}{1-v_s/v_w}
\end{align}
which is the frequency of the acoustic wave in medium. Correspondingly, using the formula $v_w=f_w\lambda_w$, the wavelength of the acoustic wave in medium is
\begin{align}\label{eq:Doppler-FS7-277}
\lambda_w=\frac{v_w-v_s}{f}
\end{align}

The following analysis focuses on specific situations that
differentiate acoustic waves from EM-waves. In relativistic scenarios,
no object with mass can travel faster than $c$ of the speed of light
in a vacuum. In stark contrast, acoustic waves propagate with a much
lower, medium-dependent speed. Consequently, the wave source,
observer, and even the medium itself may move at a speed higher than
the acoustic wave's propagation speed, leading to phenomena like shock
waves (e.g., sonic booms), as analyzed below.

First, when a medium moves at a supersonic velocity relative to its
stationary propagation speed, i.e., $|v'_m|\geq v$, acoustic signals
become unable to propagate against the flow. In such cases, all sound
energy is swept downstream, as the medium's speed exceeds the wave's
ability to travel upstream. However, it is worth noting that $v_w$
will never be negative. Mathematically, the actual propagation speed
$v_w$ in/against the direction of the flow is more accurately
represented as $v_w=\max\{0,v+v'_m\}$.

Second, when a stationary source emits signals and an observer moves
away at a speed $v_o\geq v_w$, the observer will never receive the
acoustic signals according to
\eqref{eq:Doppler-FS7-275}. Mathematically, $v_o> v_w$ results in an
observed frequency $f'<0$, which is physically meaningless for a
traveling wave. Thus, \eqref{eq:Doppler-FS7-275} is subject to the
physical constraint of $v_o\leq v_w$.

\begin{figure}[tb]
  \begin{center}
 \includegraphics[angle=0,width=.45\linewidth]{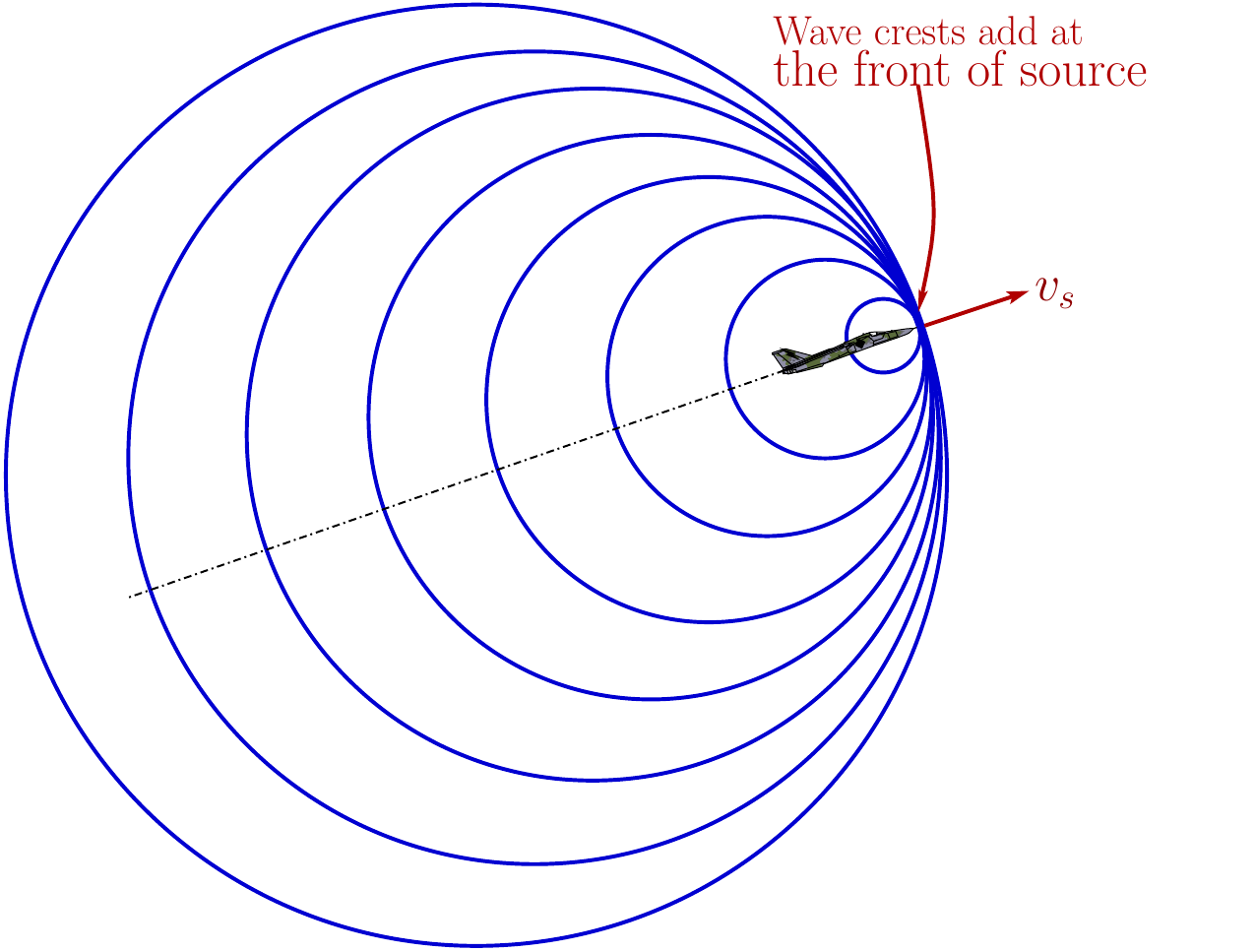}
  \end{center}
  \caption{When the velocity of source is close to the propagation speed of acoustic signals, wave crests add in the front of source, forming sound barrier. }
  \label{figure-Acoustic-vseqvw}
\end{figure}

Third, according to \eqref{eq:Doppler-FS7-275}, as $v_s\rightarrow
v_w$, $f'\rightarrow \infty$ and hence $\lambda'\rightarrow 0$ due to
$f'\lambda'=(v_m-v_s)$. This physically means that all emitted waves
by source are constructively added at the head of the source, as shown
in Fig.~\ref{figure-Acoustic-vseqvw}. The ``bunching up'' of wave
crests creates the extremely compressed medium in the front of the
source. By Newton’s third law, the medium will generate an equal force
(referred to as drag force) on the source.  This phenomenon is known
as the ``sound barrier''.  In the case of $v_s\rightarrow v_w$, an
observer in front of the source hears nothing until the source
arrives, at which point the observer experiences a near-instantaneous
pressure discontinuity (like a thump), known as a ``shock wave''.

\begin{figure}[tb]
  \begin{center}
 \includegraphics[angle=0,width=.5\linewidth]{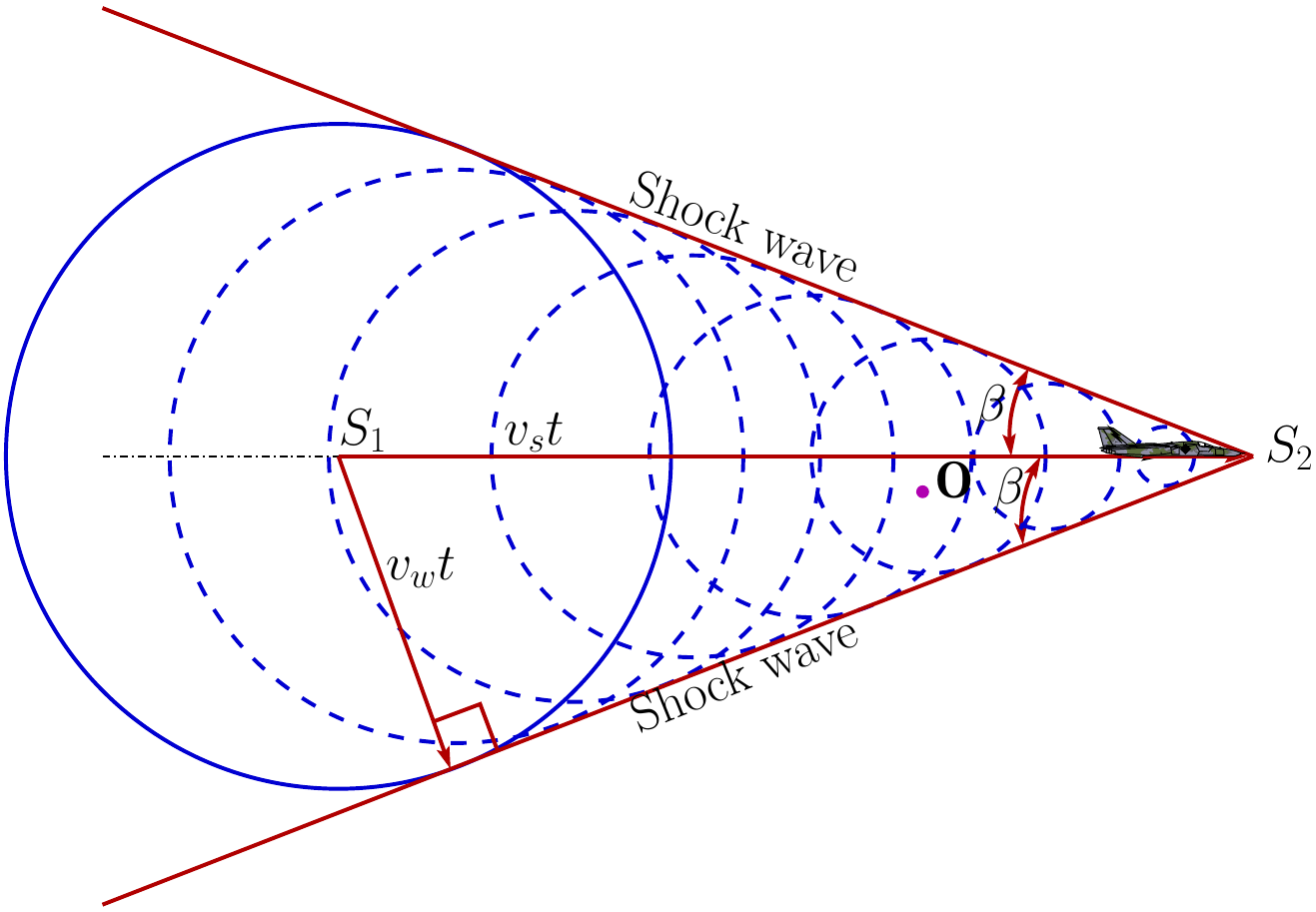}
  \end{center}
  \caption{When the velocity of source is larger than the propagation speed of acoustic signals,  source's sound generates a shock wave cone. }
  \label{figure-Acoustic-vsgeqvw}
\end{figure}

Fourht, in the supersonic regime of $v_{s}>v_{w}$, the formula of
\eqref{eq:Doppler-FS7-275} no longer describes the Doppler effect in
front of the source, as the denominator of \eqref{eq:Doppler-FS7-275}
becomes negative. In this regime, the source ``outruns'' its own
emitted waves, leading to the formation of a so-called Mach-cone, as
shown in Fig.~\ref{figure-Acoustic-vsgeqvw}. On the surface of the
Mach-cone, wave crests are constructively added, forming shock waves.
As this cone passes a stationary observer, it experiences a {\em sonic
  boom}. After the sonic boom and when the observer enters inside the
Mach-cone, it measures a sudden frequency dropping and finally a
minimum frequency of $f'=fv_w/{(v_w+|v_s|)}$, when the observer
reaches the rotational symmetric axis of the Mach-cone.

According to the geometry illustrated in
Fig.~\ref{figure-Acoustic-vsgeqvw},
\begin{align}\label{eq:Doppler-FS7-278}
\sin\beta=\frac{v_w}{v_s}
\end{align}
Hence, $(\sin\beta)^{-1}$ is equal to the {\em Mach number} that is defined
as $M=v_s/v_w$.

Finally, in the supersonic regime of $v_{s}>v_{w}$, as seen in
Fig.~\ref{figure-Acoustic-vsgeqvw}, for the signals that the source
emits before its passing of the observer, the observer receives these
signals reversely, i.e., the signals emitted later are received
earlier than the signals emitted earlier. Analogously, if the source
counts `1, 2, 3, ..., 10' before passing the observer, the observer,
once inside the Mach cone, will hear '10, 9, 8,..., 1' with each
individual number also played in reverse. For the signals that the
source emits after its passing of the observer, these signals are
received in the normal order with a time-varying frequency $f'$, until
it converges to the frequency $f'=fv_w/{(v_w+|v_s|)}$, when the
observer arrives at the rotational symmetric axis of the
Mach-cone. Furthermore, the signals emitted by the source before and
after its passing of the observer interfere with each other at the
observer.

\section{Concluding Remarks}\label{section-Doppler-FS.8}

The analyses of the Doppler effect in previous sections assumed a
single, pure sinusoidal frequency.  However, real-world wireless
systems are operated under specific bandwidth constraints. In
particular, signals in narrowband, wideband, and ultra-wideband
wireless systems are distributed over a frequency range. Because the
Doppler shift is proportional to the operating frequency, different
frequency components of a transmitted signal experience distinct
frequency shifts. This results in a frequency-varying Doppler effect
where the received signal exhibits a Doppler spectrum rather than a
single frequency offset, introducing complex distortions, such as,
inter-carrier interference in multi-carrier signaling systems.

While Section~\ref{section-Doppler-FS.5} analyzed the Doppler effect
in the atmosphere, it focused on time-varying propagation paths rather
than the direct Doppler shifts from random atmospheric events,
including the fluctuation of air temperature, pressure and humidity in
the troposphere, and the variation of electron density in the
ionosphere due to the various environment changes.  Instead, only the
Doppler effect by the time-varying propagation paths was addressed. In
addition to the random events in the atmosphere, in practical wireless
systems, there are many other physical phenomena, including
refraction, reflection, diffraction, scattering by various types of
environmental objects, which result in multiple propagation paths from
transmitter to receiver. In these systems, different propagation paths
may be associated with different Doppler shifts, resulting in Doppler
spread. Therefore, even when a single-tone signal is transmitted, the
received signal may contain many frequency components around the
emitted frequency, forming a Doppler spectrum.

In wireless sensing and communications systems, Doppler spread
presents significant design challenges but can also be exploited to
improve performance. In wireless communications, Doppler spread
results in time-selective fading channels, which can be exploited via
appropriate design of signalling and channel
coding~\ncite{Lie-Liang-MC-CDMA-book,Lie-Liang-Optimization-book,book-Simon-Alouini-2nd-Ed,Shu_Lin-Ryan},
to attain the time-domain diversity gain and, hence, improve systems'
performance. On the other hand, in wireless sensing, if the knowledge
about the Doppler effect, such as its first-, second-order statistics,
is available, this knowledge can be exploited to improve the accuracy
of sensing. This constitutes an important research issue, in
particular, in the regime of integrated sensing and communications
(ISAC), whereas most existing research works assume the Doppler effect
without spread~\ncite{10726912,9705498,9585321,9924202}.


\addcontentsline{toc}{chapter}{Bibliography}
\bibliographystyle{ieeetr}



\end{document}